\documentclass[11pt,a4paper]{scrartcl}
\usepackage{CLICdp}
\usepackage{CLICdp_definitions}
\title{Updated baseline for a staged Compact Linear Collider}
\clicdpdraft{2016}{010}  
\date{\today}
\collName{The CLIC and CLICdp collaborations}
\addauthor{
M.J.~Boland,
U.~Felzmann,
P.J.~Giansiracusa,
T.G.~Lucas,
R.P.~Rassool\\
\emph{University of Melbourne, Melbourne, Australia}
}

\addauthor{
C.~Balazs,
T.K.~Charles\\
\emph{Monash University, Melbourne, Australia}
}

\addauthor{
K.~Afanaciev,
I.~Emeliantchik,
A.~Ignatenko,
V.~Makarenko,
N.~Shumeiko$^{1}$\\
\emph{Belarusian State University, Minsk, Belarus}
% ...
}

\addauthor{
\begin{flushleft}
{$^{1}$}Deceased\\
% ...
\end{flushleft}
}

\abstract{The Compact Linear Collider (CLIC) is a multi-TeV high-luminosity linear \epem collider under development. 
For an optimal exploitation of its physics potential, CLIC is foreseen to be built and operated in a staged approach with three centre-of-mass energy stages ranging from a few hundred GeV up to 3\,TeV. 
The first stage will focus on precision Standard Model physics, in particular Higgs and top-quark measurements. 
Subsequent stages will focus on measurements of rare Higgs processes, as well as searches for new physics processes and precision measurements of new states, e.g. states previously discovered at LHC or at CLIC itself. 
In the 2012 CLIC Conceptual Design Report, a fully optimised 3 TeV collider was presented, while the proposed lower energy stages were not studied to the same level of detail.
This report presents an updated baseline staging scenario for CLIC. 
The scenario is the result of a comprehensive study addressing the performance, cost and power of the CLIC accelerator complex as a function of centre-of-mass energy and it targets optimal physics output based on the current physics landscape. 
The optimised staging scenario foresees three main centre-of-mass energy stages at 380\,GeV, 1.5\,TeV and 3\,TeV for a full CLIC programme spanning 22 years.
For the first stage, an alternative to the CLIC drive beam scheme is presented in which the main linac power is produced using X-band klystrons.
}

\graphicspath{ {./logos/}{./Figures/} }

\newlength{\abc}
\AtBeginDocument{%
\renewcommand{\ref}[1]{\mbox{\autoref{#1}}}

}

\begin{document}

% title pages
\pagenumbering{roman}
\begin{titlepage}
\noindent
\setlength{\unitlength}{1mm}

\begin{picture}(0.001,0.001)
\put(125,3){CERN--2016--004}
\put(125,-2){12 August 2016}

\put(0,-50){\includegraphics[width=15cm]{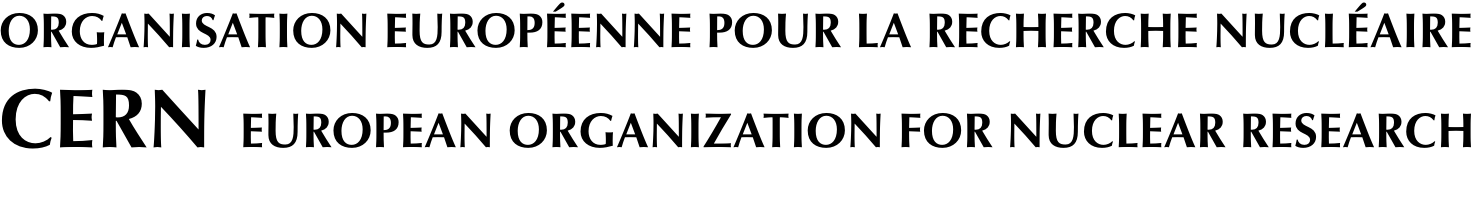}}
\put(23,-170){\includegraphics[width=11.25cm]{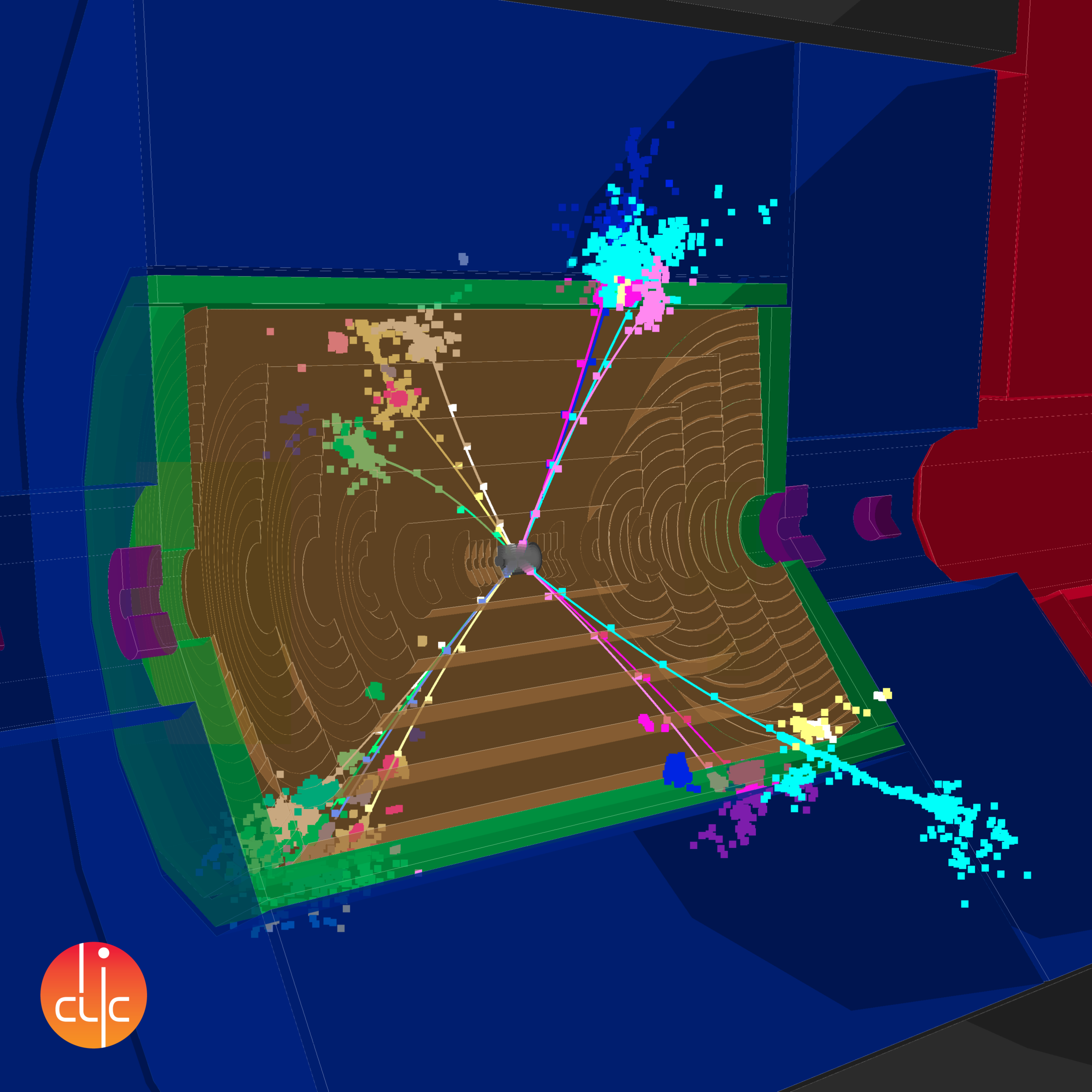}}

\put(22,-195){\huge\bfseries
             {\scshape Updated baseline for a staged}}
\put(30,-205){\huge\bfseries
             {\scshape Compact Linear Collider}}

\put(77,-235){\makebox(0,0){GENEVA}}
\put(77,-240){\makebox(0,0){2016}}
\end{picture}

\newpage
\thispagestyle{empty}
\mbox{}\\
\vfill
\begin{flushleft}%\large
\begin{tabular}{@{}l@{~}l}
ISBN & 978--92--9083--432--8 (paperback)\\
ISBN & 978--92--9083--433--5 (PDF)\\
ISSN & 0007--8328\\
DOI  & \texttt{\href{http://dx.doi.org/10.5170/CERN-2016-004}{http://dx.doi.org/10.5170/CERN-2016-004}}
\end{tabular}\\[3mm]
Available online at \texttt{\href{https://publishing.cern.ch}{https://publishing.cern.ch}} and \texttt{\href{https://cds.cern.ch}{https://cds.cern.ch}}\\[3mm]
Copyright \copyright{} CERN, 2016\\
\raisebox{-1mm}{\includegraphics[height=12pt]{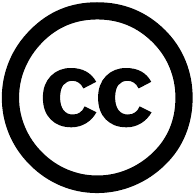}}
 Creative Commons Attribution 4.0\\
Knowledge transfer is an integral part of CERN's mission.\\
CERN publishes this report Open Access under the Creative Commons Attribution 4.0 license (\texttt{\href{http://creativecommons.org/licenses/by/4.0/}{http://creativecommons.org/licenses/by/4.0/}})
in order to permit its wide dissemination and use.\\
The submission of a contribution to a CERN Yellow Report shall be deemed to constitute the contributor's agreement to this copyright and license statement. Contributors are requested to obtain any clearances that may be necessary for this purpose.\\[3mm]
This report is indexed in: CERN Document Server (CDS), INSPIRE.\\[3mm]
This report should be cited as:\\
Updated baseline for a staged Compact Linear Collider, edited by P.N.\ Burrows, P.\ Lebrun, L.\ Linssen, D.\ Schulte, E.\ Sicking, S.\ Stapnes, M.A.\ Thomson, CERN--2016--004 (CERN, Geneva, 2016),  \texttt{\href{http://dx.doi.org/10.5170/CERN-2016-004}{http://dx.doi.org/10.5170/CERN-2016-004}} \\[3mm]

\end{flushleft}

\cleardoublepage

%Abstract
\thispagestyle{empty}
\vspace*{5cm}
\begin{center}
  \large{\bfseries\sffamily Abstract}
\end{center}
\begin{quotation}
\noindent\MyAbstract
\end{quotation}
\vspace*{5cm}

% Editors
\begin{center}
{\large {\bfseries\sffamily Corresponding editors}} 
\vspace*{0.25cm}

{
Philip N.\ Burrows (University of Oxford), Philippe Lebrun (CERN), \\Lucie Linssen (CERN), Daniel Schulte (CERN), Eva Sicking (CERN), \\Steinar Stapnes (CERN), Mark A.\ Thomson (University of Cambridge)
}
\end{center}
\end{titlepage}

% clear page with even number
\cleardoublepage
\textcolor{white}{ }
\thispagestyle{empty}
\newpage

% Collaboration Name
\begin{center}%
  \Large\bfseries\sffamily\MyCollName%
\end{center}%

% List of authors
{
\setcounter{footnote}{0}\def\@currentlabel{}%
\begingroup\def\thefootnote{\arabic{footnote}}
\def\@makefnmark{\hbox{$^{\@thefnmark)}$}}
\large
{
\MyAuthors
\par}
\endgroup

\vspace{-5mm}
\clearpage
\normalsize
\pagebreak

% clear page with even number
\cleardoublepage
\textcolor{white}{ }
\thispagestyle{empty}
\newpage

% Table of Contents
\tableofcontents

\clearpage
\textcolor{white}{ }
\thispagestyle{empty}
\clearpage
  
% sections start
\pagenumbering{arabic}
\section{Introduction}
\label{sec:Introduction}
The Compact Linear Collider (CLIC) is a multi-TeV high-luminosity linear \epem collider under development by the international CLIC collaboration~\cite{clic-study}.
It is based on a novel two-beam acceleration technique with accelerating gradients at the level of 100\,MV/m. Feasibility studies for the CLIC accelerator have systematically and successfully addressed the main technical challenges of the accelerator concept, while detailed detector and physics studies, carried out by the CLIC detector and physics (CLICdp) collaboration~\cite{clicdp}, have confirmed the ability to perform high-precision measurements at CLIC. The results have been documented in the CLIC Conceptual Design Report (CDR)~\cite{CLICCDR_vol1,cdrvol2, CLICCDR_vol3}.

The CDR was finalised in 2012, just before the discovery of the Higgs boson at the Large Hadron Collider (LHC)~\cite{Aad:2012tfa, Chatrchyan:2012ufa}.
Hence, for the CDR the Higgs mass was not fully taken into account in the choice of the CLIC energy staging. The CDR presented an accelerator complex optimised for 3\,TeV with  two initial lower-energy stages, projected at 500\,GeV and 1.4/1.5\,TeV~\cite{CLICCDR_vol3}, which were not technically optimised to the same degree.  

Since then comprehensive studies have addressed performance, cost and power optimisation of the accelerator complex as a function of energy. 
In parallel further Higgs and top-quark physics studies have been carried out~\cite{CLICHiggsPaper,Thomson:2015jda,topLC}, in particular for the initial-energy stage. 

Recent implementation studies for CLIC have converged towards a staged approach offering a comprehensive physics programme that may span several decades of exploration. 
In this scheme, CLIC would provide high-luminosity \epem collisions covering a centre-of-mass energy range from 380\,GeV to 3\,TeV. 
For a given nominal centre-of-mass energy, the energy can be tuned down by up to a third, with limited loss of luminosity performance~\cite{CLICCDR_vol3}. 
As additional LHC data are collected, new physics results may call for further adjustment of the two envisaged higher energy stages of CLIC.

In line with the European Strategy for Particle Physics, the ongoing CLIC study aims to provide for the next Strategy update (expected to take place in 2019--2020) a project implementation plan for CLIC construction.

This report presents an updated CLIC staging scenario, where the initial stages are adapted to the currently-known physics landscape and for which a more thorough technical optimisation has been carried out. 

The first stage is proposed to be at 380\,GeV. It gives access to Higgs boson measurements through the Higgsstrahlung and $\PW\PW$-fusion production processes, thereby providing accurate model-independent measurements of Higgs couplings to both fermions and bosons~\cite{CLICHiggsPaper}. This stage also addresses precision top-quark physics and includes devoting approximately 15\% of the running time to a threshold scan of top pair production in the vicinity of 350\,GeV.

The second stage around 1.5\,TeV opens the energy frontier, allowing for the discovery of new physics phenomena, while also giving access to additional Higgs and top-quark properties such as the top-Yukawa coupling, the Higgs self-coupling and rare Higgs branching ratios. 
The third stage at 3\,TeV further enlarges the CLIC physics potential. 
It will give direct access to the discovery and accurate measurements of pair-produced particles, typically with a mass up to 1.5\,TeV or single particles with a mass up to 3\,TeV. 
New electroweak particles or dark matter candidates are of special interest, as they may be easier to observe at CLIC than at the LHC. 
Furthermore, the 3\,TeV stage provides the best sensitivity to new physics processes at much higher energy scales via indirect searches.
Beam polarisation (80\% $\Pem$ polarisation) can help to constrain the underlying physics.
 
A staged implementation of CLIC as described here offers an impressive energy frontier physics programme that reaches beyond the LHC. It is an excellent option for a post-LHC facility at CERN.

\ref{sec:physics} of this report summarises the physics potential of CLIC for Higgs, top quark and Beyond-Standard Model (BSM) physics and justifies the choice of 380\,GeV for the initial energy stage.
\ref{sec:AcceleratorOptimisation} describes the methodology of the accelerator optimisation study performed after the CDR.
\ref{sec:StagingBaseline} describes the implementation of the improved CLIC staging baseline with 380\,GeV as the first energy stage. It includes detailed descriptions of the accelerator layout and estimates of the energy consumption and cost. 
Further plans towards additional optimisations of the accelerator design are also outlined.
\ref{sec:alternative} describes an alternative klystron-based scenario for the first CLIC energy stage.
\ref{sec:summary} concludes with a summary and outlook.

\section{CLIC physics}
\label{sec:physics}

\subsection{CLIC detector and experimental conditions}
\label{sec:physics:detector}

CLIC has a broad physics potential covering precision Higgs boson  and top-quark physics, direct searches for BSM physics as well as indirect BSM searches through precision observables. This potential has been assessed through the simulation and reconstruction of benchmark physics processes in two dedi\-cated CLIC detector concepts and results have been reported in several summary documents~\cite{cdrvol2,CLICCDR_vol3,Abramowicz:1563377}. These detector concepts, \clicsid\ and \clicild, are based on the
SiD~\cite{Aihara:1216387} and ILD~\cite{Abe:1272772} detector concepts for the
International Linear Collider (ILC)~\cite{Behnke:2013xla} and have been adapted to 
the experimental environment at CLIC, which is characterised by challenging conditions imposed by the high centre-of-mass energies and by the CLIC accelerator technology. 
In the multi-TeV region this leads to an environment with beam-induced background levels that are relatively high for a lepton collider. 

CLIC will deliver bunch trains with a repetition rate of 50\,Hz. Each bunch train consists of 312 individual bunches, with 0.5\,ns between bunch crossings at the 
interaction point. The beam-crossing angle is 20\,mrad in the horizontal plane. On average there is less than one $\Pep\Pem$ interaction per bunch-train crossing. However, for CLIC operation at $\sqrt{s}\,>\,1\,$\tev the 
crossing of highly-focussed intense-beams leads to significant radiation of photons. This ``beamstrahlung'' process 
results in high rates of incoherent electron-positron pairs and low-$\pT$ $t$-channel \gghadrons events.
In addition, the energy loss through beamstrahlung generates a long lower-energy tail to the luminosity spectrum~\cite{CLICCDR_vol1,cdrvol2}. 
The CLIC detector design and the event reconstruction techniques employed are both optimised to mitigate the influence of these backgrounds. The accelerator design foresees 80\% electron polarisation. Provisions are made for adding the equipment for positron polarisation if required.

The CLIC detector concepts are optimised for the precise reconstruction of complex final states in the multi-TeV region and in an environment with high background levels. 
For most sub-detector systems, the 3\,TeV detector design is suitable also for the lower-energy stages, with one notable exception being the inner tracking and vertex detectors. 
The lower backgrounds at sub-TeV energies enable these detectors to be placed at a smaller radius with respect to the beam~\cite{cdrvol2}, offering enhanced capabilities for jet flavour tagging. 
The key performance parameters of the CLIC detector concepts are:
\begin{itemize}
\item excellent track-momentum resolution, at the level of $\sigma_{\pT}/\pT^2 \lesssim 2\cdot 10^{-5}~\textrm{GeV}^{-1}$\,;
\item precise impact-parameter resolution, at the level of $\sigma^{2}_{d_{0}}=(5~\micron)^{2}+(15~\micron \,\text{GeV})^{2}/p^{2}\sin^{3}{\theta}$, to provide accurate vertex reconstruction, enabling flavour-tagging with clean $\PQb$-, $\PQc$- and light-quark jet separation;
\item jet-energy resolution $\sigma_{E}/E \lesssim 3.5\%$ for jet energies in the range 100\,GeV to 1\,TeV ($\lesssim 5\%$ at 50\,GeV);
\item detector coverage for electrons and photons extending to very low polar angles ($\sim$10\,mrad) with respect to the outgoing beam axes\,, e.g. to maximise background rejection.
\end{itemize}

A main design driver for the CLIC and ILC detector concepts is the required jet-energy resolution. As a result, the CLIC detector concepts \clicsid\ and \clicild 
are based on fine-grained electromagnetic and hadronic calorimeters, optimised for particle-flow analysis techniques~\cite{Thomson:2009rp,Marshall:1664557}. 
In the particle-flow approach, the aim is to reconstruct the individual, visible final-state particles within a jet using information from the tracking detectors combined with that from the highly granular calorimeters. In addition, particle-flow event reconstruction provides a powerful tool for the rejection of beam-induced 
backgrounds~\cite{cdrvol2}. The CLIC detector concepts employ strong central solenoid magnets, located outside the hadronic calorimeter, providing an axial magnetic field of 5\,T in \clicsid\ and 4\,T in \clicild. The \clicsid\ concept employs central silicon-strip tracking detectors, whereas \clicild\ assumes a large
central gaseous Time Projection Chamber. In both concepts the central tracking system is augmented with silicon-based inner vertex and tracking detectors. 
The LumiCal and BeamCal, two compact electromagnetic calorimeters, allow the measurement of electrons and photons down to approximately 10\,mrad in polar angle.
The two detector concepts each have an overall height of 14\,m and a length of 13\,m; they are described in more detail in~\cite{cdrvol2}. 

A study is almost completed to define a single optimised CLIC detector. 
The new concept incorporates lessons learnt from \clicsid\ and \clicild and will be used for future benchmark CLIC physics studies.

\subsection{CLIC energy stages and their impact on the physics potential}
\label{sec:stages_physics_impact}

For optimal exploitation of its physics capabilities CLIC is foreseen to follow a staged construction and operation scenario~\cite{CLICCDR_vol3}. 
The aim is to choose the energy stages such that the luminosity performance and physics potential are maximised. Therefore a comprehensive set of physics benchmark studies at the energies listed in \autoref{tab:BenchmarkEnergies} was performed.
Based on the current physics landscape and the results of the physics benchmark studies and accelerator optimisation studies, the optimal choice for the CLIC energy stages is shown to be 380\,GeV, 1.5\,TeV and 3\,TeV (see \autoref{sec:physicssummary}), with some additional running time at the first stage devoted to a $\PQt\PAQt$ threshold scan near 350\,GeV. 
\begin{table}
\caption{
CLIC energy stages investigated in  the physics benchmark studies discussed in \autoref{sec:Higgs_physics}, \autoref{sec:top} and \autoref{sec:bsm}. 
The optimal staging scenario resulting from these studies is presented in \autoref{sec:physicssummary}, \autoref{tab:stagingproposal}.}
\centering
 \begin{tabular}{lrrr}
\toprule
Stage                 &       $\sqrt{s}$ (GeV)         &   \LumiInt (\fbinv)\\
\midrule
1                     & $250 / 350 / 380 / 420 / 500$  &         $500+100$  \\
2                     &                  1400          &              1500  \\
3                     &                  3000          &              2000  \\
\bottomrule
 \end{tabular}
 \label{tab:BenchmarkEnergies}
\end{table}

The first energy stage should give access to Standard Model (SM) Higgs physics and top-quark physics and should provide the possibility of performing a $\PQt\PAQt$ threshold scan. 
The second energy stage should offer direct sensitivity to many BSM models. It provides higher Higgs statistics, allowing for measurements of rare Higgs decays, $\PQt\PAQt\PH$ production and double-Higgs production. 
The third energy stage gives the best sensitivity to new physics and double-Higgs production, therefore allowing for improved measurements of the Higgs self-coupling and $\PH\PH\PW\PW$ quartic coupling. 

For the physics benchmark studies integrated luminosities of $500\,\invfb$, $1.5\,\invab$ and $2\,\invab$ were assumed for the three energy stages (\autoref{tab:BenchmarkEnergies}). Near the $\PQt\PAQt$ production threshold, around 350\,\gev, $100\,\invfb$ was assumed for an energy scan for an accurate measurement of the top-quark mass. These integrated luminosities correspond to CLIC operation for 4 to 5 years at each stage, with additional time reserved at each stage for accelerator commissioning and luminosity ramp up in the first years~\cite{CLICCDR_vol3}.

To characterise the CLIC physics performance, physics events were generated using the \clicsid and \clicild detectors, taking realistic experimental conditions into account. These include the luminosity spectrum at different collision energies and the pileup of \gghadrons events according to the time structure of the CLIC beam~\cite{cdrvol2}.
The WHIZARD 1.95 event generator~\cite{Kilian:2007gr,Moretti:487440} was employed in most cases, followed by PYTHIA 6.4~\cite{Sjostrand:2006za} for hadronisation and TAUOLA~\cite{Was:2000st} for tau decays. The effects of Initial State Radiation (ISR) are included in WHIZARD. GEANT4~\cite{Agostinelli:602040,Allison2006} was used to simulate the detector response. The MARLIN~\cite{MarlinLCCD} framework was used for the digitisation and reconstruction of events simulated with the \clicild detector, while the org.lcsim~\cite{Graf:2011zzc} framework was used to digitise detector hits and perform track reconstruction in the \clicsid detector. Particle flow reconstruction was performed using PandoraPFA~\cite{Thomson:2009rp,Marshall:1664557}. FASTJET was used for jet clustering~\cite{Cacciari:2011ma}, where the longitu\-dinally invariant $k_{\text{T}}$ algorithm~\cite{Catani:1993hr,Ellis:1993tq} in the exclusive mode, which includes two additional beam jets, was found to give good performance. LCFIPlus~\cite{Suehara:2015ura} was used for flavour-tagging. The event simulation and reconstruction of the large data samples used in the studies was performed using the iLCDirac grid submission tools~\cite{ILCDIRAC}.

The results of these physics benchmark studies are presented in \autoref{sec:Higgs_physics} (Higgs physics), in \autoref{sec:top} (top physics) and in \autoref{sec:bsm} (BSM scenarios). The collective results allowed us to select an optimal staging scenario based on our current knowledge. This is presented in \autoref{sec:physicssummary} and in \autoref{sec:StagingBaseline}. The staging scenario can be further adapted to future discoveries from the LHC or elsewhere.

\subsection{Higgs physics}
\label{sec:Higgs_physics}

A high-energy $\Pep\Pem$ collider provides a clean experimental environment to study the properties of the Higgs boson with high precision~\cite{CLICHiggsPaper}. 
The evolution of the leading-order $\epem$ SM Higgs production cross sections with the centre-of-mass energy is shown in \autoref{fig:higgs:cross} for a Higgs boson mass of 126\,GeV~\cite{PDG2014}.

\begin{figure}[t]
  \centering
 \includegraphics[width=0.5\columnwidth]{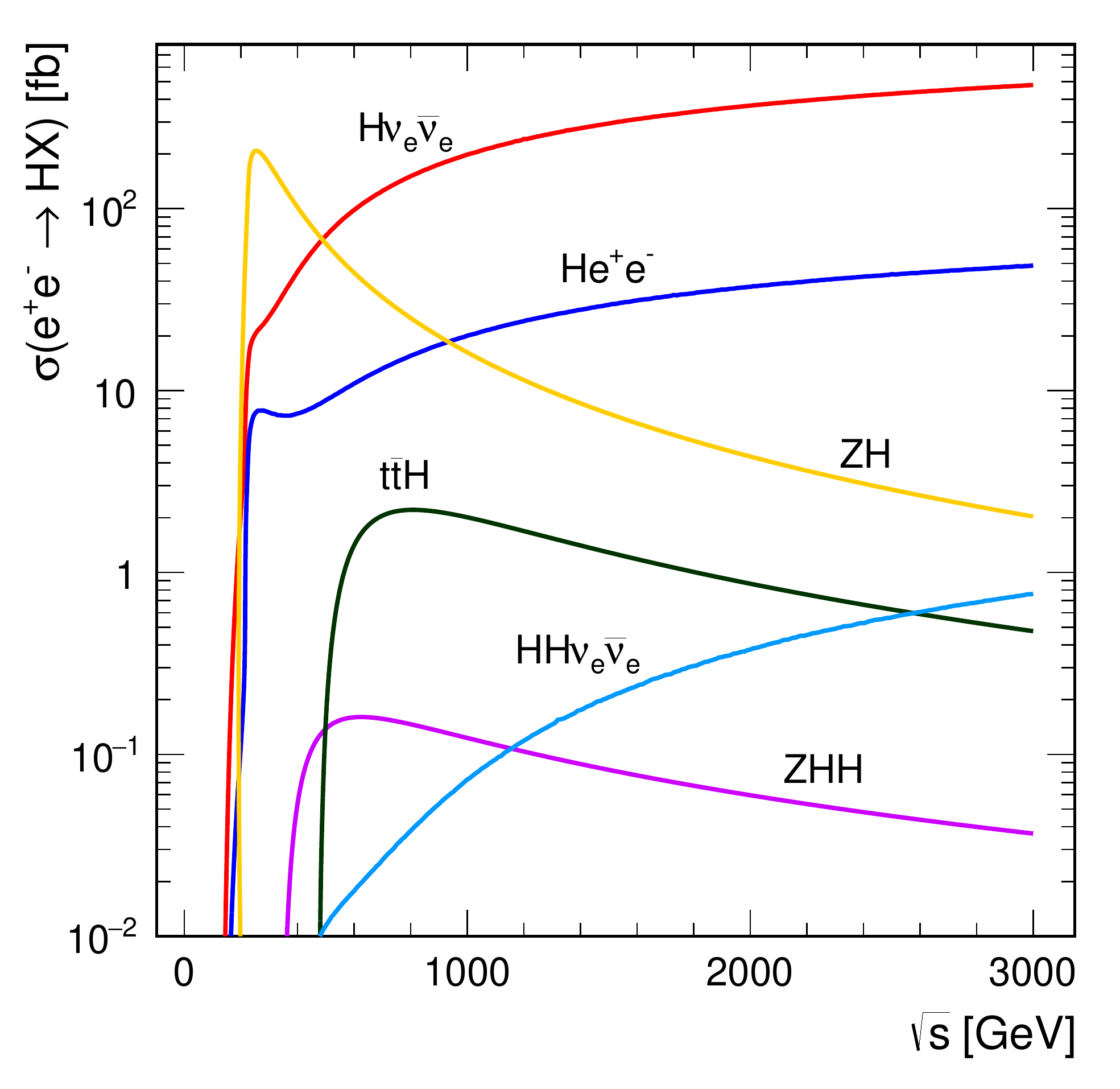}
 \caption{The centre-of-mass dependencies of the cross sections for the main Higgs production processes at an $\Pep\Pem$ collider. The values shown correspond to unpolarised beams and do not include the effect of beamstrahlung.
 \label{fig:higgs:cross}}
\end{figure}

The Feynman diagrams for the three highest cross section Higgs production processes are shown in \autoref{fig:higgs:eezh}. 
Around $\roots = 350\,\GeV$ the \higgsstrahlung process ($\Pep\Pem\to\PZ\PH$) has the
largest cross section, but the $\PW\PW$-fusion process ($\Pep\Pem\to\PH\PGne\PAGne$) is also significant. The combined study of these two processes probes the Higgs boson properties (width and branching ratios) in a model-independent manner.   
Around $\roots = 1.4$\,TeV and at 3\,TeV Higgs production is dominated 
by the $\PW\PW$-fusion process, with the $\PZ\PZ$-fusion process ($\Pep\Pem\to\PH\Pep\Pem$) also becoming significant. 
Here the relatively large $\PW\PW$-fusion cross section, combined with the high luminosity of CLIC, results in large data samples, 
allowing precise (${\cal{O}}(1\%)$) measurements of the couplings of the Higgs boson to both fermions and gauge bosons.
In addition, rarer processes such as
$\Pep\Pem\to\PQt\PAQt\PH$ and $\Pep\Pem\to\PH\PH\PGne\PAGne$ (\autoref{fig:higgs:lambda}) 
provide access to the top Yukawa coupling and the Higgs trilinear self-coupling as determined by the parameter $\lambda$ in the Higgs potential. 
In the SM, the Higgs boson originates from a doublet of complex scalar fields $\phi$ described by the potential
\begin{equation}
       V(\phi) = \mu^2\phi^\dagger\phi + \lambda(\phi^\dagger\phi)^2 \,,
\label{eq:HiggsPotential}       
\end{equation}
where $\mu$ and $\lambda$ are the parameters of the Higgs potential.

\begin{figure}[t]
\unitlength = 1mm
\vspace{6mm}
\centering
\begin{fmffile}{./Figures/Physics/Higgs/higgs_production/eezh}
\begin{fmfgraph*}(25,20)
\fmfstraight
\fmfleft{i1,i2}
\fmfright{o1,o2}
\fmflabel{$\Pem$}{i1}
\fmflabel{$\Pep$}{i2}
\fmflabel{$\PZ$}{o2}
\fmflabel{$\PH$}{o1}
\fmf{photon,tension=1.0,label=$\PZ$}{v1,v2}
\fmf{fermion,tension=1.0}{i1,v1,i2}
\fmf{photon,tension=1.0}{o2,v2}
\fmf{dashes,tension=1.0}{v2,o1}
\fmfdot{v1}
\fmfdot{v2}
\end{fmfgraph*}
\end{fmffile}
\hspace{10mm}
\begin{fmffile}{./Figures/Physics/Higgs/higgs_production/eevvh}
\begin{fmfgraph*}(25,20)
\fmfstraight
\fmfleft{i1,i2}
\fmfright{o1,oh,o2}
\fmflabel{$\Pem$}{i1}
\fmflabel{$\Pep$}{i2}
\fmflabel{$\PAGne$}{o2}
\fmflabel{$\PH$}{oh}
\fmflabel{$\PGne$}{o1}
\fmf{fermion, tension=2.0}{i1,v1}
\fmf{fermion, tension=1.0}{v1,o1}
\fmf{fermion, tension=1.0}{o2,v2}
\fmf{fermion, tension=2.0}{v2,i2}
\fmf{photon, lab.side=right,lab.dist=1.5,label=$\PW$,tension=1.0}{v1,vh}
\fmf{photon, lab.side=right, lab.dist=1.5,label=$\PW$,tension=1.0}{vh,v2}
\fmf{dashes, tension=1.0}{vh,oh}
\fmfdot{vh}
\end{fmfgraph*}
\end{fmffile}
\hspace{10mm}
\begin{fmffile}{./Figures/Physics/Higgs/higgs_production/eeeeh}
\begin{fmfgraph*}(25,20)
\fmfstraight
\fmfleft{i1,i2}
\fmfright{o1,oh,o2}
\fmflabel{$\Pem$}{i1}
\fmflabel{$\Pep$}{i2}
\fmflabel{$\Pep$}{o2}
\fmflabel{$\PH$}{oh}
\fmflabel{$\Pem$}{o1}
\fmf{fermion, tension=2.0}{i1,v1}
\fmf{fermion, tension=1.0}{v1,o1}
\fmf{fermion, tension=1.0}{o2,v2}
\fmf{fermion, tension=2.0}{v2,i2}
\fmf{photon, lab.side=right,lab.dist=1.5,label=$\PZ$,tension=1.0}{v1,vh}
\fmf{photon, lab.side=right, lab.dist=1.5,label=$\PZ$,tension=1.0}{vh,v2}
\fmf{dashes, tension=1.0}{vh,oh}
\fmfdot{vh}
\end{fmfgraph*}
\end{fmffile}
\vspace{5mm}
\caption{The three highest cross section Higgs production processes at CLIC.\label{fig:higgs:eezh}}
\end{figure}
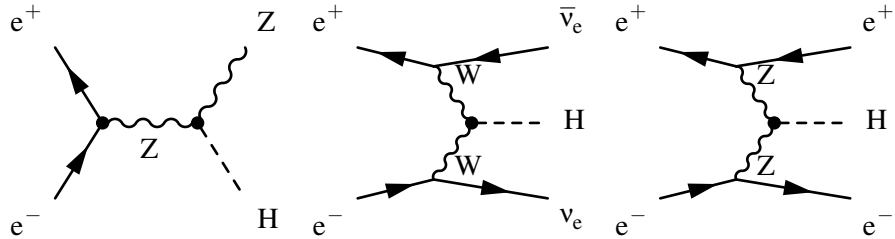

\begin{figure}[t]
\unitlength = 1mm 
\vspace{5mm}
\centering
     \begin{fmffile}{./Figures/Physics/Higgs/higgs_production/eetth} 
     \begin{fmfgraph*}(25,20)
        \fmfstraight
         \fmfleft{i1,i2}  
         \fmfright{o1,oh,o2}
         \fmflabel{$\Pem$}{i1}
        \fmflabel{$\Pep$}{i2}  
         \fmflabel{$\PQt$}{o2} 
         \fmflabel{$\PH$}{oh}
         \fmflabel{$\PAQt$}{o1}
          \fmf{photon,label=$\PZ$,tension=2.0}{v1,v2}
          \fmf{fermion,tension=1.0}{i1,v1,i2}
           \fmf{phantom,tension=1.0}{o1,v2,o2}
            \fmffreeze
           \fmf{fermion,tension=1.0}{o1,v2}
           \fmf{fermion,tension=1.0}{v2,vh,o2}
           \fmf{dashes,tension=0.0}{vh,oh}
           \fmfdot{vh}
	   \end{fmfgraph*}
	   \end{fmffile}
	   \hspace{10mm} 
	 \begin{fmffile}{./Figures/Physics/Higgs/higgs_production/eevvhh}
        \begin{fmfgraph*}(25,20)
            \fmfleft{i1,i2}  
            \fmfright{o1,oh1,oh2,o2}
            \fmflabel{$\Pem$}{i1}
            \fmflabel{$\Pep$}{i2}  
            \fmflabel{$\PAGne$}{o2} 
            \fmflabel{$\PH$}{oh1}            
            \fmflabel{$\PH$}{oh2}
            \fmflabel{$\PGne$}{o1}
             \fmf{fermion, tension=2.0}{i1,v1}
              \fmf{fermion, tension=1.0}{v1,o1}
             \fmf{fermion, tension=1.0}{o2,v2}
                          \fmf{fermion, tension=2.0}{v2,i2}
             \fmf{photon,  label=$\PW$,label.dist=1.5,tension=1.0}{v1,vh}
             \fmf{photon,  label=$\PW$,label.dist=1.5,tension=1.0}{v2,vh}
             \fmf{dashes, label=$\PH$, label.dist=1.5,tension=2.0}{vh,vh1}
             \fmf{dashes,  tension=1.0}{oh1,vh1,oh2}
             \fmfdot{vh1}
        \end{fmfgraph*}
    \end{fmffile}
	   \hspace{10mm} 
	 \begin{fmffile}{./Figures/Physics/Higgs/higgs_production/ghhww}
        \begin{fmfgraph*}(25,20)
            \fmfleft{i1,i2}  
            \fmfright{o1,oh1,oh2,o2}
            \fmflabel{$\Pem$}{i1}
            \fmflabel{$\Pep$}{i2}  
            \fmflabel{$\PAGne$}{o2} 
            \fmflabel{$\PH$}{oh1}            
            \fmflabel{$\PH$}{oh2}
            \fmflabel{$\PGne$}{o1}
             \fmf{fermion, tension=2.0}{i1,v1}
              \fmf{fermion, tension=1.0}{v1,o1}
             \fmf{fermion, tension=1.0}{o2,v2}
              \fmf{fermion, tension=2.0}{v2,i2}
             \fmf{photon, label=$\PW$,label.side=right,label.dist=1.5,  tension=1.0}{vh,v1}
              \fmf{photon, label=$\PW$,label.sde=right,label.dist=1.5, tension=1.0}{v2,vh}
             \fmf{dashes,  tension=1.0}{oh1,vh,oh2}
             \fmfdot{vh}
        \end{fmfgraph*}
    \end{fmffile}
    \vspace{5mm}
\caption{The main processes at CLIC involving the  top Yukawa coupling $g_{\PH\PQt\PQt}$, the Higgs boson trilinear self-coupling $\lambda$ and the quartic
              coupling $g_{\PH\PH\PW\PW}$.
 \label{fig:higgs:lambda}}
\end{figure}
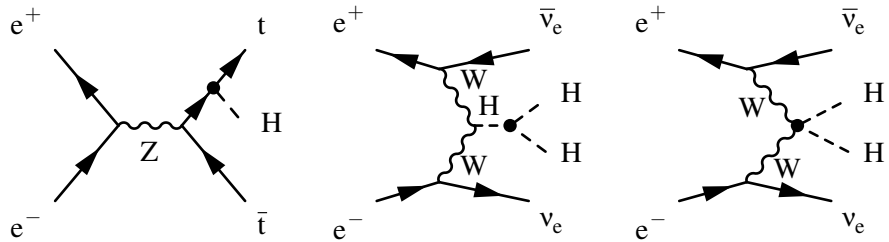

\autoref{tab:higgs:events} shows the cross sections and the expected numbers of $\PH\PZ$, $\PH\PGne\PAGne$ and $\PH\Pep\Pem$ events for the 
three studied energy stages. These numbers account for the effect of 
beamstrahlung and ISR, which result in a tail in the distribution of the effective centre-of-mass energy $\rootsprime$.

%%%%%%

\begin{table}[t]\centering
\caption{The leading-order Higgs unpolarised cross sections for the \higgsstrahlung, $\PW\PW$-fusion, and $\PZ\PZ$-fusion processes for $\mH=126\,\GeV$ at the three centre-of-mass energies studied. 
The quoted cross sections include the effects of ISR but do not include the effects of beamstrahlung. 
Also listed are the numbers of expected events including the effects of the CLIC beamstrahlung spectrum and ISR. 
The cross sections and expected numbers do not take into account the  enhancements possible with polarised beams.
}
\label{tab:higgs:events}
 \begin{tabular}{lrrr}
   \toprule 
                  $\roots$                     & \tabt{350\,GeV} & \tabt{1.4\,TeV} & \tabt{3\,TeV} \\ \midrule
    \LumiInt                             & 500\,\fbinv     & 1.5\,\abinv    & 2\,\abinv  \\
    $\sigma(\Pep\Pem\to\PZ\PH)$          & 133\,fb         & 8\,fb           & 2\,fb         \\
    $\sigma(\Pep\Pem\to\PH\PGne\PAGne)$  & 34\,fb          & 276\,fb         & 477\,fb       \\
    $\sigma(\Pep\Pem\to\PH\Pep\Pem)$     & 7\,fb           & 28\,fb          & 48\,fb        \\
    \# $\PH\PZ$ events                   & 68,000         & 20,000         & 11,000       \\
    \# $\PH\PGne\PAGne$ events           & 17,000         & 370,000        & 830,000      \\
    \# $\PH\Pep\Pem$ events              & 3,700          & 37,000         & 84,000       \\
    \bottomrule
  \end{tabular}

\end{table}

\autoref{tab:higgs:events} does not take beam polarisation into account. In fact, the majority of CLIC Higgs physics studies have been performed assuming unpolarised $\Pep$ and $\Pem$ beams. 
However, for the baseline CLIC design the electron beam can be 80\% polarised. The accelerator design is compatible with introducing positron polarisation at a lower level but this is not currently part of the baseline design.  
By selecting different beam polarisations it is possible to enhance or suppress different physical processes. The chiral nature of the weak coupling can offer 
significant enhancements in $\PW\PW$-fusion Higgs production, as indicated in \autoref{tab:higgs:polarisation}. 
The potential gains for the \higgsstrahlung process $\Pep\Pem\to\PZ\PH$ and the $\PZ\PZ$-fusion process $\Pep\Pem\to\PH\Pep\Pem$ however are less significant (\autoref{tab:higgs:polarisation}).
In practice, the balance between operation with different beam polarisations will depend on the CLIC physics programme taken as a whole, including the searches for BSM particles. For example, based on current knowledge one could imagine similar running times with both negative and positive longitudinal electron polarisation at the first energy stage in order to optimise top-quark asymmetry measurements (see \autoref{sec:top}). At the higher energy stages, principal running with negative electron polarisation is likely to be favoured in view of the significant enhancement of Higgs statistics.

\begin{table}[tb]\centering
\caption{The dependence of the event rates for the $s$-channel $\Pep\Pem\to\PZ\PH$ process and the pure $t$-channel $\Pep\Pem\to\PH\PGne\PAGne$ and $\Pep\Pem\to\PH\Pep\Pem$ processes for three example beam polarisations. 
The numbers are only approximate as they do not account for interference between $\Pep\Pem\!\to\PH\PZ\!\to\PH\PGne\PAGne$ and $\Pep\Pem\!\to\PH\PGne\PAGne$.
}
\label{tab:higgs:polarisation}
  \begin{tabular}{cccc}\toprule
    Polarisation                              & \tabttt{Enhancement factor}                        \\ \cmidrule(l){2-4}
    $P(\Pem):P(\Pep)$                  & $\!\!\Pep\Pem\!\to\PZ\PH\!\!$ & $\!\!\Pep\Pem\!\to\PH\PGne\PAGne\!\!$& $\!\!\Pep\Pem\!\rightarrow\PH\epem\!\!$ \\ \midrule
    unpolarised                                  & 1.00                & 1.00        &   1.00        \\
    $-80\%\,:\phantom{+3}\,0\%$   & 1.12                & 1.80        &    1.12             \\
    $+80\%\,:\phantom{+3}\,0\%$  & 0.88                & 0.20        &    0.88             \\
    \bottomrule
   \end{tabular}

\end{table}

Polar-angle distributions for single Higgs production at the three studies CLIC energy stages are shown in \autoref{fig:higgs:theta}. At $350\,\GeV$ most Higgs bosons are produced in the central parts of the detector, whereas at centre-of-mass energies above $1\,\TeV$ they are mainly produced in the forward regions, therefore requiring good forward detection capabilities.

\begin{figure}[t]
\centering
\includegraphics[width=0.5\columnwidth]{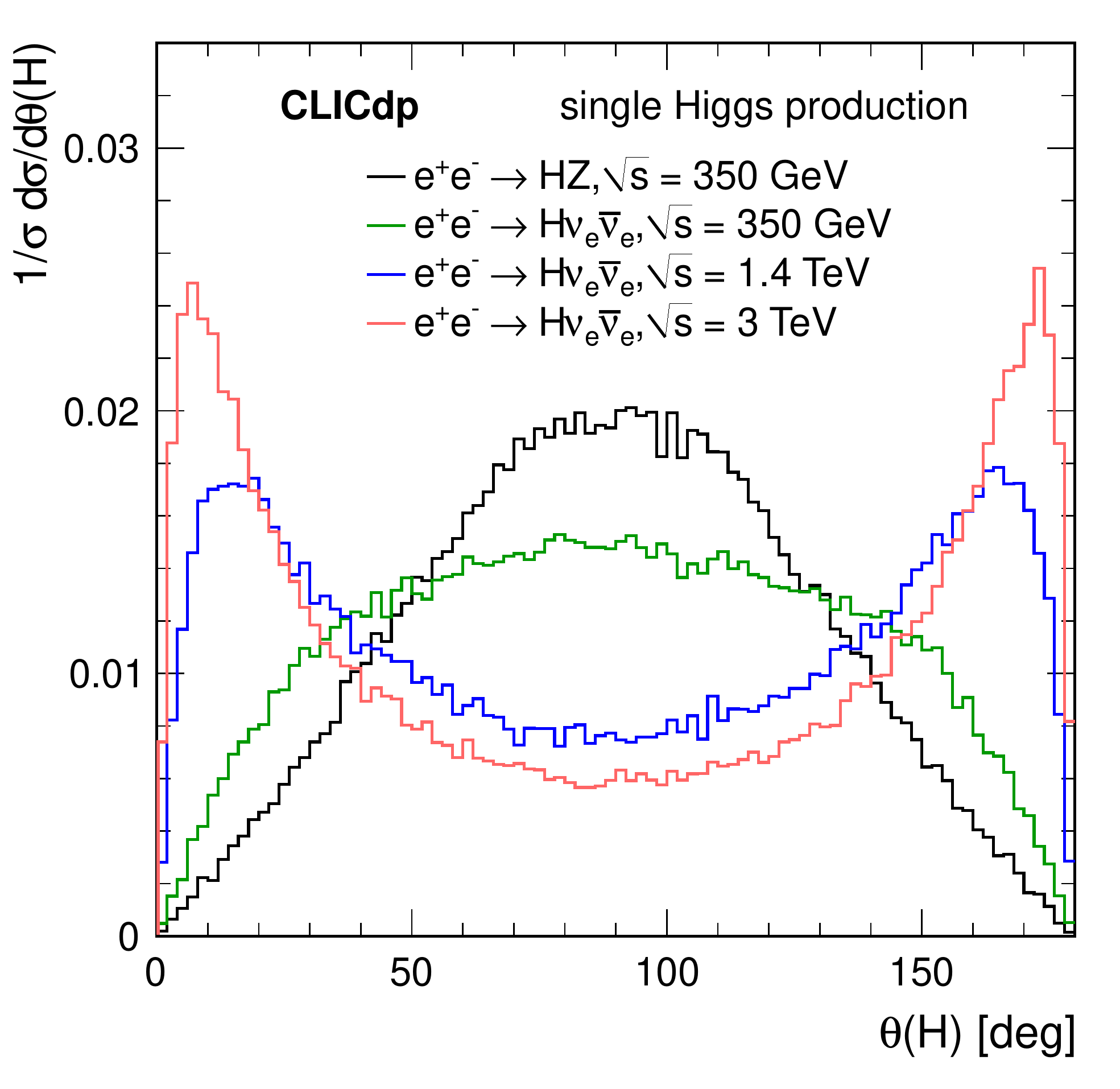}
\caption{Polar angle distributions for single Higgs production at the three studied CLIC energy stages. All distributions include the effects of the CLIC beamstrahlung spectrum and ISR. All distributions are normalised to unit area. \label{fig:higgs:theta}}
\end{figure}

%%%%%%%%%%%%%%%%%%%%%%%%%%%%%%%%%%%%%
The choice of the CLIC energy stages is motivated by the desire to pursue a programme of precision Higgs and top physics, as well as to operate the accelerator above 1\,TeV at the earliest possible time.  
The lower-energy operation is partly motivated by the \higgsstrahlung process, as it provides the opportunity to measure the couplings of the Higgs boson in a
model-independent manner, as further described below. This is unique to an electron-positron collider. The
clean experimental environment, and the relatively low SM cross sections for
background processes, allow $\Pep\Pem\to\PZ\PH$ events to be selected based solely on the measurement of the 
four-momentum of the $\PZ$ boson through its decay products. 

The most distinct event topologies occur for $\PZ\to\epem$ and $\PZ\to\mpmm$ decays, each with a branching ratio of about 3.4\%~\cite{PDG2014}, which can be identified by requiring that the di-lepton invariant mass is consistent with $\mZ$. 
The four-momentum of the system recoiling against the $\PZ$ boson can be obtained from $E_\mathrm{rec} = \roots - E_{\PZ}$ and $\vec{p}_\mathrm{rec} = -\vec{p}_{\PZ}$.  
In $\Pep\Pem\to\PZ\PH$ events the invariant mass of this recoiling system will peak at $m_{\PH}$, allowing the $\PH\PZ$ events to be selected based only on the observation of the leptons from the $\PZ$ boson decay, providing a model-independent measurement of the Higgs coupling to the $\PZ$ boson, $g_{\PH\PZ\PZ}$. 

The hadronic decay channel $\PZ\rightarrow\PQq\PAQq$ with a branching ratio of $\sim$70\%~\cite{PDG2014} has a less distinct event topology, but it provides a statistically more precise measurement of recoil mass analysis.
The combination of the leptonic and hadronic decay channels allows $g_{\PH\PZ\PZ}$ to be determined in a model-independent way with 
a precision of 0.8\% at\,350 GeV~\cite{Thomson:2015jda}. In addition, the recoil mass from $\PZ\rightarrow\PQq\PAQq$ decays provides a direct search for possible Higgs decays to invisible final states,
and can be used to constrain the invisible decay width of the Higgs, $\Gamma_\text{invis}$, to less than 1\% at 90\% confidence level~\cite{Thomson:2015jda}. 

As the HZ \higgsstrahlung production process with the Z boson decaying to $\PQq\PAQq$ provides the best accuracy for the $g_{\PH\PZ\PZ}$ measurement, it is important to understand its dependence on the CLIC centre-of-mass energy. The study was hence performed for three different centre-of-mass energies: 250\,GeV, 350\,GeV and 420\,GeV for the same integrated luminosity of 500\,\fbinv at each energy~\cite{Thomson:2015jda}. Among these, the 350\,GeV case provides the best precision, as shown in \autoref{tab:hadronicHZ}. 

\begin{table}[t]
\caption{Statistical precision achievable on $\sigma(\PH\PZ)$ from the hadronic recoil mass analysis for $\sqrt{s}=250$\,GeV, 350\,GeV and 420\,GeV, assuming for each energy 500\,\fbinv and unpolarised beams~\cite{Thomson:2015jda}.}
\label{tab:hadronicHZ}
\centering
\begin{tabular}{ccc}
\toprule 
$\roots$ & $\sigma(\PH\PZ)$ & $\Delta \,\sigma(\PH\PZ)$ \\  
\midrule       
250\,GeV &  136\,fb         & $\pm3.7\,\%$   \\  
350\,GeV &   93\,fb         & $\pm1.8\,\%$   \\ 
420\,GeV &   68\,fb         & $\pm2.6\,\%$   \\                                                                                                                                                                                                                                                                                         
\bottomrule
\end{tabular}
\end{table}

This can be understood as a trade-off between recoil mass resolution, HZ production cross section and signal-to-background ratio. 
In the hadronic channel, the recoil mass resolution degrades with increasing centre-of-mass energy, due to the $\rootsprime$ dependence of the recoil mass itself, combined with detector resolution trends. On the other hand, the signal-to-noise ratio is significantly less favourable near the HZ production threshold at 250\,\gev, where one of the dominant background processes ($\Pep\Pem\rightarrow\PQq\PAQq\PQq\PAQq$) can easily mimic HZ events. Therefore, despite the fact that the HZ production cross section is higher at 250\,\gev, the optimal $\sqrt{s}$ choice for this measurement is in the region near 350\,\gev.

The statistical accuracy of the Higgs mass determined through the recoil mass measurement for leptonic $\PZ$-decays in \higgsstrahlung events is 110\,MeV. This accuracy will be improved further by including the Higgs invariant mass reconstruction for Higgs decay channels with large branching fractions.

By identifying the individual final states for different Higgs decay modes, and by taking advantage of the accessibility of both \higgsstrahlung and Higgs production through $\PW\PW$-fusion, precise measurements of the Higgs boson branching ratios (\BR) can be made. Although the cross section is lower, the $t$-channel $\PW\PW$-fusion process $\Pep\Pem\to\PH\PGne\PAGne$ is an important part
of the Higgs physics programme around $\roots = 350\,\GeV$.

Once the Higgs coupling to the $\PZ$ boson, $g_{\PH\PZ\PZ}$, is known, the Higgs coupling to the $\PW$ boson can be determined from, for example, 
the ratios of \higgsstrahlung to $\PW\PW$-fusion cross sections,
\begin{equation*}
    \frac{\sigma(\epem\to\PZ\PH)\times \BR(\PH\to\PQb\PAQb) }{ \sigma{(\epem\to\PGne\PAGne\PH)} \times \BR(\PH\to\PQb\PAQb)  } \propto \left(\frac{g_{\PH\PZ\PZ}}{g_{\PH\PW\PW}} \right)^2 \,.
\end{equation*}

In order to determine absolute measurements of the other Higgs couplings, the Higgs total decay width needs to be inferred from the data. 
For the Higgs boson mass of 126\,GeV, the total Higgs decay width in the SM ($\Gamma_{\PH}$) is less than $5\,\MeV$ and
cannot be measured directly.  However, given that the absolute 
couplings of the Higgs boson to the $\PZ$ and $\PW$ bosons can be obtained as described above, the total decay width of the 
Higgs boson can be determined from $\PH\to\PW\PW^*$ or $\PH\to\PZ\PZ^*$ decays. For example, the measurement of the 
Higgs decay to $\PW\PW^*$ in the $\PW\PW$-fusion process determines
\begin{equation*}
                 \sigma(\PH\PGne\PAGne)\times \BR(\PH\to\PW\PW^*)   \propto \frac{g^4_{\PH\PW\PW}}{\Gamma_{\PH}}\,,
\end{equation*}
and thus the total width can be determined utilising the model-independent measurement of $g_{\PH\PW\PW}$. In practice a fit (see \autoref{sec:combined_fits})
was performed to all of the experimental measurements involving Higgs boson couplings. For the first CLIC energy stage the fit is based on the 350 GeV data set. At the higher  energies significantly more Higgs data from the $\PW\PW$-fusion process become available, profiting from the fact that this Higgs production process as well as the CLIC luminosity increase with \roots. These data also provide a more accurate measurement of the Higgs mass, for example through invariant mass reconstruction in $\PH\to\PQb\PAQb$ decays. This allows determination of the Higgs mass with a statistical accuracy of $47\,\MeV$ at 1.4\,TeV centre-of-mass energy and $44\,\MeV$ at 3\,TeV. 
Combining these results yields a statistical accuracy of $32\,\MeV$, or $24\,\MeV$ when using 80\% electron beam polarisation~\cite{CLICHiggsPaper}.

Although the $\PW\PW$-fusion process has the largest cross section for Higgs production above 1\,TeV, other processes are also important.
For example, measurements of the $\PZ\PZ$-fusion process provide further constraints on the $g_{\PH\PZ\PZ}$ coupling. 
Furthermore, operation at $\roots = 1.4\,\tev$ enables a determination of 
the top Yukawa coupling from the process $\Pep\Pem\to\PQt\PAQt\PH\to\PQb\PW^+\PAQb\PW^-\PH$ in which a top quark radiates a Higgs boson. In the SM this coupling is the strongest Higgs-fermion coupling, as the top quark is the most massive fermion. At 1.4\,TeV this coupling can be determined with a statistical accuracy of 
4.1\%~\cite{Redford:1690648} when using 80\% electron beam polarisation.

Finally, the self-coupling of the Higgs boson at the $\PH\PH\PH$ vertex is measurable in $1.4\,\tev$ and $3\,\tev$ operation. 
After spontaneous symmetry breaking the form of the Higgs potential given in \autoref{eq:HiggsPotential} gives rise to a trilinear Higgs self-coupling of strength proportional to $\lambda v$, where 
$v$ is the vacuum expectation value of the Higgs potential. The measurement of the strength of the Higgs self-coupling therefore provides direct access to 
the coupling $\lambda$ assumed in the Higgs mechanism. This measurement is 
an essential part of experimentally establishing the Higgs mechanism as described by the SM. 
The measurement of the Higgs boson self-coupling at the LHC will be extremely challenging even with $3000\,\fbinv$ of data (see for example~\cite{Dawson:2013bba}). 
At a linear collider, the trilinear Higgs self-coupling can be measured through the $\Pep\Pem\to\PZ\PH\PH$ and $\Pep\Pem\to\PH\PH\PGne\PAGne$ processes.
The achievable precision  has been studied for the $\Pep\Pem\to\PZ\PH\PH$ process at 
$\roots=500\,\GeV$ in the context of the ILC, where the results show that a very large integrated luminosity 
is required to reach even a moderate precisions of approximately 30\%~\cite{Fujii:2015jha}. 
For this reason, the most favourable channel for the measurement of the Higgs self-coupling is the $\Pep\Pem\to\PH\PH\PGne\PAGne$ process at 
$\roots\ge 1\,\TeV$. 
In this channel, the sensitivity increases with increasing centre-of-mass energy.
The measurement of the Higgs boson self-coupling forms a central part of the CLIC Higgs physics programme; ultimately a precision of approximately 10\% on $\lambda$ can be achieved~\cite{CLICHiggsPaper}.

\subsubsection{Combined Higgs fits}
\label{sec:combined_fits}

From the Higgs cross section and branching ratio measurements at the three studied CLIC energy stages, the Higgs coupling parameters and
total width are extracted by a global fit~\cite{Simon:1603687,CLICHiggsPaper}. 
Here, a $-80\%$ electron polarisation is assumed at $1.4\,\TeV$ and $3\,\TeV$, while no beam polarisation is assumed at 350\,GeV. 
The corresponding increase in cross section is taken into account by multiplying the event rates by a factor of 1.8 (see \autoref{tab:higgs:polarisation}).
This approach is conservative since it assumes that all backgrounds including those from $s$-channel processes, which do not receive the same enhancement by polarisation, scale by the same factor.

Some of the Higgs cross section and branching ratio analyses in the $\PW\PW$-fusion and $\PZ\PZ$-fusion processes were so far only performed for 1.4\,TeV centre-of-mass energy. 
For these channels, the uncertainties on the measurements estimated at 1.4\,TeV were extrapolated to 3\,TeV, based on scaling of the integrated luminosity.
In addition to this scaling, the signal fraction within the detector acceptance, which changes between the centre-of-mass energies, is taken into account for each Higgs decay channel individually.

Two types of fits were used: a model-independent fit making no additional assumptions, and a model-dependent fit following the strategies used for the interpretation of LHC Higgs results.
To obtain the precision expected at CLIC, it was assumed that the value expected in the SM has been observed throughout.
In the fit only statistical uncertainties were used.
Correlations between measurements were taken into account in cases where they are expected to be large. 
This applies to the measurements of $\sigma \times \BR$ for $\PH \to \bb, \cc, \Pg\Pg$ in \higgsstrahlung and $\PW\PW$-fusion events at 350\,GeV, which were extracted simultaneously from the data in a combined fitting procedure.

The model-independent fit was performed with eleven free parameters: the Higgs couplings $\gHZZ$,
$\gHWW$, $\gHbb$, $\gHcc$, $\gHTauTau$, $\gHMuMu$, $\gHtt$, the Higgs total width,
$\Gamma_{\PH}$, as well as the three effective couplings
$g^\dagger_\mathrm{\PH\Pg\Pg}$, $g^\dagger_{\PH\PGg\PGg}$ and $g^\dagger_{\PH\PZ\PGg}$. The
latter three parameters were treated in the same way as the physical
Higgs couplings in the fit.

\begin{table}[t]\centering
\caption{
Results of the model-independent Higgs fit. Values marked ``$-$'' can not be measured with sufficient precision at the given energy. 
The three effective couplings $g^\dagger_{\PH\Pg\Pg}$, $g^\dagger_{\PH\PGg\PGg}$ and $g^\dagger_{\PH\PZ\PGg}$ are also included in the fit. 
}
\label{tab:MIResults}
\begin{tabular}{lrrr}
\toprule
Parameter & \multicolumn{3}{c}{Relative precision}\\
\midrule
                          & $350\,\GeV$   & + $1.4\,\TeV$   & + $3\,\TeV$   \\
                          & $500\,\fbinv$ & + $1.5\,\abinv$ & + $2\,\abinv$ \\
\midrule
$\gHZZ$                   & 0.8\%         &  0.8\%          & 0.8\%         \\
$\gHWW$                   & 1.3\%         &  0.9\%          & 0.9\%         \\
$\gHbb$                   & 2.8\%         &  1.0\%          & 0.9\%         \\
$\gHcc$                   & 6.0\%         &  2.3\%          & 1.9\%         \\
$\gHTauTau$               & 4.2\%         &  1.7\%          & 1.4\%         \\
$\gHMuMu$                 & $-$           & 14.1\%          & 7.8\%         \\
$\gHtt$                   & $-$           &  4.4\%          & 4.4\%         \\
\midrule
$g^\dagger_{\PH\Pg\Pg}$   & 3.6\%         &  1.7\%          & 1.4\%         \\
$g^\dagger_{\PH\PGg\PGg}$ & $-$           &  5.7\%          & 3.2\%         \\
$g^\dagger_{\PH\PZ\PGg}$  & $-$           & 15.6\%          & 9.1\%         \\
\midrule
$\Gamma_{\PH}$            & 6.4\%         &  3.7\%          & 3.6\%         \\
\bottomrule
\end{tabular}
\end{table}

\begin{figure}[t]
  \centering
  \includegraphics[width=0.55\columnwidth]{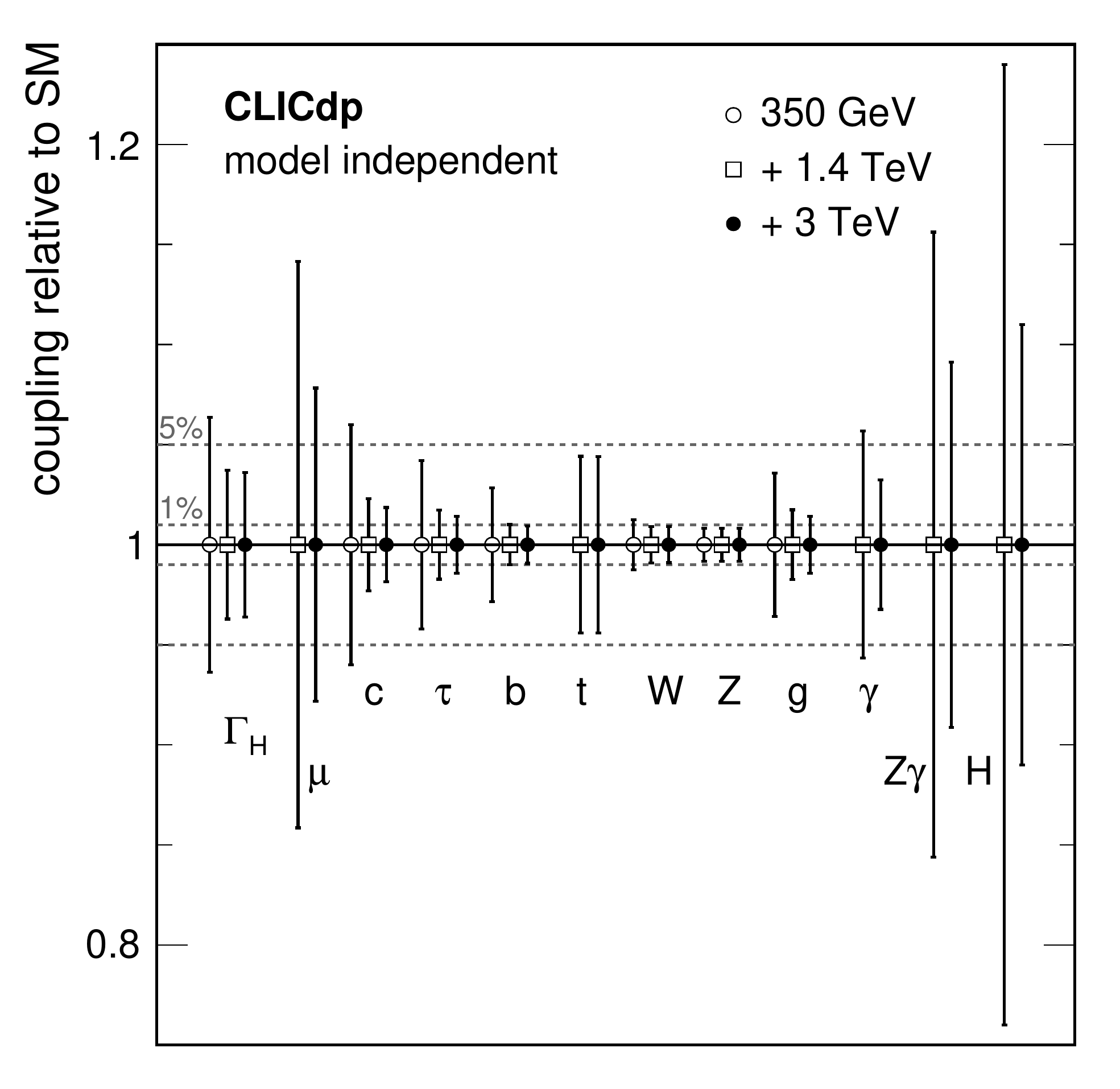}
  \caption{Illustration of the precision of
    the Higgs couplings of the studied three-stage CLIC programme determined
    in a model-independent fit. \label{fig:combinedFit:MI}}
\end{figure}

The fit was performed in three stages, taking the statistical uncertainties obtainable from CLIC at the three considered energy stages successively into account. 
The fits for subsequent stages also include all measurements of the previous stages. 
\autoref{tab:MIResults} summarises the results. 
They are graphically illustrated in \autoref{fig:combinedFit:MI}. 
Since the model-independence of the analysis hinges on the absolute measurement of $\sigma(\PZ\PH)$ at $350\,\GeV$, which provides the coupling $\gHZZ$, the precision of all other couplings is ultimately limited by this uncertainty.

For the model-dependent fit, it is assumed that the Higgs decay
properties can be described by ten independent parameters
$\kappa_{\PH\PZ\PZ}$, $\kappa_{\PH\PW\PW}$, $\kappa_{\PH\PQb\PQb}$,
$\kappa_{\PH\PQc\PQc}$, $\kappa_{\PH\PGt\PGt}$,
$\kappa_{\PH\PGm\PGm}$, $\kappa_{\PH\PQt\PQt}$, $\kappa_{{\PH\Pg\Pg}}$, $\kappa_{\PH\PGg\PGg}$
and $\kappa_{\PH\PZ\PGg}$. These factors are defined by the ratio of
the Higgs partial width divided by the partial width expected in the
Standard Model as
\begin{equation}
\kappa_i^2 = \Gamma_i/\Gamma_i^{\text{SM}}\,.
\end{equation}

In this scenario the total width is given by the sum of the ten
partial widths considered, which is equivalent to assuming no
invisible Higgs decays. The variation of the total width from its SM
value is thus given by
\begin{equation}
\frac{\Gamma_{\PH,\text{model dependent}}}{\Gamma_{\PH}^{\text{SM}}} = \sum_i \kappa_i^2 \ \BR_i, \label{eq:KappaWidth}
\end{equation}
where $\BR_i$ is the SM branching ratio for the respective final state.
To obtain these branching ratios, a fixed value for the Higgs mass has to be imposed. 
The theoretical uncertainties on the branching ratios taken from \cite{Dittmaier:2012vm} are ignored. 
To exclude effects from numerical rounding errors, the sum of the $\BR$s was normalised to unity.

Since at the first energy stage of CLIC no significant measurements of
the $\PH\to\mpmm$, $\PH\to\PGg\PGg$ and $\PH\to\PZ\PGg$ decays are possible, the fit is
reduced to six free parameters with an
appropriate rescaling of the branching ratios used in the total width
for $350\,\GeV$.

\begin{table}[t]\centering
\caption{Results of the model-dependent global Higgs fit. Values marked ``$-$'' cannot be measured with sufficient precision at the given energy. 
The uncertainty of the total width is calculated from the fit results, taking the parameter correlations into account. }
\label{tab:MDResults}
\begin{tabular}{lrrr}
\toprule
Parameter & \multicolumn{3}{c}{Relative precision}\\
\midrule
                                 & $350\,\GeV$   & + $1.4\,\TeV$   & + $3\,\TeV$   \\
                                 & $500\,\fbinv$ & + $1.5\,\abinv$ & + $2\,\abinv$ \\
\midrule
$\kappa_{\PH\PZ\PZ}$             &  0.57\%       &  0.37\%         & 0.34\%        \\
$\kappa_{\PH\PW\PW}$             &  1.1\%        &  0.21\%         & 0.14\%        \\
$\kappa_{\PH\PQb\PQb}$           &  2.0\%        &  0.41\%         & 0.24\%        \\
$\kappa_{\PH\PQc\PQc}$           &  5.9\%        &  2.2\%          & 1.7\%         \\
$\kappa_{\PH\PGt\PGt}$           &  3.9\%        &  1.5\%          & 1.1\%         \\
$\kappa_{\PH\PGm\PGm}$           &  $-$          & 14.1\%          & 7.8\%         \\
$\kappa_{\PH\PQt\PQt}$           &  $-$          &  4.3\%          & 4.3\%  \\
$\kappa_{\PH\Pg\Pg}$             &  3.2\%        &  1.6\%          & 1.2\%         \\
$\kappa_{\PH\PGg\PGg}$           &  $-$          &  5.6\%          & 3.1\%       \\
$\kappa_{\PH\PZ\PGg}$            &  $-$          & 15.6\%          & 9.1\%       \\
\midrule
$\Gamma_{\PH,\text{md,derived}}$ &  1.6\%        &  0.41\%         & 0.28\%        \\
\bottomrule
\end{tabular}
\end{table}

\begin{figure}[t]
  \centering
  \includegraphics[width=0.55\columnwidth]{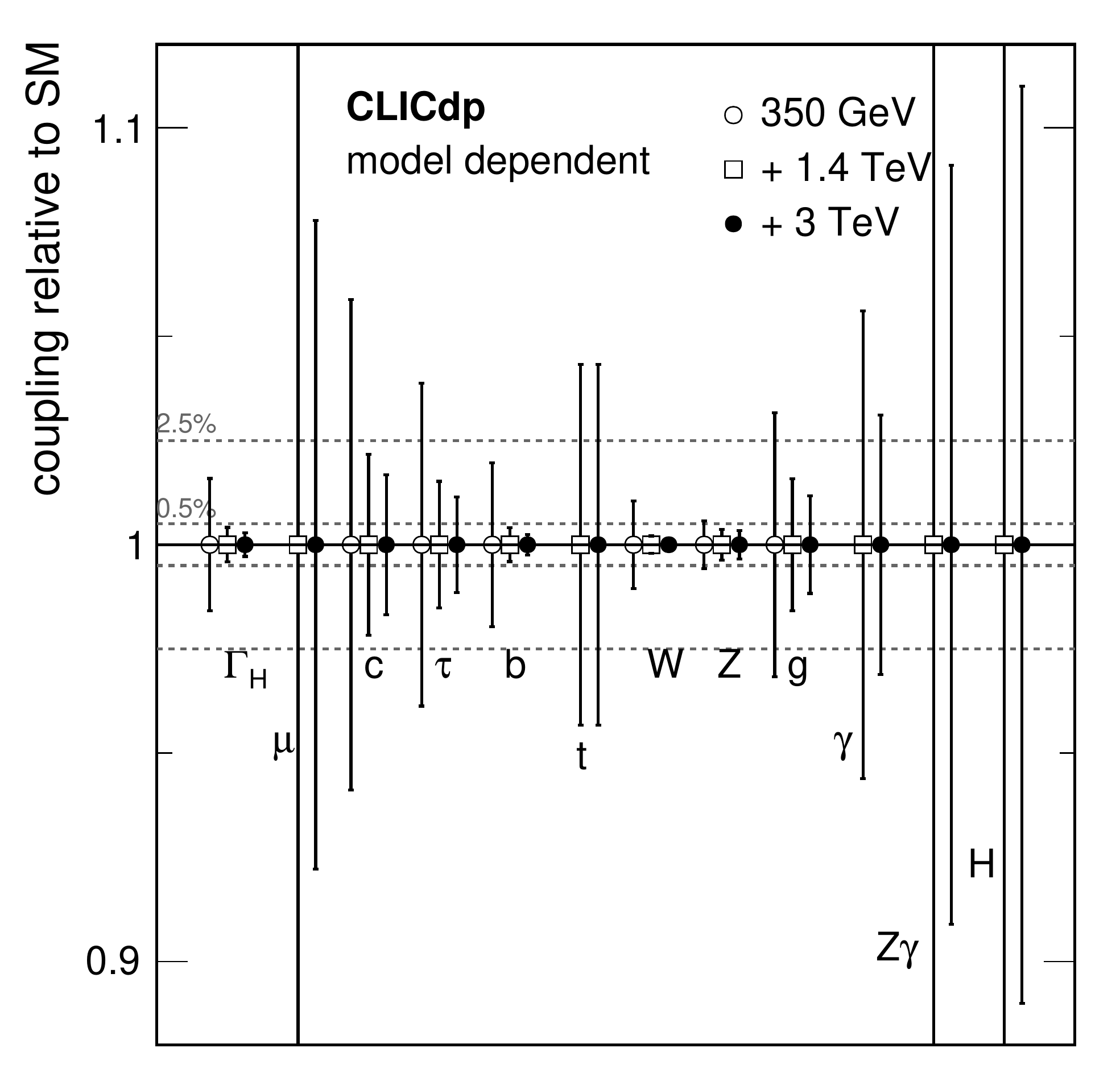}
  \caption{Illustration of the precision of
    the Higgs couplings determined
    in a model-dependent fit in the studied three-stage CLIC programme. 
    Note the reduced $y$-axis range with respect to \autoref{fig:combinedFit:MI}.
    \label{fig:combinedFit:MD}
    }
\end{figure}

As in the model-independent case the fit was performed in three stages,
taking the statistical errors at the three considered energy
stages ($350\,\GeV$, $1.4\,\TeV$, $3\,\TeV$) successively into
account. Each new stage also includes all measurements of the previous
stages. The total width is not a free parameter of the fit. Instead,
its uncertainty, based on the assumption given in
\autoref{eq:KappaWidth}, is calculated from the fit results, taking
the full correlation of all parameters into
account. \autoref{tab:MDResults} summarises the results of the model-dependent fit,
and \autoref{fig:combinedFit:MD} illustrates the evolution of the
precision over the full CLIC programme.

These results show that the CLIC Higgs physics programme, interpreted with a combined
fit of the couplings to fermions and gauge bosons as well as the total
width, and combined with the measurement of the self-coupling, will
provide a comprehensive picture of the Higgs properties. Each of the CLIC stages 
contributes significantly to the total
precision, with the first stage near $350\,\GeV$ providing the
model-independent ``anchor'' of the coupling to the $\PZ$ boson as well
as a first measurement of the total width and coupling measurements to
most fermions and bosons. 
The higher-energy stages add direct measurements of the coupling to top quarks, to muons and photons as well as overall improvements of the branching ratio measurements.
Hence, they improve the uncertainties of the total widths and all couplings except the one to the $\PZ$ boson already measured with best precision in the first stage. 
They also provide a measurement of the self-coupling of the Higgs boson. 
In a model-dependent analysis, the improvement with increasing energy is
even more significant than in the model-independent fit, since the
overall limit on all couplings imposed by the model-independent
measurement of the $\PH\PZ$ recoil process is removed. With these results on Higgs couplings at the percent level, CLIC will have strong capabilities to detect possible deviations from the SM Higgs properties. It will thereby have a powerful handle on pinning down influences of possible BSM phenomena~\cite{Gupta:2012mi,Gupta:2013zza}.

\subsection{Top-quark physics}
\label{sec:top}
In addition to a broad and comprehensive Higgs physics programme, CLIC also provides the opportunity to study the top quark with unprecedented precision via production of $\PQt\PAQt$ pairs in $\Pep\Pem$ collisions.
Measurements of the top-quark properties are of special interest as it is the heaviest elementary particle of the SM.
Due to its mass, the top quark couples strongest to the Higgs field and it has a central role in many BSM models.
Precision measurements of top-quark properties at $\Pep\Pem$ colliders promise therefore to be highly sensitive to physics beyond the SM.

In the following, two areas of top-quark physics at CLIC are described.
The first set of measurements will focus on the top-quark mass.
The second will explore the potential of measurements of other top-quark properties as probes for BSM physics, for instance the top-quark coupling to $\PZ$ and $\PGg$.

\subsubsection{Top-quark mass}
\label{sec:topmass}
Together with the Higgs mass, the top-quark mass is a key input parameter to studies of the SM vacuum stability.
With the precision of the Higgs mass as measured by ATLAS and CMS~\cite{Aad:2015zhl}, the uncertainty on the top-quark mass~\cite{ATLAS:2014wva,AtlasTop, Aaboud:2016igd, Khachatryan:2015hba} currently remains the leading uncertainty in tests of the SM vacuum stability~\cite{Degrassi:2012ry,Buttazzo:2013uya}.
While the statistical top-quark mass precision expected for High-Luminosity LHC (HL-LHC) is in the order of 10\,MeV, significant systematic uncertainties of the direct mass measurement are expected in the order of 600\,MeV~\cite{Agashe:2013hma}, for instance due to the ambiguity in the interpretation of the top-quark mass parameter in the Monte Carlo event generators in terms of the field theoretical mass scheme~\cite{Vos:2016til}.

For CLIC two complementary techniques to measure the top-quark mass were investigated.
The first and more precise option is a scan of the top pair-production cross section at several centre-of-mass energies close to the threshold of approximately $\sqrt{s} = 350$\,\gev, from which the top-quark mass can be extracted using a fit to the threshold shape.
The second option is the reconstruction of the top invariant mass from its decay products.

Both techniques have been studied using full CLIC detector simulations~\cite{Seidel:2013sqa} including beam-induced and non-$\PQt\PAQt$ physics backgrounds as well as a realistic CLIC luminosity spectrum.
An input top-quark mass of $m_{\PQt} = 174.0$\,\gev and a width of $\Gamma_{\PQt}=1.37$\,\gev were assumed.
The two dominant $\PQt\PAQt$ decay modes corresponding to a branching ratio of 75\% were considered.
These are the fully-hadronic decay mode $\PQt\PAQt \to \PWp\PQb \PWm\PAQb \to \PQq\PAQq\PQb \PQq\PAQq\PAQb$ and the semi-leptonic decay mode $\PQt\PAQt \to \PWp\PQb \PWm\PAQb \to \PQq\PAQq\PQb \Pl\PGn\PAQb$ excluding $\PGt$ final states.
The top pair reconstruction and identification used flavour tagging, kinematic fitting, and multivariate background rejection, resulting in clean signal event samples.

\paragraph{Threshold scan}
\begin{figure}[t!]
  \centering
  \begin{subfigure}[b]{0.48\textwidth}
    \includegraphics[width=\textwidth]{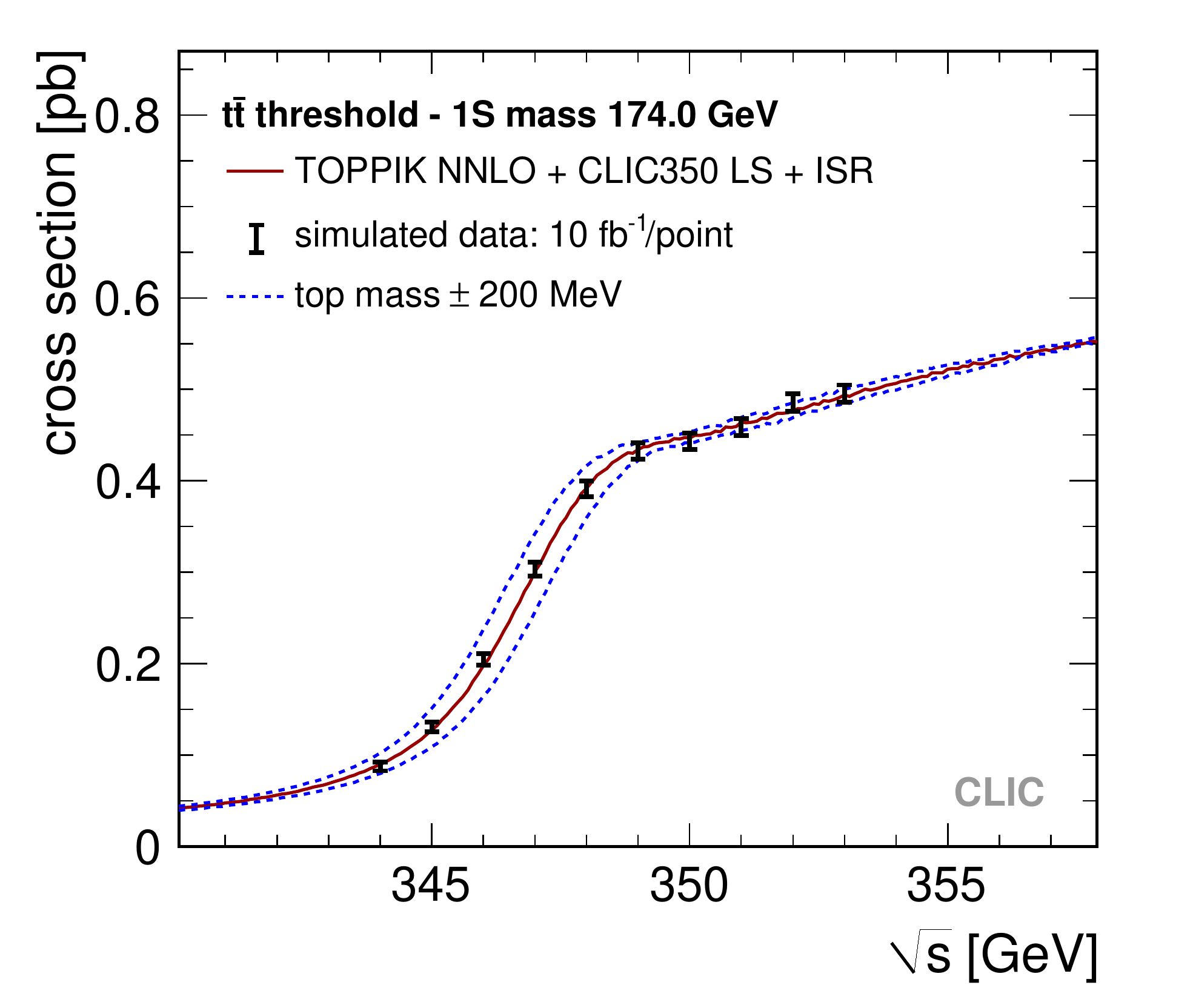}
    \caption{}
    \label{fig:threshold}
  \end{subfigure}
  \hfill
  \begin{subfigure}[b]{0.48\textwidth}
    \includegraphics[width=\textwidth]{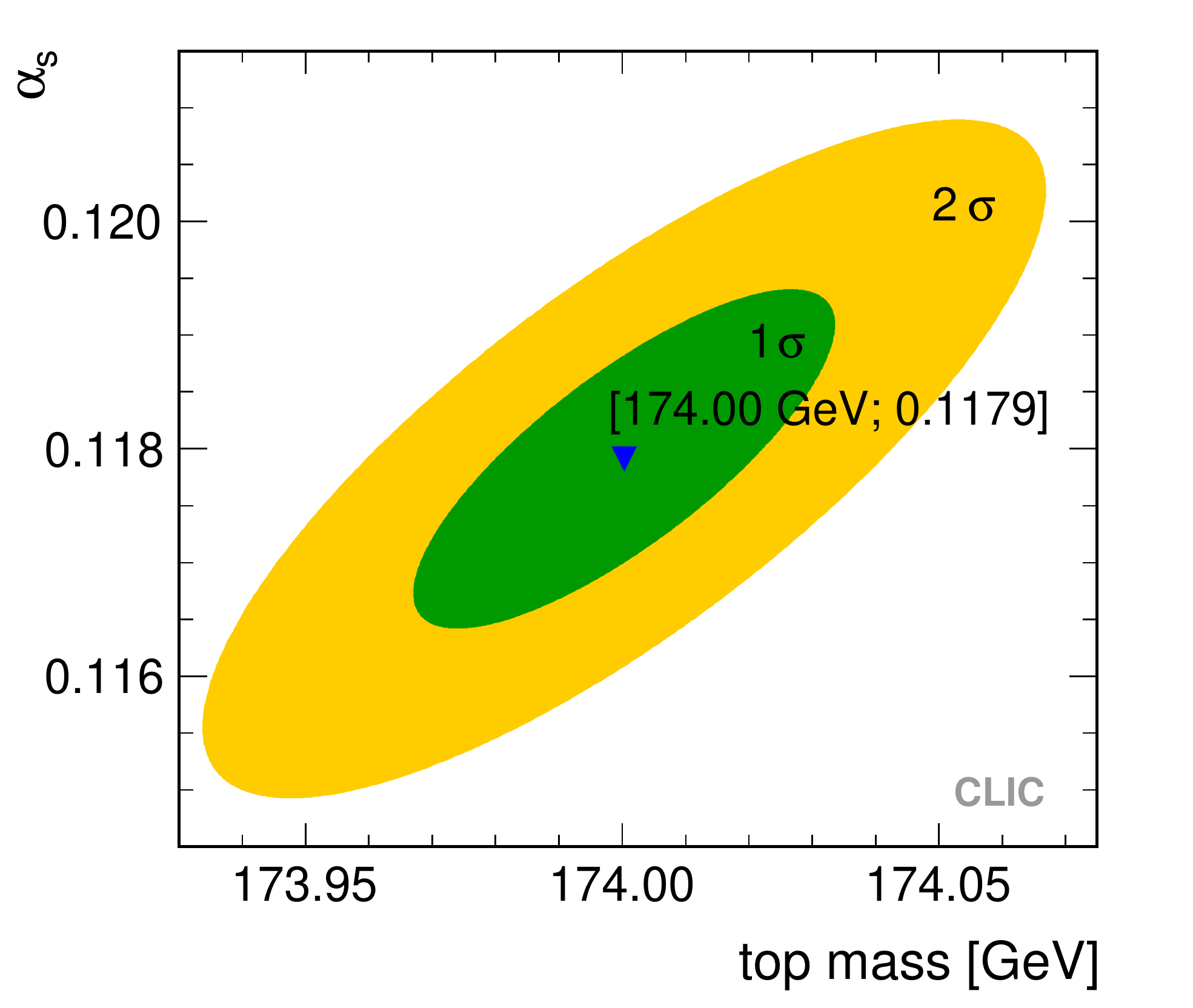}
    \caption{}
    \label{fig:alphas}
  \end{subfigure}
  \caption{
  a: $\PQt\PAQt$ cross section as a function of centre-of-mass energy for $\PQt\PAQt$ production simulated in a scan of ten data sets of $10\,\invfb$ each in steps of 1\,\gev around $\sqrt{s} = 350\,\gev$~\cite{Seidel:2013sqa}.
  b: Correlation between the top-quark mass $m_{\PQt}$ and the strong coupling constant $\alpha_{s}$ extracted from the threshold scan~\cite{Seidel:2013sqa}.}
  \label{fig:topplots}
\end{figure}

Dedicated operation during the first CLIC energy stage would enable a threshold scan of $\PQt\PAQt$ production~\cite{Seidel:2013sqa}.
By collecting ten data sets of $10\,\invfb$ each in steps of 1\,\gev around $\sqrt{s} = 350\,\gev$, the $\PQt\PAQt$ production cross section can be measured as a function of centre-of-mass energy as shown in \autoref{fig:threshold}. 
In this centre-of-mass energy range, the $\PQt\PAQt$ cross section rises from 50\,fb to about 500\,fb amounting to several tens of thousands of $\PQt\PAQt$ events. 
A study of the threshold shape as discussed here can not be performed at hadron colliders, where the parton-parton centre-of-mass energy is not known.

Theoretical evaluations of the evolution of the top pair production cross section close to the threshold are since recently available at next-to-next-to-next-to leading order (NNNLO), allowing for an accurate extraction of the top-quark mass in the theoretically well-defined 1S mass scheme with an uncertainty of the order of tens of MeV~\cite{Beneke:2015kwa,Simon:2016htt}.
Using a two-dimensional template fit to the threshold, the 1S mass of the top quark and the strong coupling constant $\alpha_s$ can be extracted simultaneously (see \autoref{fig:alphas}).
The statistical uncertainty on the measured 1S top-quark mass using this method is 33\,MeV.
The total uncertainty, including also the theoretical uncertainty and systematic uncertainties on the beam energy, the luminosity spectrum and the background subtraction, amounts to approximately 50\,MeV. 

The 1S mass of the top quark extracted in this threshold scan can be transformed into the $\overline{\text{MS}}$ mass scheme commonly used in precision calculations.
Even with the current world average value of $\alpha_s$ \cite{PDG2014} taken as external input for this conversion, this results only into an additional theory uncertainty of the order of 10\,MeV~\cite{Marquard:2015qpa}.

\paragraph{Invariant mass}
The invariant mass measurement of the top quark has been studied for CLIC at $\sqrt{s}=500\,$\gev, for an integrated luminosity of 100\,fb$^{-1}$~\cite{Seidel:2013sqa}.
With a top pair-production cross section of 530\,fb at $\sqrt{s}=500\,$\gev this results in 53000 $\PQt\PAQt$ events.
By using maximum likelihood fits to the reconstructed invariant mass distributions, shown in \autoref{fig:TopMass} for fully-hadronic events, a top-quark mass compatible with the input value was extracted that had a statistical precision of $80$\,MeV~\cite{Seidel:2013sqa}. 
Relevant systematic uncertainties, for instance including the uncertainty on the jet energy scale, are limited to a similar level as the statistical uncertainty.
The extracted top width is compatible with the input value and it has a statistical uncertainty of 220\,MeV.

\begin{figure}[t!]
\centering
\includegraphics[width=0.5\textwidth]{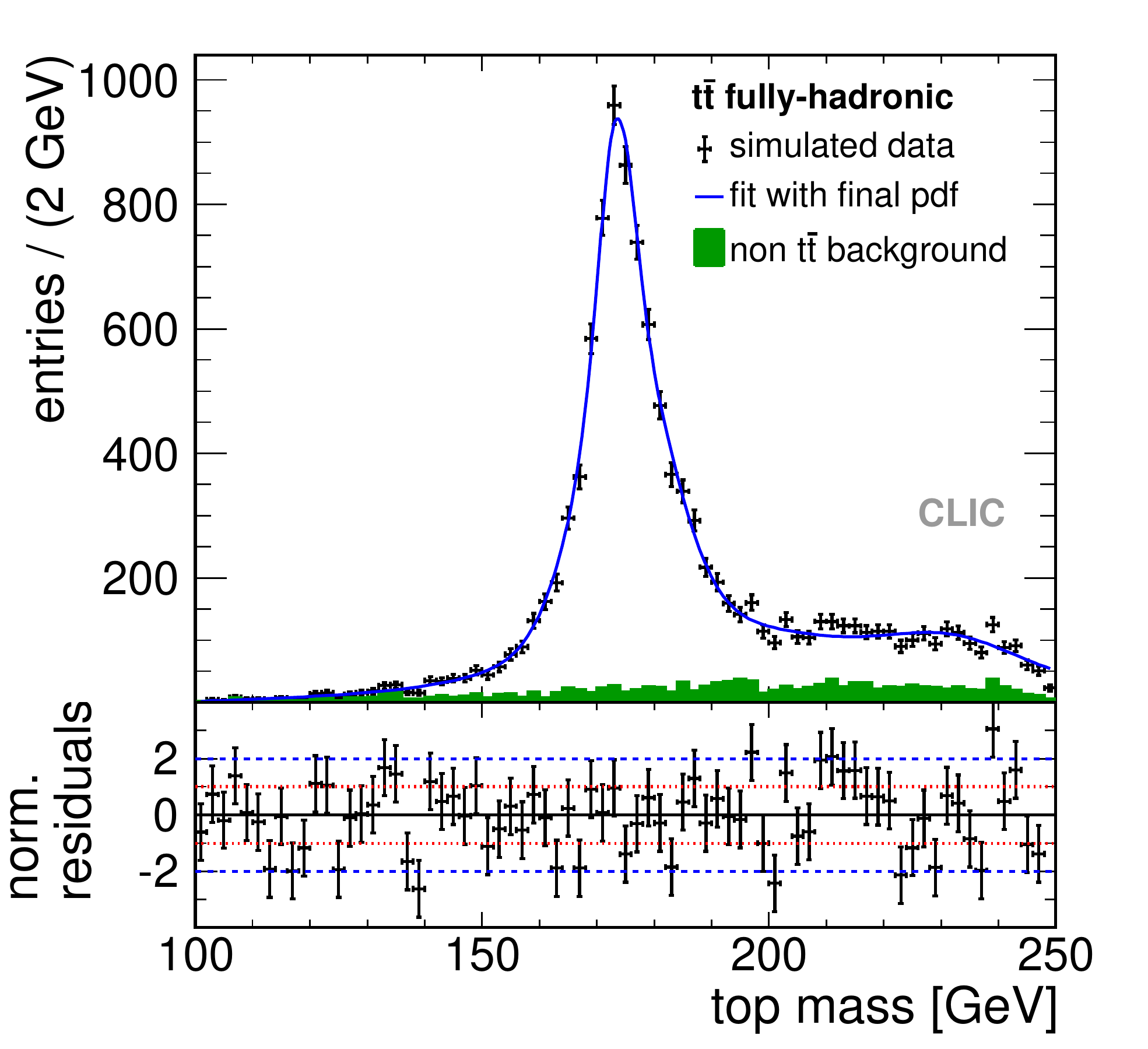}
\caption{
Distribution of reconstructed top-quark mass for events classified as fully-hadronic \cite{Seidel:2013sqa}. 
The data points include signal and background events for an integrated luminosity of 100\,fb$^{-1}$ at $\sqrt{s}=500$\,\gev. 
The pure background contribution is shown by the green solid histogram. 
The top-quark mass is extracted through a fit to a probability density function (pdf) that includes detector resolution effects.}
\label{fig:TopMass}
\end{figure}

An advantage of the top-quark invariant-mass measurement is that it can be performed at any centre-of-mass energy above the top pair production threshold.
The top-quark mass is then obtained in the context of the event generator used in the comparison to data.
Uncertainties when translating the measured mass to the $\overline{\text{MS}}$ scheme could however be large.
These systematic uncertainties are expected to be larger than the experimental uncertainties listed above~\cite{Seidel:2013sqa,Vos:2016til}.

\subsubsection{Top quark as a probe of BSM physics}
For top-quark measurements as probes of BSM physics, a trade-off has to be made between centre-of-mass energies optimised for available statistics, or for small uncertainties on theoretical predictions, or for the expected magnitude of BSM effects on the top sector.
Details of these considerations are described in the following, with example measurements at the first stage of CLIC and an outlook on measurements at the subsequent energy stages.

\paragraph{Impact of initial CLIC energy stage on top BSM physics reach}
The production of $\PQt\PAQt$ events in $\Pep\Pem$ collisions sets on at approximately $\sqrt{s}=350$\,\gev, and the cross section exhibits a steep rise until a maximum of approximately 640\,fb is reached near 420\,\gev.
Above this energy the cross section decreases following an $1/s$ dependence.
To record $\PQt\PAQt$ events with high statistics, operation close to the maximum in the $\PQt\PAQt$ production cross section is favourable.

In the threshold region, non-relativistic QCD leads to an uncertainty of 3\%~\cite{Beneke:2015kwa}.
Far away from the threshold, the uncertainty of NLO predictions is much lower, for example 0.5\% at $\sqrt{s}=500$\,GeV~\cite{Amjad:2013tlv}.
Hence, a comparison of data with theory predictions is more stringent at higher energies.
In the transition region around 380\,GeV the matching between both approaches is currently being worked on.

Taking into account that, in addition, relative contributions from many BSM effects in the top sector are expected to increase as a function of the centre-of-mass energy, operation at energies substantially above the top pair production threshold is favourable.

The choice of the centre-of-mass energy of the initial stage of CLIC has to take these considerations into account, while also considering the Higgs physics reach discussed in \autoref{sec:Higgs_physics}.

\paragraph{Top-quark form-factor measurement at 380\,GeV}
At energies above the top pair production threshold, $\PQt\PAQt$ pairs are produced through s-channel processes via $\PZ$ boson or $\PGg$ exchange.
The relative contributions of $\PZ$ boson and $\PGg$ exchange to the $\PQt\PAQt$ production depend on the beam polarisation as well as on the top-quark polarisation.
Hence the couplings to $\PZ$ boson and $\PGg$ can be extracted from forward-backward and polarisation asymmetries of the produced top quarks.
Since CLIC provides the possibility of $\pm80$\% $\Pem$ beam polarisation and since the top quarks decay before hadronisation, such that their polarisation can be measured by studying their decay products, the top-quark couplings to the electroweak gauge bosons can be studied in detail.

\begin{figure}[t!]
\centering
\includegraphics[width=0.47\textwidth]{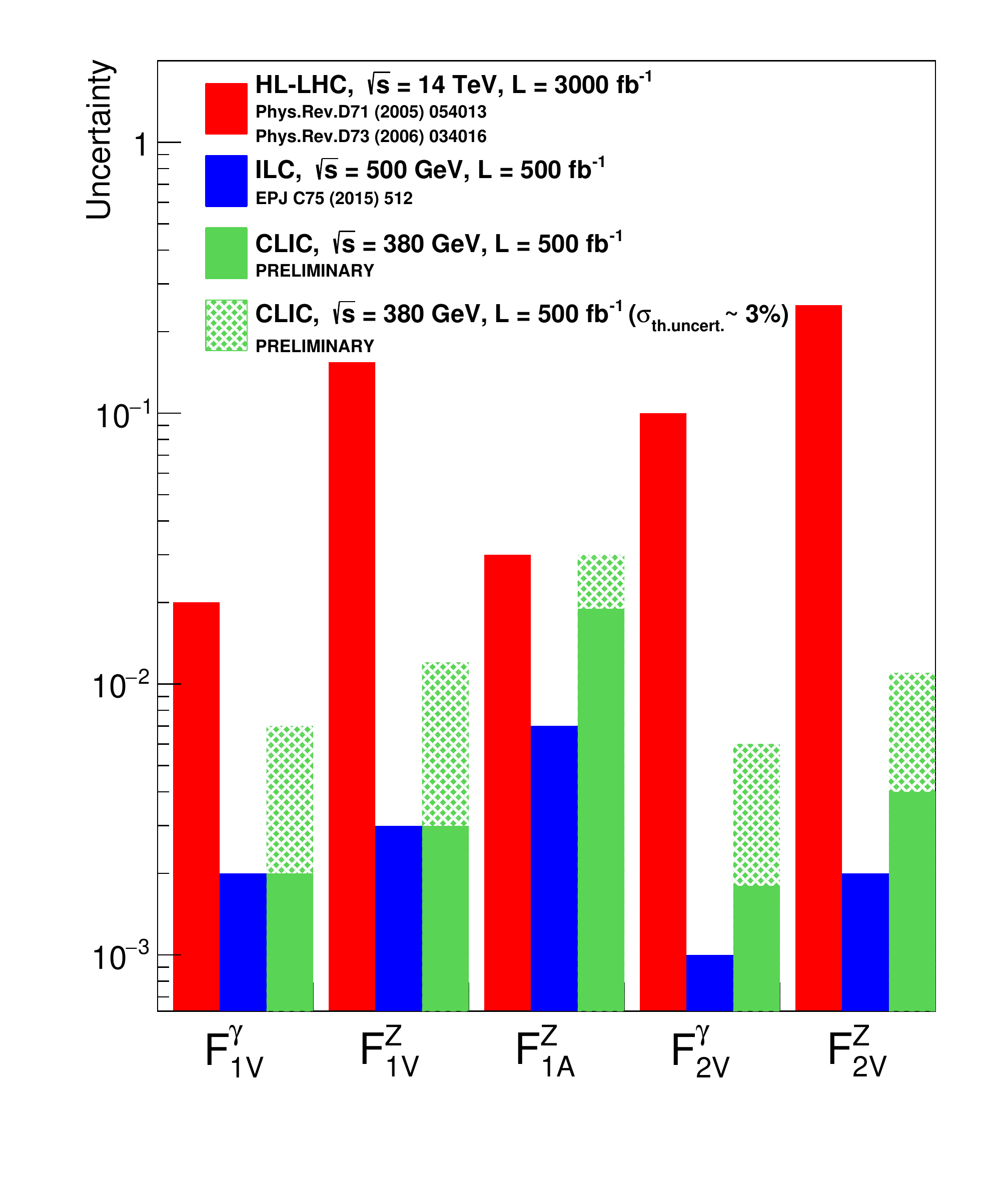}
\caption{Uncertainties on the top-quark form factors (assuming SM values for the remaining form factors), comparing estimations for HL-LHC, ILC and CLIC~\cite{topLC}.
The form factors are extracted from the measured forward backward asymmetry and cross section.
For the ILC, $\pm80\%$ $\Pem$ polarisation and $\mp30\%$ $\Pep$ polarisation are considered and for CLIC, $\pm80\%$ $\Pem$ polarisation is considered.
}
\label{fig:TopFormFactors}
\end{figure}

Simulation studies of the measurements of the top-quark form factors~\cite{Amjad:2013tlv}, extracted from the top pair-production cross section and top forward-backward asymmetries, have been performed for CLIC at 380\,\gev ~\cite{topLC}.
\autoref{fig:TopFormFactors} shows a comparison of the uncertainties on the top-quark form factors between HL-LHC at 14\,\tev~\cite{HL-LHC-top1,HL-LHC-top2}, ILC at 500\,\gev~\cite{Doublet:2012wf,Fujii:2015jha,Amjad:2015mma} and CLIC at 380\,\gev.
At ILC and CLIC, the top couplings to the electroweak gauge bosons can be extracted with precisions at the percent level, well beyond those projected for HL-LHC. 
Since the theory uncertainties are not yet calculated at 380\,\gev, the CLIC uncertainties include a theory uncertainty of 3\%, corresponding to the maximal theory uncertainty on the $\PQt\PAQt$ production cross section at the threshold~\cite{Amjad:2013tlv}. This is a very conservative upper limit.

As all couplings between top quarks and electroweak gauge bosons are predicted to high precision by the SM, a stringent comparison between the observed and the predicted couplings can be performed.
In many BSM models the top couplings to the electroweak interaction are substantially modified. This is the case, for example, in composite Higgs models or extra dimension models, where coupling modifications are of the order of 10\%~\cite{Richard:2014upa,Amjad:2015mma,Vos:2016til}, resulting in a large discovery potential for future $\Pep\Pem$ colliders.
Additional information can also be extracted from an analysis of the top-quark decays~\cite{Janot:2015yza}.

\paragraph{Top BSM physics outlook for subsequent CLIC stages}
In addition to the measurements performed at the initial centre-of-mass energy stage, CLIC allows for measurements at centre-of-mass energies of $\sim$1.4\,\tev and 3\,\tev.
This offers an extended BSM physics potential in the top sector at CLIC in comparison to ILC.
\\
For operations above approximately 420\,GeV, where the $\PQt\PAQt$ production has its maximum, the $\PQt\PAQt$ production cross section decreases, resulting in a decreasing statistical precision in the measurement of the top-quark properties.
However, this might be compensated by two effects.
At higher collision energies the top quarks have an increased boost resulting in a better separation between the decay products of the two top quarks, leading most likely to an improved reconstruction of the top-quark properties.
Moreover, the relative contribution from new physics in many BSM models increases with the centre-of-mass energy following $s/\Lambda^2$, where $\Lambda$ is the scale of the new physics.
\\
Dedicated simulation studies focussing on top-quark measurements at the CLIC higher-energy stages are performed at the moment.

\subsection{Physics beyond the Standard Model}
\label{sec:bsm}
Strategies for studying BSM physics at CLIC follow two different approaches, depending on the nature of the new physics involved~\cite{Abramowicz:1563377}. 
Direct detection of new particles is possible up to the kinematic limit, typically up to $\sqrt{s} / 2$ for particles produced in pairs and $\sqrt{s}$ for single particle production. 
In this area CLIC is particularly sensitive to electroweak states and therefore complementary to the LHC. 
Much higher mass scales are attainable through indirect searches, where precision measurements of observables are compared with SM expectations. Both approaches have been assessed for CLIC and the current findings are summarised below.

\subsubsection{Direct searches for BSM physics}
Based on the detector concepts described in \autoref{sec:physics:detector} direct studies of the BSM physics potential were performed at 1.4\,\tev and 3\,\tev . Several of these studies make use of SUSY models containing sparticles with masses within the kinematic limit. Whether addressing sleptons, gauginos, squarks or heavy Higgs-like particles, the studies confirm that CLIC can measure them accurately, typically to 1\% accuracy on their mass and up to the kinematic limit of typically $\sqrt{s} / 2$ ~\cite{cdrvol2,CLICCDR_vol3,Contino:2013gna,WEUSTE:1499132,Barklow:1443499,Srivastava:2012hn}.
These direct searches use SUSY particles as an example. As SUSY provides a very wide range of states, all with different signatures in the detector, these studies have a wide applicability to physics scenarios in addition to SUSY. 

\subsubsection{Sensitivity of precision measurements to BSM physics}
In many cases new physics beyond the kinematic limit of CLIC can still be detected due to its effects on SM observables. Given their high masses, precision measurements of Higgs and top are very powerful in this context, as already indicated in the previous subsections. Various other indirect studies are described in~\cite{cdrvol2,CLICCDR_vol3,Abramowicz:1563377}, among which are the two examples illustrated in \autoref{fig:BSMplots}.
In \autoref{fig:BSMZp} the process $\Pep\Pem \rightarrow \PGmp\PGmm$ is studied at 1.4\,\tev and 3\,\tev. This process is sensitive to the presence of a high-mass $\PZpr$ boson. Observables such as the total cross section, the forward-backward asymmetry and the left-right asymmetry with $\pm80\%$ electron polarisation are used in the study. Depending on the effective parameters of the underlying theory (in this case the minimal anomaly-free $\PZpr$ model), a discovery reach up to tens of \tev is attainable at CLIC. Alternatively, if a high-mass $\PZpr$ boson were discovered earlier at the LHC, CLIC would be able to pin down its properties to high accuracy~\cite{Abramowicz:1563377,Blaising:1471941}.

Another possibility concerns composite Higgs models, in which the Higgs boson exists as a composite state of bound fermions. 
Using measurements of double-Higgs production and single-Higgs production, CLIC could significantly reduce the parameter space. At 3\,\tev with $1\,\invab$ of data it would provide an indirect probe up to a Higgs composite scale of 20\,\tev and 70\,\tev respectively~\cite{Abramowicz:1563377,Contino:2013gna}, (see \autoref{fig:BSMcomp}). 

An additional example is the study of triple gauge couplings in $\PWp\PWm$ production, where coupling accuracies at the level of 10$^{-4}$ are predicted~\cite{cdrvol2}, thereby providing relevant input on BSM physics. 
Furthermore vector boson scattering (\PW, \PZ) can be studied where the sensitivity rises quickly with $\sqrt{s}$~\cite{Fleper}.

Similarly, large numbers of $\PW$ bosons will be produced in $\Pe\Pe \rightarrow \PW\PW$ and $\Pe\Pe \rightarrow \Pe\PW\PGn$ events at CLIC. For example, it was found in a generator-level study of $\Pe\Pe \rightarrow \Pe\PW\PGn$ events that the number of hadronic $\PW$ decays within the detector volume is expected to be around $10^{7}$ at 1.4\,\tev and $1.5 \times 10^{7}$ at 3\,\tev. These samples provide the potential for a measurement of the $\PW$ boson mass from its hadronic decays with a few-MeV statistical accuracy. A full simulation study is foreseen to study the impact of systematic effects, such as the uncertainty of the jet energy scale, on this measurement~\cite{Abramowicz:1563377}.

\begin{figure}[t!]
  \centering
  \begin{subfigure}[b]{0.58\textwidth}
    \includegraphics[width=\textwidth]{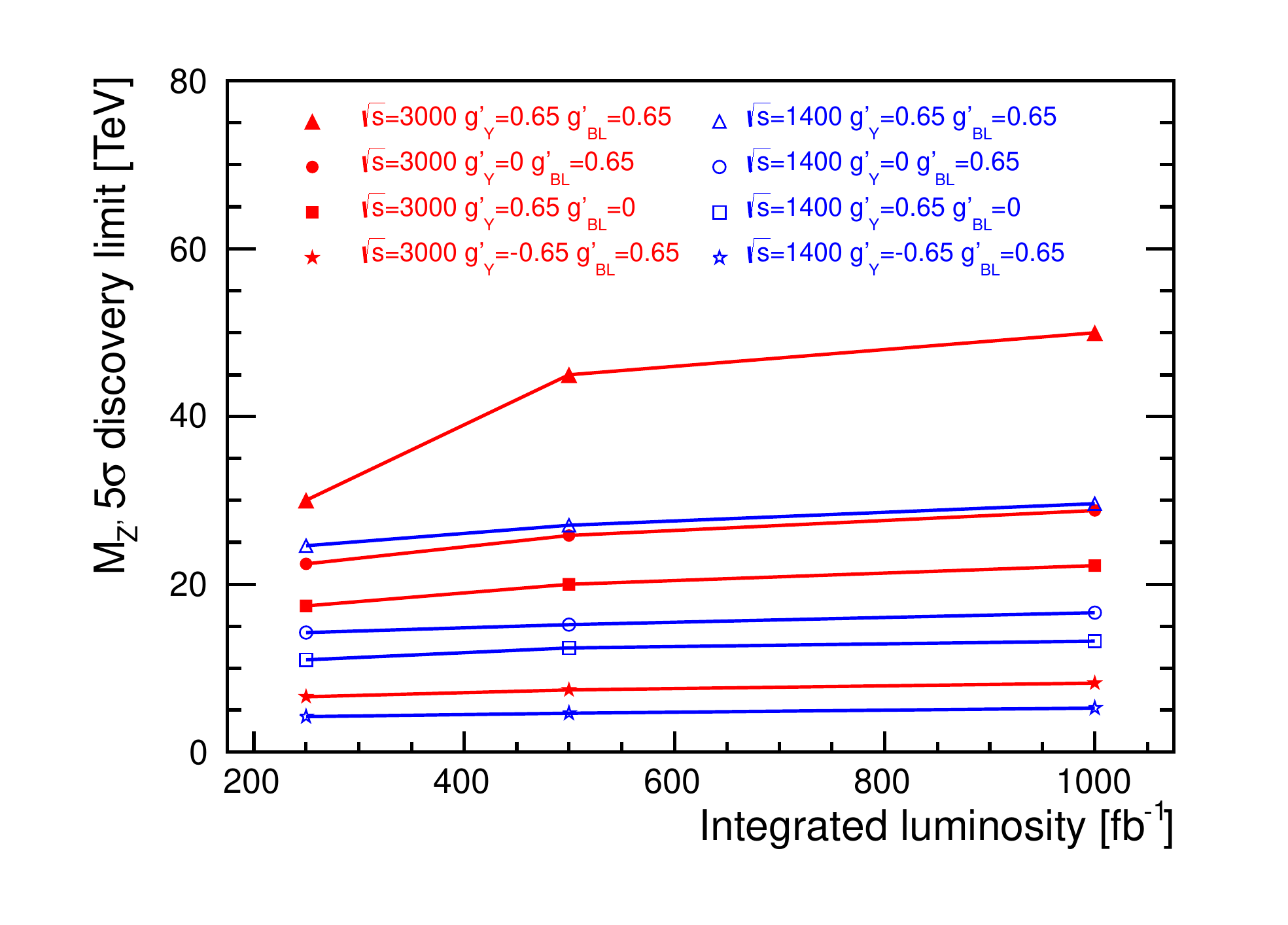}
    \caption{}
    \label{fig:BSMZp}
  \end{subfigure}
  \hfill
  \begin{subfigure}[b]{0.4\textwidth}
    \includegraphics[width=\textwidth]{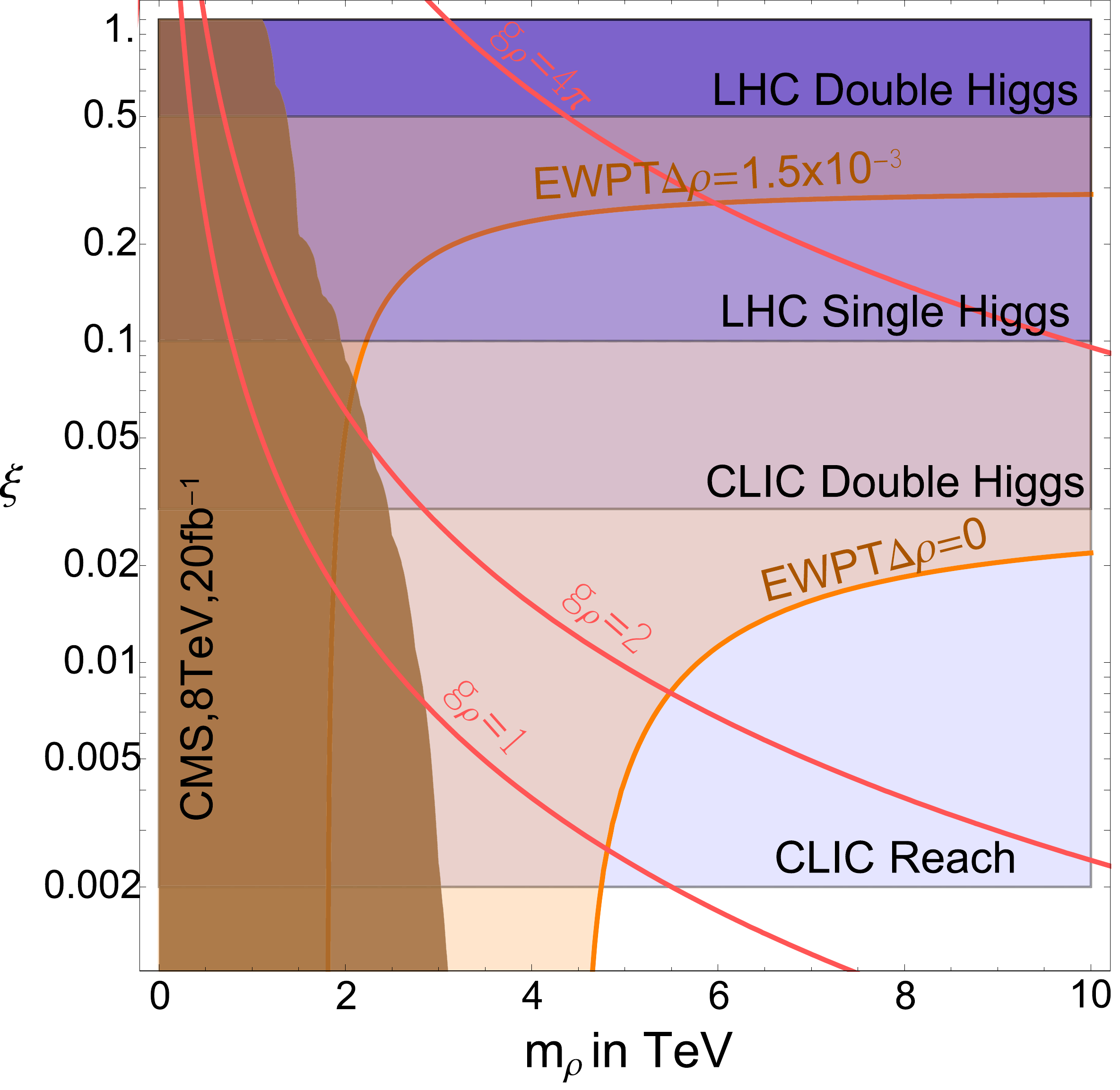}
    \caption{}
    \label{fig:BSMcomp}
  \end{subfigure}
  \caption{
  a: $\PZpr$ mass discovery limit at $5\sigma$ from the measurement of $\Pep\Pem \rightarrow \PGmp\PGmm$ as a function of the integrated luminosity and for different coupling values in the minimal anomaly-free $\PZpr$ model. 
  For more details see~\cite{Abramowicz:1563377,Blaising:1471941}. 
  b: Summary plot of the current constraints (orange curves and brown region) and prospects for direct and indirect probes at LHC and CLIC (horizontal regions) of the strong interactions triggering electroweak symmetry breaking. $m_{\rho}$ is the mass of the vector resonances and $\xi = (v/f)^{2}$ measures the strengths of the Higgs interactions.  
  For more details see~\cite{Abramowicz:1563377,Contino:2013gna}.}
  \label{fig:BSMplots}
\end{figure}

\subsection{Summary of physics requirements for the CLIC energy stages}
\label{sec:physicssummary}
With the above assessment of the CLIC physics potential for Higgs, top-quark and BSM physics, one has the ingredients to reflect on the optimal choice for the future CLIC energy stages. 

A choice of 350\,\gev for the first energy stage turns out to be close to optimal for initial precision Higgs studies. As explained in \autoref{sec:Higgs_physics} it gives access to the Higgs boson in the $\PH\PZ$ production process, thereby providing a model-independent measurement of the $\gHZZ$ coupling of the Higgs to the $\PZ$ boson. This measurement forms the cornerstone of all other Higgs coupling measurements and determines the ultimate accuracy with which Higgs couplings can be measured in a model-independent way at CLIC. As discussed in \autoref{sec:Higgs_physics} the $\gHZZ$ coupling is best measured in events where the $\PZ$ decays to $\PQq\PAQq$ at centre-of-mass energies near 350\,\gev. This centre-of-mass energy is also favourable for the high-accuracy measurement of the top-quark mass through a threshold scan, as shown in \autoref{sec:top}. However, a $\sqrt{s}$ choice above the $\PQt\PAQt$ threshold, e.g. near 400\,\gev, offers clear advantages for the measurement of top-quark kinematic variables and forward-backward asymmetry due to the additional boost of the produced top quarks. Taking all these arguments together, a centre-of-mass energy of 380\,\gev appears to be a more optimal choice for the first CLIC energy stage, as it combines favourable conditions for both Higgs and top-quark physics. At this energy CLIC would collect $500 \invfb$ of data, complemented with $100\,\invfb$ for a $\PQt\PAQt$ threshold scan near 350\,\gev. 

At the higher energies, above 1\,\tev, Higgs and top-quark physics continue to provide guaranteed precision physics cases. Profiting from the increased cross sections at higher centre-of-mass energies, combined with larger instantaneous luminosities, Higgs production through WW and ZZ fusion yields significantly improved statistical accuracies, thereby also giving access to rare Higgs decay modes. The measurement of the top Yukawa coupling, through $\PQt\PAQt\PH$ production, provides optimal accuracy in the range 1\,\tev to 1.5\,\tev \cite{Price:2014oca,Redford:1690648}. The measurement of the Higgs self-coupling profits greatly from increasing rates at high $\sqrt{s}$, providing a unique opportunity for CLIC to measure the self coupling down to the 10\% level at $\sqrt{s} = 3\,\tev$. 
Top pair production at the higher CLIC energies has the potential of being a sensitive probe for BSM physics (see \autoref{sec:top}).
Overall the high-energy reach of CLIC provides significant potential for studying BSM phenomena, either through accurate measurements of new states previously discovered at LHC or by acting as a discovery machine in its own right. For new particles produced in pairs direct detection is possible up to the kinematic limit of $\sqrt{s} / 2$. Indirect detection through precision observables profits from high $\sqrt{s}$ in many cases as well, as illustrated with the $\PZpr$ and composite Higgs examples of \autoref{sec:bsm}.

Based on the current knowledge, CLIC delivers optimal physics potential when constructed and operated in three main energy stages: 380\,\gev, 1.5\,\tev and 3\,\tev.
Here, the low-energy stage is chosen for optimal Higgs and top-quark physics reach, while 3\,TeV is the maximum which can presently be envisaged.
The choice of the intermediate energy stage at 1.5\,\tev is driven by the fact that this is the maximum energy that can be reached with a single CLIC drive-beam complex.
Realistically, one can assume that CLIC will operate for the equivalent of 125 days per year at 100\% efficiency (see \autoref{sec:PowerEnergy}). 
A period of luminosity ramp-up will be necessary at each stage of CLIC. 
Together with the expected peak luminosity at the different energies (see \autoref{sec:StagingBaseline}) this results in the integrated luminosities listed in \autoref{tab:stagingproposal}.

\begin{table}
\caption{Proposed CLIC energy staging scenario for optimal physics performance, assuming between 5 and 7 years of running including luminosity ramp-up at each of the three energy stages as described in \autoref{sec:operationScenario}.}
\centering
 \begin{tabular}{lrrr}
\toprule
Stage                 & $\sqrt{s}$ (GeV) &   \LumiInt (\fbinv)\\
\midrule
\multirow{ 2}{*}{1}   &             380  &               500  \\
                      &             350  &               100  \\
\midrule
2                     &            1500  &              1500  \\
\midrule
3                     &            3000  &              3000  \\
\bottomrule
 \end{tabular}
 \label{tab:stagingproposal}
\end{table}

\clearpage
\section{CLIC post-CDR accelerator optimisation}
\label{sec:AcceleratorOptimisation}
\subsection{Overview}

A first optimisation of the parameters for a 3\,TeV CLIC accelerator complex was performed as early as 2008, based on performance and cost models developed at that time. During the years leading to the CDR in 2012~\cite{CLICCDR_vol1,CLICCDR_vol3}, a large number of simulation studies and R\&D tests validated most aspects of the CLIC design. In parallel, more detailed models of power consumption and cost of a 3\,TeV CLIC facility were developed for the CDR. The results of those studies, together with physics scenarios envisaged at the time, provided the basis for the proposal to build CLIC in energy stages. In the CDR~\cite{CLICCDR_vol3}, an example of the implementation and operation of CLIC in three energy stages is described.
During the past years more high-gradient tests of the main linac accelerating structures have been made, which allow a review of the performance limitations that are used in the optimisation. The gradient $G$ which can be achieved in these structures is largely limited by vacuum arcing, otherwise known as radio frequency (RF) breakdowns. For a reliable operation of CLIC, RF breakdowns must occur during less than 1\% of the RF pulses in any of the installed structures along the entire length of the main linacs. This translates into the specification of ``fewer than $3\times10^{-7}$ breakdowns per pulse and per metre of accelerating structure''. 

Designing structures with the required low breakdown rates at the design gradient is done by constraining certain RF parameters, such as the maximum local power along the surface of the structure when an RF pulse is sent through. Experiments and long-term testing of CLIC RF structures are ongoing. To first approximation, these tests confirm the assumptions for the RF constraints used for the CDR. For example, an RF structure which does not have features necessary for higher-order-mode damping (damping will be required for the operation of the structures with intense beam trains) reached a gradient well above the nominal 100\,MV/m for an RF pulse of 180\,ns, complying with the parameters assumed in the CDR. Other RF structures that include the features needed for damping fell slightly short of the goal stated in the CDR. Presently, incremental improvements in the design and production of the RF structures are being implemented, aiming to have ``damped'' structures performing at the same high level as ``undamped'' ones.
Given the observations on performance of the RF structures, a scale $S$\footnote{also referred to as safety factor or safety margin} describing the RF limits has been introduced as a parameter in the optimisation models used for the CLIC re-baselining. A value of $S=1.0$ corresponds to the already achieved performance using RF structures with damping features, whereas e.g. $S=1.1$ assumes that there will be a 10\% improvement of the gradient while maintaining the low breakdown rates (i.e. corresponding approximately to the current performance of ``undamped'' structures)\footnote{occasionally, a range of values from $S=0.8$ to $S=1.2$ is used to demonstrate parameter dependencies}.

In a scenario of construction of CLIC in energy stages, a reasonably conservative assumption is $S=1.0$ for the first stage, when limited experience on RF structures is available. In addition we will see in subsequent sections that the optimum gradient for lower energy stages is in any case lower than 100\,MV/m. For the second and third energy stages, the additional development time, production and operational experience are expected to contribute to improved structure performance.

In the present re-baselining study, the first (380\,GeV) and final (3\,TeV) stages are being considered at the outset, and the nominal gradient in the RF structures of 100\,MV/m is assumed for CLIC at 3\,TeV. The parameters chosen for the intermediate energy stage are then, to a large extent, given by the initially defined parameters for the 380\,GeV and the 3\,TeV CLIC accelerator. Within this approach, three different strategies emerge for designing a staged construction and operation of CLIC:

In the first strategy, priority is given to optimising the cost of the 380\,GeV energy stage. In such a strategy, one would use an RF structure design and RF pulse length fully optimised for the first energy stage, without taking parameter optimisation for the higher energy stages into account. As a consequence, for the upgrades to higher energies, all RF structures would have to be replaced by new ones, i.e. by the structures optimised for the 3\,TeV energy stage. In such an ``unconstrained'' scenario, one would be free to change the length of drive beam decelerating sectors, and hence the RF pulse length, when moving from the first energy stage to higher energies. Note that this scenario would allow optimisation of the first energy stage (i.e. having the lowest initial cost), however with the consequence of the highest overall cost for the three energy stages of the overall CLIC project.

In the second strategy, priority is given to the overall optimisation of the CLIC project with its three energy stages. In this strategy, structure design and RF pulses are optimised for the final energy stage. This implies that the initial 380\,GeV stage will already be built using RF structures designed for the final high energy stage, using also the gradient and RF pulse length designed for 3\,TeV. While this scenario minimises the total CLIC cost for the three stages, it compromises the performances and increases the cost of the initial low-energy stage. Moreover, since high-performance high-gradient RF structures are required from the outset, this scenario does not allow advantage to be taken of continuing improvements in the design and performance of RF structures during the operation of the first energy stage.

The third strategy compromises between the other two. Different compromise solutions between the two strategies outlined above may be considered. 

As detailed below, we have chosen a staging scenario where an optimised design of RF structures is chosen for the first CLIC stage at 380\,GeV, and structures of a different design are added for the energy upgrades (i.e. the structures used for the first stage can be re-used in the later stages). The same RF pulse length has been chosen for the first and the later energy stages, allowing one to avoid major modifications to the drive beam generation when upgrading to later stages.

The following subsections describe the details of the methodology used in the recent CLIC optimisation for a staged construction. A full description of the staging implementation chosen, including the corresponding parameters, is presented in \autoref{sec:StagingBaseline}.

\subsection{Ingredients for the optimisation procedure}

A systematic methodology was developed for determining the main beam parameters and subsequently the cost and power consumption of an optimised CLIC design, (a) for a given physics performance goal and (b) for a given main linac accelerating structure.

The physics performance goal is parametrised by the centre-of-mass energy,
the luminosity and the fraction of luminosity above 99\% of the nominal centre-of-mass energy.
Physics benchmark studies have shown that the impact of beam-induced backgrounds can efficiently be minimised through optimised detector design and the application of selection criteria during event reconstruction~\cite{cdrvol2}. Therefore background considerations are not part of the accelerator optimisation process.

The accelerating structure is parameterised by the number of cells, the phase advance per cells, the loaded gradient and the iris radius and thickness of the first and the last cell. The iris radius and thickness are assumed to decrease linearly along the structure.

\subsubsection{Beam parameters}

The possible choices of the beam parameters and their interdependence have been established by detailed beam dynamics studies.  
This allows one to select the optimum beam parameters for a given main linac structure design.
The main constraints are:
\begin{itemize}
\item The bunch charge and length at the collision point and the bunch spacing are limited by wakefield effects in the main linac, in particular by the accelerating structure design.
\item The horizontal emittance is determined mainly by the damping ring as a function of the bunch charge; the ring to main linac transport system and main linac slightly increase this value.
\item The vertical emittance is given by the damping ring, the ring to main linac transport systems,
the main linac and the beam delivery system.
\item The vertical beta-function at the interaction point is determined by the final focus system.
\item The horizontal beta-function has two sources for the lower limit. One limit arises from the final focus system. The other limit
arises from the beamstrahlung. It depends on the bunch charge and the required quality of the luminosity spectrum.
\item The minimum spacing between the bunches is determined by the long-range wakefields in the main linac.
\item  The number of bunches in each train is finally determined by the luminosity target and the integrated luminosity per bunch. A repetition rate of 50\,Hz for the pulses is assumed. This allows to lock the RF to the frequency of the mains power and reduces pulse-to-pulse variations of the RF and of potential stray fields.

\end{itemize}

\subsubsection{RF structure design}
One of the key elements which has made the overall optimisation of the CLIC accelerator complex possible is the parameterisation of the full RF properties
of the accelerating structures \cite{Sjobak:2014nta,KyrreThesis} including fundamental RF parameters (the ratio of impedance over stored energy $R/Q$, the group velocity, the beam loading, the efficiency, etc.) and higher-order-mode properties. 
The high-gradient parameterisation makes a gradient and breakdown rate performance prediction based on the structure geometry for a given fabrication technology. 
The prediction is based on geometry and RF pulse length through a number of quantities including peak surface electric field, peak surface magnetic field, overall power-flow and a local power flow.  
The limiting values for these quantities are based on many test results from prototype structures with different geometries used in the high-gradient
testing programme carried out in the context of the CLIC study.
The highest achieved values in different tests indicate that the baseline CLIC 3\,TeV structure CLIC\_G described in the CDR (its properties will be listed in \autoref{tab:CLICG_DB244}) should perform above 100\,MV/m including the power overhead needed for beam loading.
Currently, 3\,TeV-type prototype baseline structures (TD26CC in the test structure nomenclature) are being tested and the results are being analysed.
The results for the prototypes give gradients very close to 100\,MV/m at the target breakdown rate of 
$3\times10^{-7}\,\text{pulse}^{-1}\,\text{m}^{-1}$ when power overhead for beam loading is included~\cite{Degiovanni:1742280}. 
The high-gradient field values derived from these tests alone imply that a minor re-design would give the full 100\,MV/m. 
Higher gradients at low breakdown rates have already been achieved with undamped structures (the T24 structure).
In addition, a number of fabrication issues have been uncovered in the prototypes, such as excessive radii on cell walls for waveguide damped structures~\cite{Degiovanni:1742280}. 
Moreover, a number of improvements to the individual cells have been identified, e.g. optimised damping waveguide width~\cite{Zha:2157079},
and the RF conditioning process is far better understood~\cite{PhysRevAccelBeams.19.032001}.
A new generation of prototypes which implement these improvements is now under preparation and we expect that they will fully meet the 100\,MV/m specification.
In order to make comparison to past results as direct as possible, it was decided to maintain the current basic parameters for the test structures (e.g. iris aperture radius range and number of cells). The baseline design is also left unchanged to maintain consistency between the design and the testing programme. Finally, changing the
baseline design now, based on partial high-gradient testing results with many improvements expected for the near future, does not seem appropriate.

\subsubsection{Cost and power models}\label{sec:opt:cost}
The cost model for the re-baselining is based on the cost estimates for the CDR, with some updates for the drive beam complex.
The cost of the drive beam complex and the main linac have been included in the model.
In addition to the technical equipment,
the main linac tunnel, the turn-arounds in the main linac and the building housing the drive beam accelerator are included. 
This model reflects most of the dependence of the cost on the accelerating structure design and on the beam parameters.
For example, the cost of the beam delivery system does not depend on the structure design. The main beam injector cost will depend on
the beam parameters. However, for a given luminosity this dependence will be very weak.

The cost model includes in particular:
\begin{itemize}
\item
The cost of the main linac takes the total length into account and the impact of the accelerating structure length;
longer structures lead to lower cost.
\item
The cost of the drive beam accelerator is based on detailed cost models for the modulators, the klystrons,
the RF waveguides and the accelerating structures.
\item
The cost of the building housing the drive beam accelerator and of the associated
infrastructure are scaled according to the peak power of the drive beam klystrons.
This ensures that the building can house the RF units of the drive beam complex also when the modulators are upgraded to the full drive beam pulse length for 3\,TeV, in order to provide enough pulses. 
In contrast to the CDR, the buildings and infrastructure for additional RF units will not be provided from the start
but will be upgraded once the higher energy stages are realised.
\item
The other components, such as the combiner rings and turn-arounds are assumed to have a fixed unit cost, independent of the CLIC centre-of-mass energy.
\end{itemize}

The power consumption model is based on the CDR \cite{CLICCDR_vol1,Jeanneret:1599195}.
It includes the whole accelerator complex.
The power is calculated separately for the RF components and the magnets. In addition the power consumption of
the cooling water and air cooling are calculated based on the power consumption for RF and magnets and taking
into account the CDR models of the distribution of these power losses in each sub-system.

The following parts of the accelerator complex are covered:
\begin{itemize}
\item The drive beam generation complex.
For the drive beam RF power the model is analytic and based on the updated designs of the modulators, klystrons and
accelerating structures.
For the magnets of the drive beam complex, the power consumption is scaled according to the drive beam energy.
\item Drive beam frequency multiplication.
The power consumption is scaled according to the beam energy.
\item Drive beam distribution system.
The power consumption is scaled according to system length, number of turn-arounds and beam energy.
\item Main beam injector complex.
The RF power consumptions of the main beam injectors, the predamping rings, the damping rings and the booster linac
 are scaled with the total average beam current. The power for magnets is constant.
\item
Main Linac.
The power for the magnets is proportional to the linac length.
\item
Beam delivery system. Its power consumption does not depend on the simulation parameters.
\end{itemize}

\subsection{Optimisation procedure and results}
\label{sec:optproc}
The parameter optimisation has been performed by defining the target energy and luminosity.
A scan was performed over a large number of different parameter points,
each consisting of a specific main linac accelerating structure and loaded gradient, see \autoref{f:beamflow}.

In the optimisation procedure, each structure is described by its frequency $f$, length $L_{\text{structure}}$ and phase advance per cell $\phi$ as well as by
the initial and final cell iris aperture radii and thicknesses ($a_1$, $a_2$, $d_1$ and $d_2$, respectively) \cite{Sjobak:1712948}.
It is assumed that the structure cells are linearly tapered between these initial and final cells.
For each loaded gradient $G$, the RF parameters are used to determine the beam parameters. This allows to determine the
number of bunches per train to reach the luminosity target and hence the RF pulse length. 
A final check is performed on whether this operation point respects the RF limitations, in which case the cost of the accelerator complex is calculated and the parameter set is stored.
Here, the parameter $S$ is taken into account in the definition of the RF limits.

\begin{figure}
\centering
\includegraphics[width=0.8\textwidth]{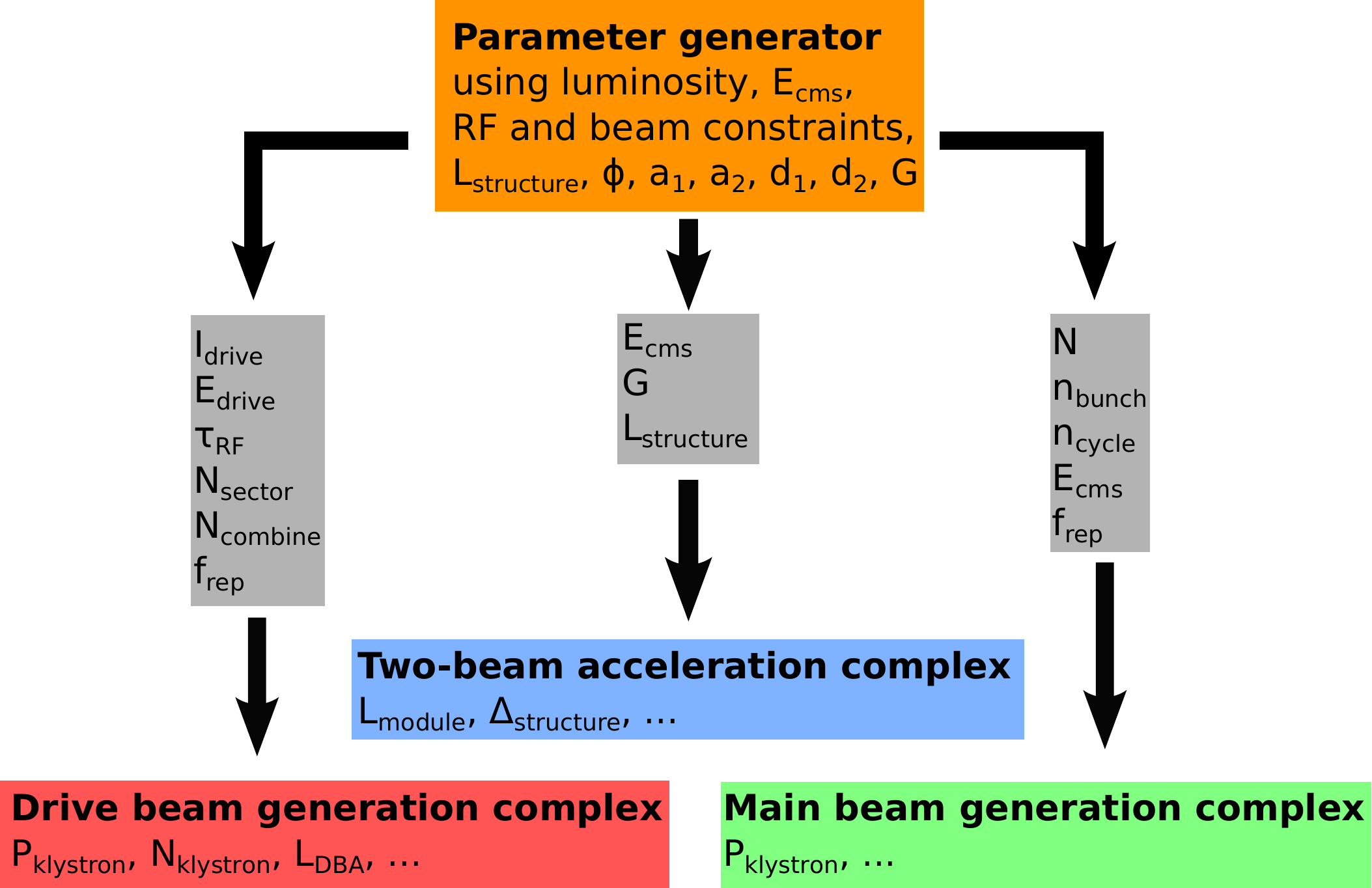}
\caption{An illustration of the accelerator parameter derivation based on the parameters of the main linac accelerating structure.}
\label{f:beamflow}
\end{figure}

In the optimisation scans, after some exploratory scans the most interesting ranges of parameters were studied. 
The gradients $G$ range typically from 50 to 120\,MV/m, the accelerating structure length $N_\text{c}$ from 20 to 60 cells, the aperture radii $a_1$ and $a_2$ each from $0.08\lambda$ to $0.18\lambda$ (where $\lambda$ is the RF wavelength) and the thicknesses $d_1$ and $d_2$ range each from 0.11 to 0.39 of the cell length.

\subsubsection{Optimum at 3\,TeV}

\begin{figure}
\centering
\includegraphics[width=0.49\textwidth]{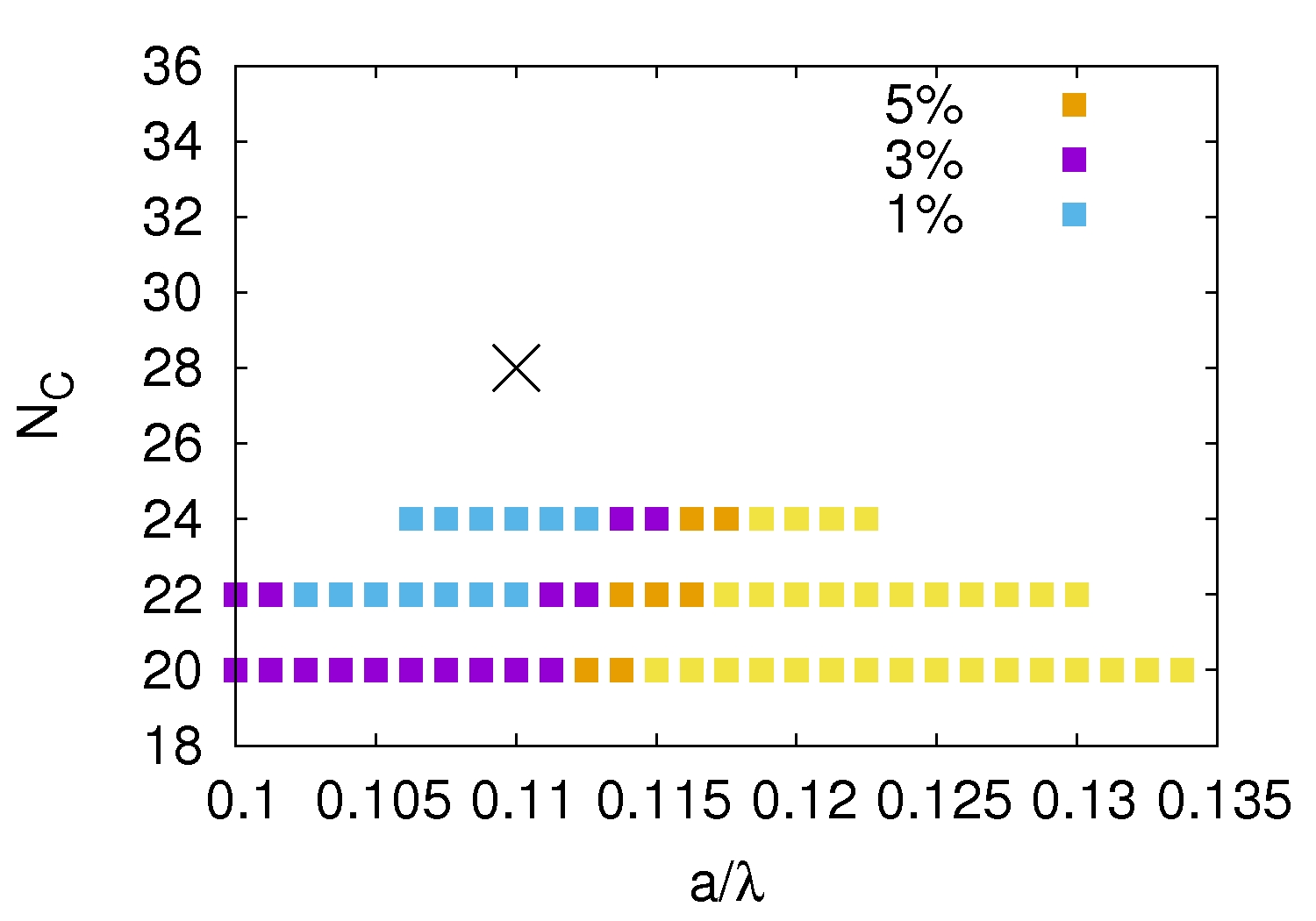}
\includegraphics[width=0.49\textwidth]{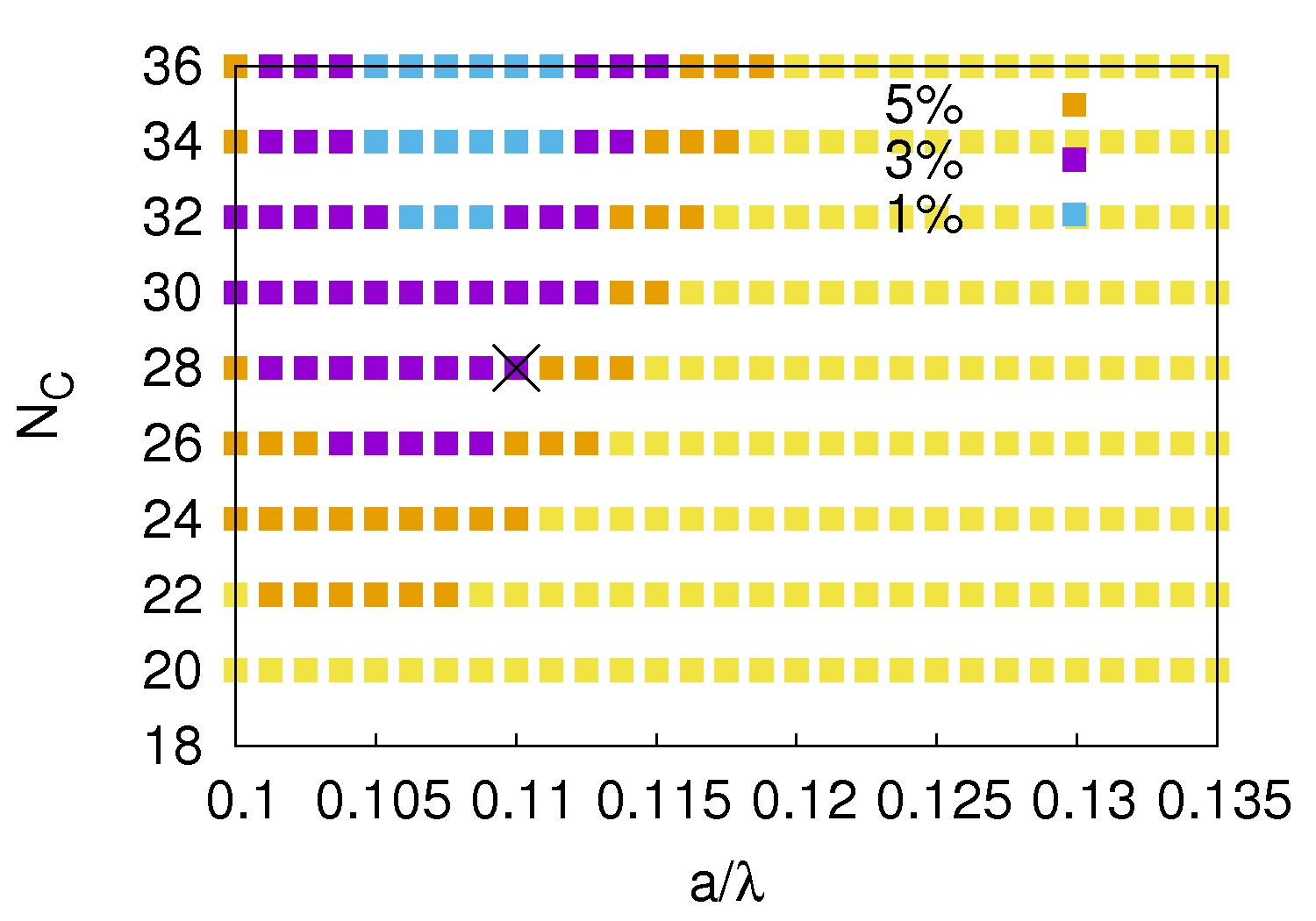}
\caption{Possible average aperture radii $a$ normalised to the RF wavelength $\lambda$ and number of cells $N_\text{c}$ for the structures that meet the luminosity goal at 3\,TeV (in yellow).
The most efficient structures, which do not require more than 1\%, 3\% or 5\% higher power consumption than the best, are indicated by blue, purple and orange colours.
For all parameter combinations without a point, no structure solution is found.
The spot ``$\times$'' indicates the parameters of the CDR structure (``CLIC\_G'').
Left: The solutions based on the current state of the conservative RF limits.
Right: The solutions for $S=1.2$.
}
\label{f:3TeV1}
\end{figure}

\begin{figure}
\centering
\includegraphics[width=0.49\textwidth]{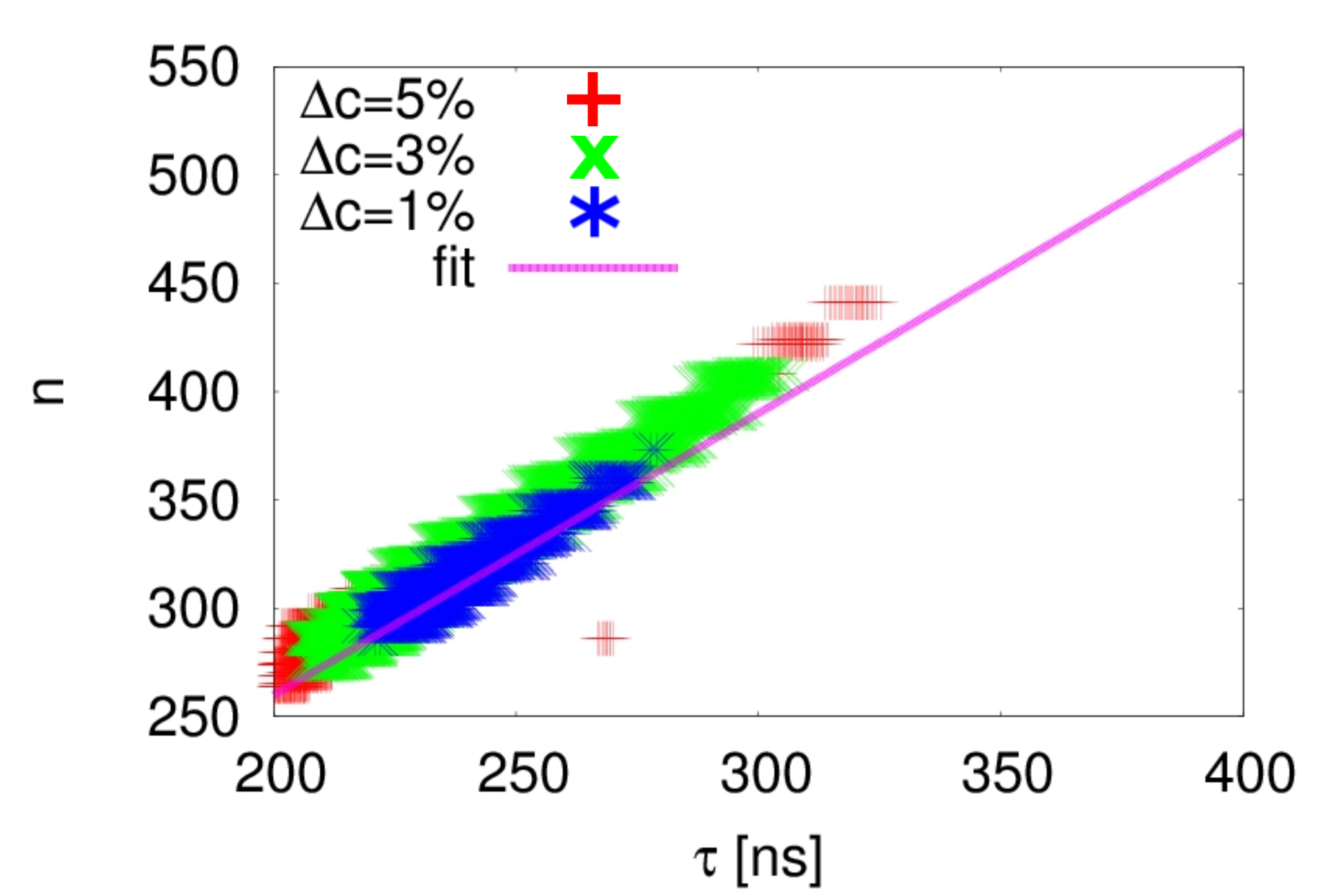}
\includegraphics[width=0.49\textwidth]{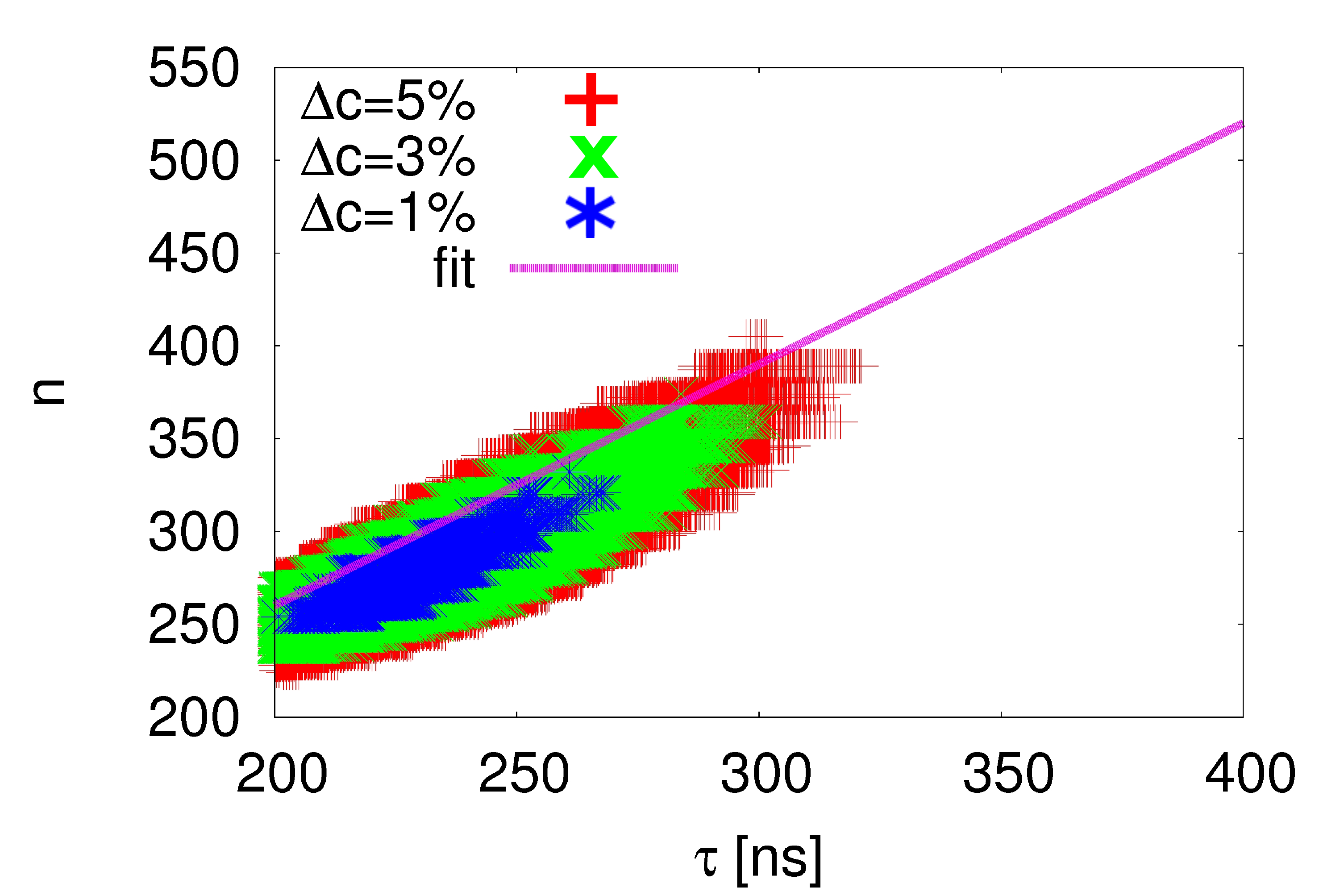}
\caption{The RF pulse length $\tau_{\text{RF}}$ and number of bunches $n_\text{b}$ for the most power efficient structures at 3\,TeV. The different colours indicate the relative cost increase $\Delta\,c$ with respect to the minimum cost.
Comparison with the reference line shows that for equal pulse length fewer bunches are needed when better RF performance is assumed.
Left: For the conservative RF limits ($S=1.0$) and a gradient of $G=100$\,MV/m.
Right: For $S=1.2$ and a gradient of $G=120$\,MV/m.
}
\label{f:3TeV2}
\end{figure}

The possible structure parameters for a 3\,TeV design are shown in \autoref{f:3TeV1}. On the left-hand side, the options for $S=1.0$ are shown.
The optimum structure has the same aperture radii as CLIC\_G, the structure of the CLIC design documented in the CDR,
but it is shorter by four cells. On the right-hand side, the options for
$S=1.2$ are shown. Again the optimum structure is similar to CLIC\_G, though slightly longer.

In \autoref{f:3TeV2}, the optimum pulse lengths and number of bunches per train are shown for the conservative RF limits and $G=100$\,MV/m.
The current pulse length of $\tau_{\text{RF}}=244\,\text{ns}$ is well within this optimum range.
If larger gradients can be reached, this optimum does not shift by very much, as can
be seen on the right-hand side of the figure. 
Here, a somewhat optimistic value of $S=1.2$ and a gradient of $G=120$\,MV/m are used for the demonstration.

While experimental results from various other structure tests are positive, it still remains to be demonstrated that CLIC\_G can reach the fully loaded gradient of 100\,MV/m. 
Other structures can reach this gradient using more pessimistic assumptions about the RF limits. 
Designs based on these structures require similar RF pulse lengths and number of bunches per train as the CDR design.
Hence an RF pulse length of $\tau_{\text{RF}}=244\,\text{ns}$ remains a good choice for the staged design. 
This remains true even if larger gradients can be reached in the future.

\subsubsection{Optimum at 380\,GeV}

\begin{figure}
\centering
\includegraphics[width=0.49\textwidth]{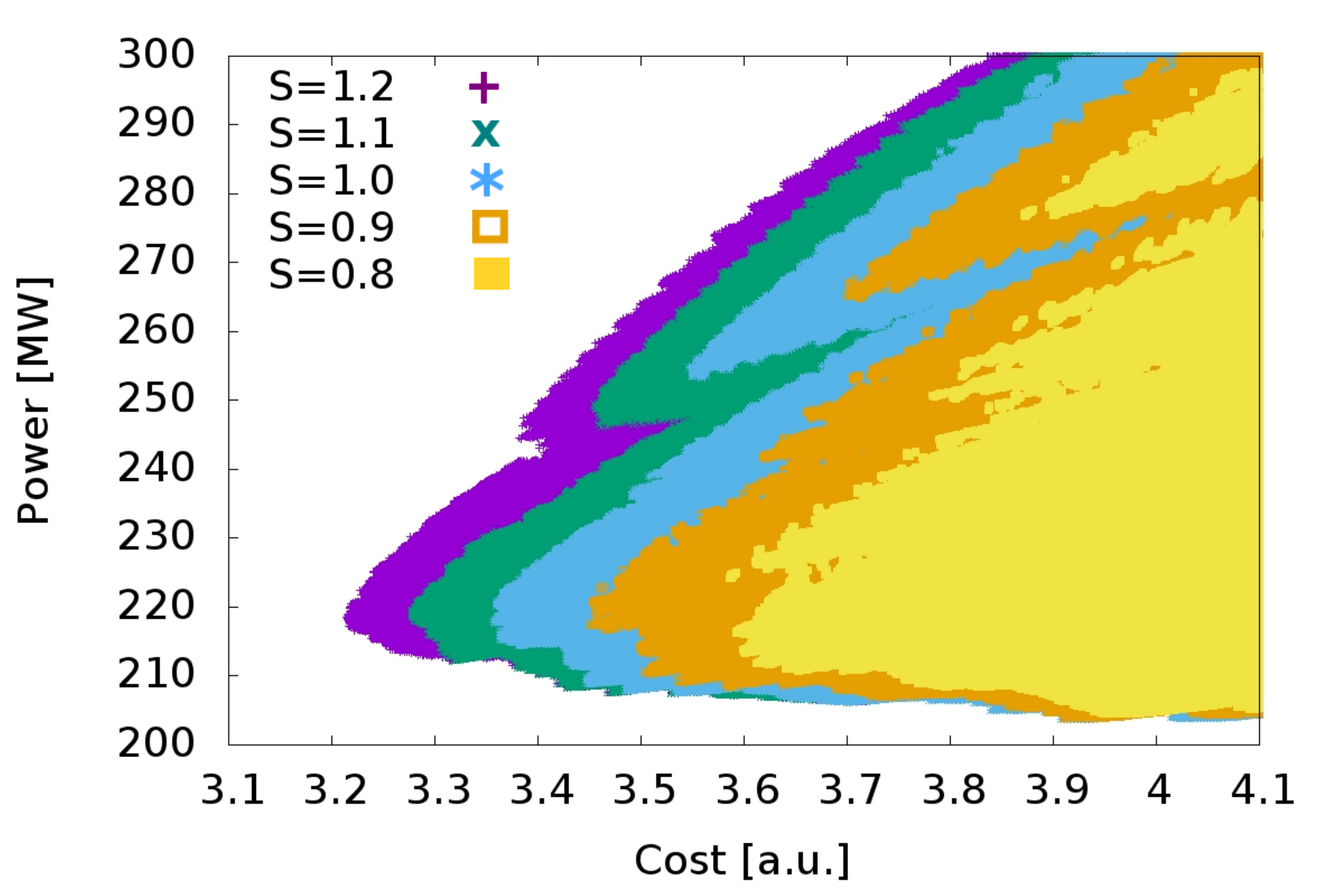}
\includegraphics[width=0.49\textwidth]{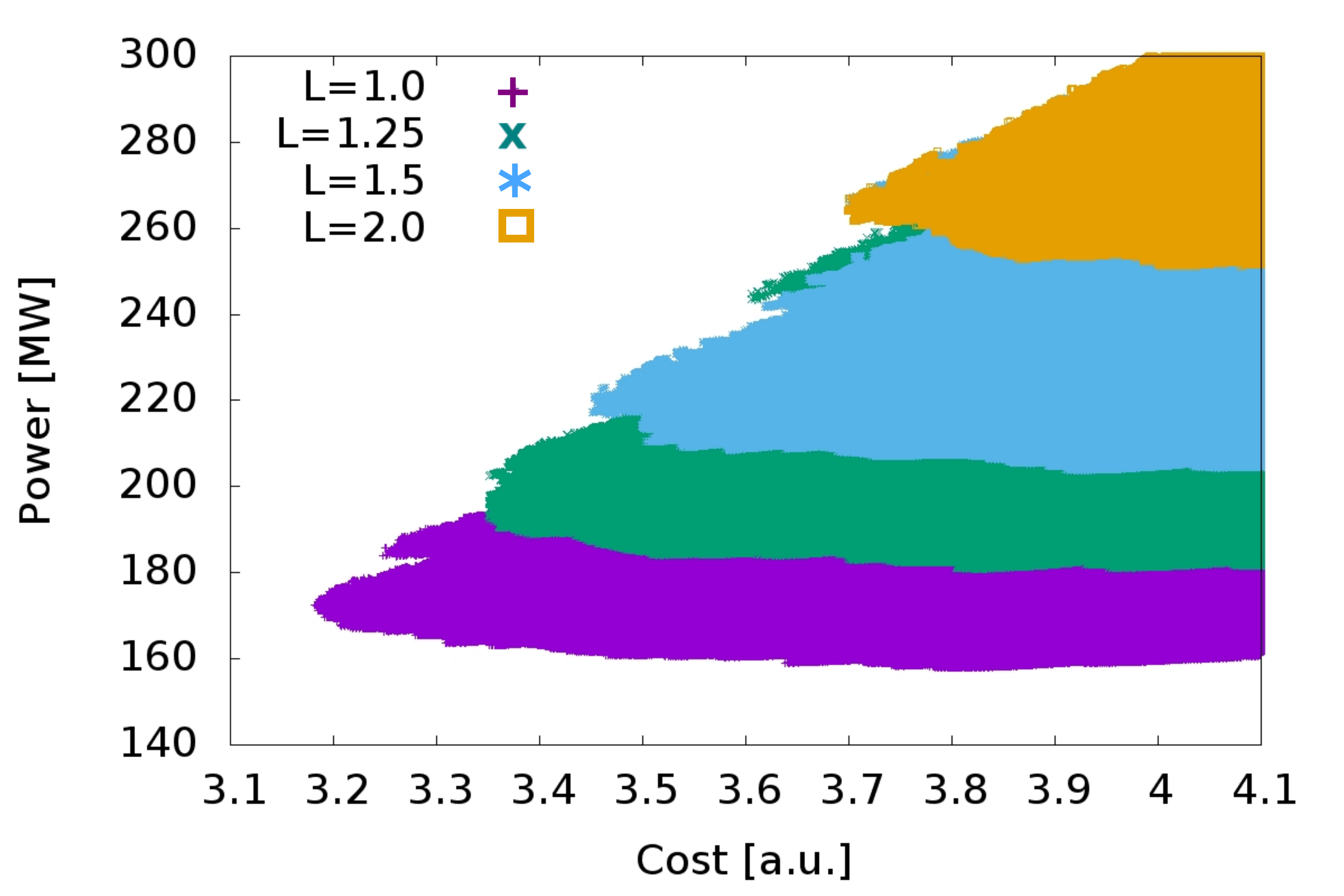}
\caption{Left: Cost and power consumption of the different possible designs at a centre-of-mass energy of 380\,GeV and a luminosity of $L=1.5\times10^{34}\,\text{cm}^{-2}\text{s}^{-1}$.
Different safety margins for the gradient are used. Right: Cost and power consumption with a gradient safety margin of 10\% ($S=1.1$) for different luminosities (in units of $10^{34}\text{cm}^{-2}\text{s}^{-1}$).}
\label{f:380GeV1}
\end{figure}

The cost and power consumption of the different designs at 380\,GeV are shown in \autoref{f:380GeV1}.
In all cases, the cheapest variants of CLIC have a power consumption close to the minimum. 
Hence the cheapest design will always be used as the optimum choice. 
The safety margin in the gradient has a small impact on the cost. 
The choice of 10\% margin therefore does not lead to a significant cost or power consumption increase.

Aiming at higher instantaneous luminosities leads to somewhat higher cost and power consumption. 
On the other hand, reducing the luminosity below the target value of $L=1.5\times10^{34}\text{cm}^{-2}\text{s}^{-1}$ would only result in minor cost saving.

\subsubsection{Staging strategy}

\begin{figure}
\centering
\includegraphics[width=0.7\textwidth]{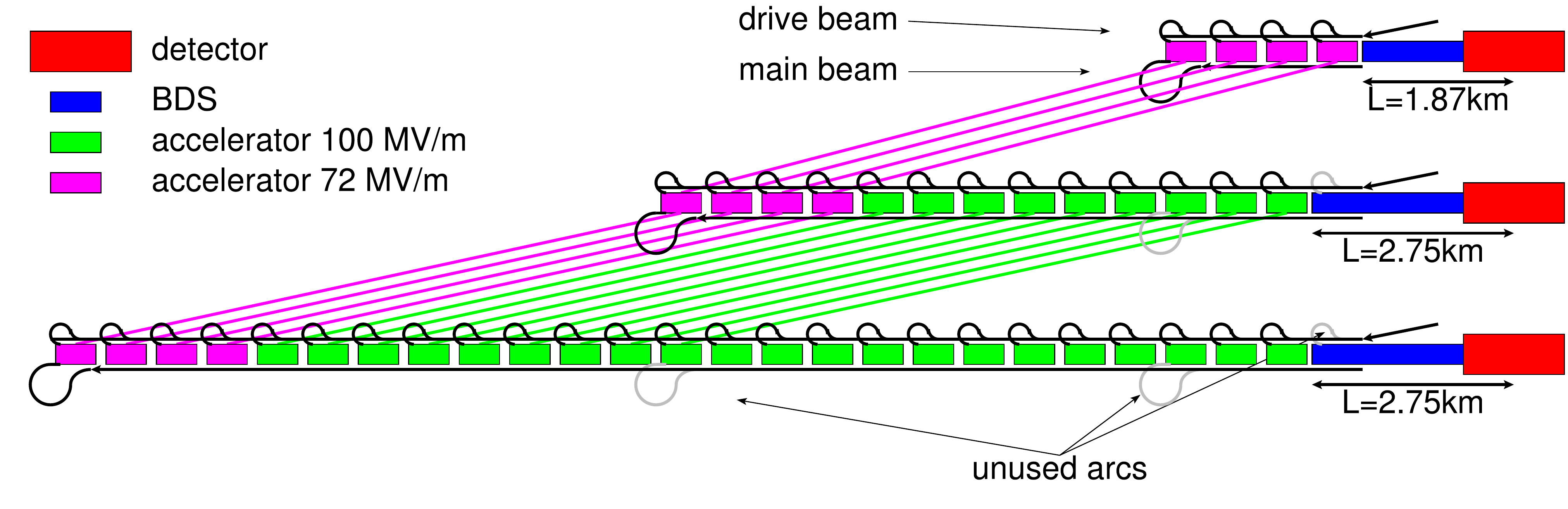}
\caption{One of the potential staging concepts. In this solution, the modules at the beginning of the previous main linac are moved to the new beginning during the upgrade.
}
\label{f:stage0}
\end{figure}
The concept of the staging is illustrated in \autoref{f:stage0}. In the first stage, the linac consists of
accelerating structures that are optimised for this energy range, while respecting the constraints for the energy upgrade.
When upgrading to higher energy, the modules containing
these structures will be moved to the beginning of the new linac and the remaining space is filled with new structures that
are optimised for 3\,TeV. Alternatively the old structures could be evenly distributed along the new linac.

This scheme places additional constraints on the first energy stage.  In order to minimise modifications of the drive beam complex, the RF pulse length of the first stage is chosen to be the same as for the subsequent energy stages.
In particular the turn-arounds in the main linac can be reused.
Within the main beam pulses the bunches have the same spacing at all energy stages to minimise the impact on the
main beam production complex. To be able to accelerate the full train of the final stage,
the fill time of the first-stage structures must be shorter and the bunch charge limit higher than in the final stage.

\begin{figure}
\centering
\includegraphics[width=0.6\textwidth]{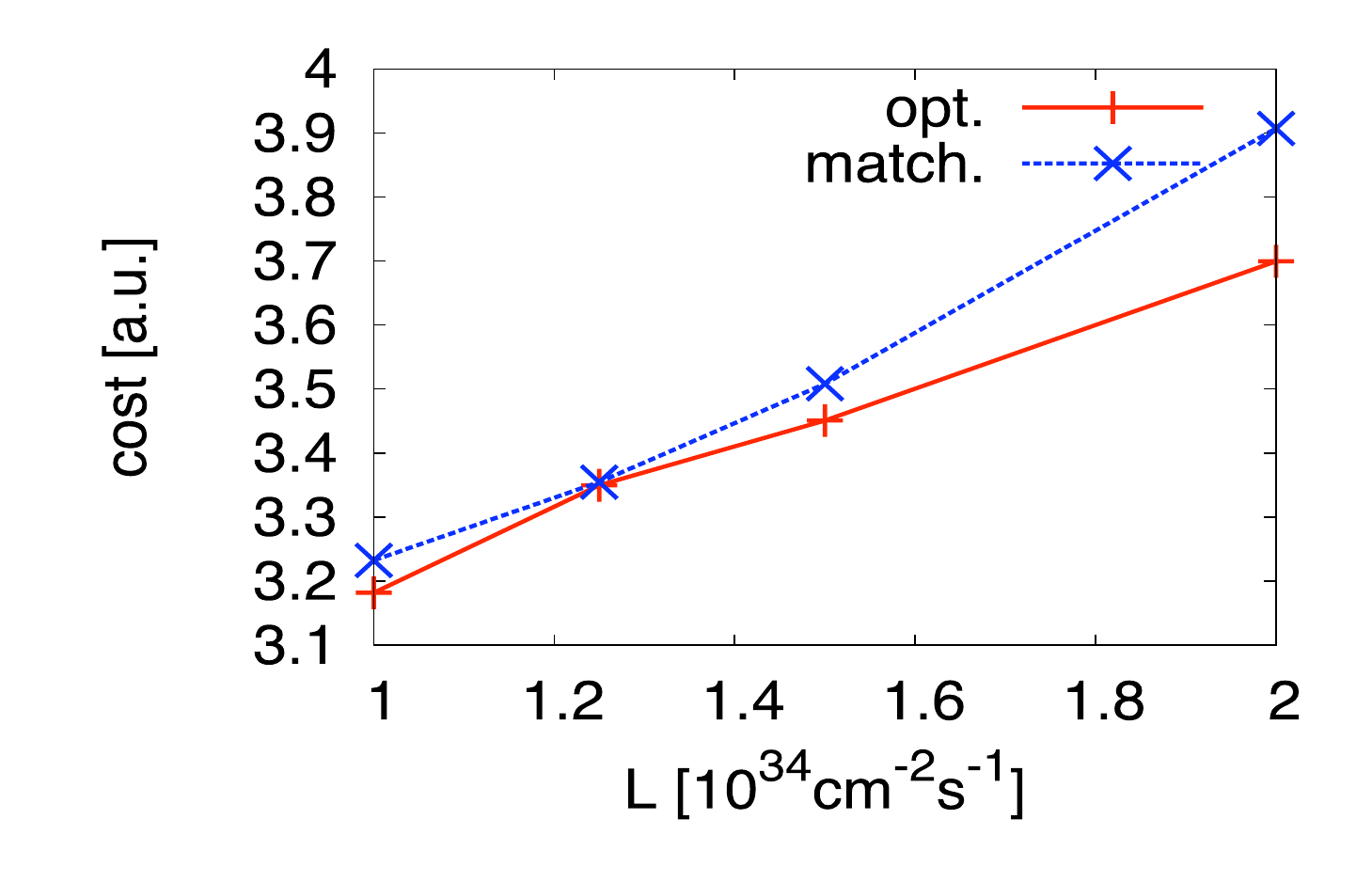}
\caption{The minimum cost of the first energy stage as a function of the luminosity target. The fully optimised (opt.) case and the one adapted for the upgrade (match.) are shown.}
\label{f:stage1}
\end{figure}

\autoref{f:stage1} illustrates the minimum cost of the first stage as a function of the luminosity. The values for the fully optimised case and for the solution
using the constraints from the 3\,TeV stage are shown. As can be seen, the cost difference is rather small, though slightly increasing towards higher luminosities.
For the target value of $L=1.5\times10^{34}\,\text{cm}^{-2}\text{s}^{-1}$ at 380\,GeV, the cost increases from 3.45 to 3.5 in arbitrary units.

The gradient of the structures for the first stage is 72\,MV/m. Consequently four decelerator stages are required per main linac in the first stage. The upgrade to 3\,TeV requires an additional 21 decelerator stages.

The structure parameters of the optimised structure for the first energy stage, in the following called ``DB244'' as it is optimised for drive beam acceleration and a 244\,ns RF pulse length, are shown in \autoref{tab:CLICG_DB244}. For comparison the parameters of the CDR structure CLIC\_G are also listed.

\begin{table}[t!]
\caption{\label{tab:CLICG_DB244}The parameters for the CLIC\_G and the DB244 structure designs.
}
\centering
\begin{tabular}{l r r r r }
\toprule
\textbf{Parameter}                    &\textbf{Symbol}        &\textbf{Unit} &\textbf{CLIC\_G} &\textbf{DB244} \\
\midrule
Frequency                             & $f$                   & GHz          & 12              & 12            \\
Acceleration gradient                 & $G$                   & MV/m         & 100             & 72            \\
\midrule
RF phase advance per cell             & $\Delta \phi$         & $^{\circ}$   & 120             & 120          \\
Number of cells                       & $N_{\text{c}}$        &              & 28              & 33            \\
First iris radius / RF wavelength     & $a_1/\lambda$         &              & 0.126           & 0.1625        \\
Last iris radius / RF wavelength      & $a_2/\lambda$         &              & 0.094           & 0.104         \\
First iris thickness / cell length    & $d_1/L_{\text{c}}$    &              & 0.200           & 0.303         \\
Last iris thickness / cell length     & $d_2/L_{\text{c}}$    &              & 0.120           & 0.172         \\
\midrule
Number of particles per bunch         & $N$                   & $10^9$       & 3.7             & 5.2           \\
Number of bunches per train           & $n_{\text{b}}$        &              & 312             & 352           \\
Pulse length                          & $\tau_{\text{RF}}$    & ns           & 244             & 244           \\
Peak input power into the structure   & $P_{\text{in}}$       & MW           & 61.3            & 59.5          \\
\bottomrule
\end{tabular}
\end{table}

\newpage
\section{CLIC staging baseline}
\label{sec:StagingBaseline}

\begin{figure}[t!]
  \centering
  \includegraphics[width=0.8\textwidth]{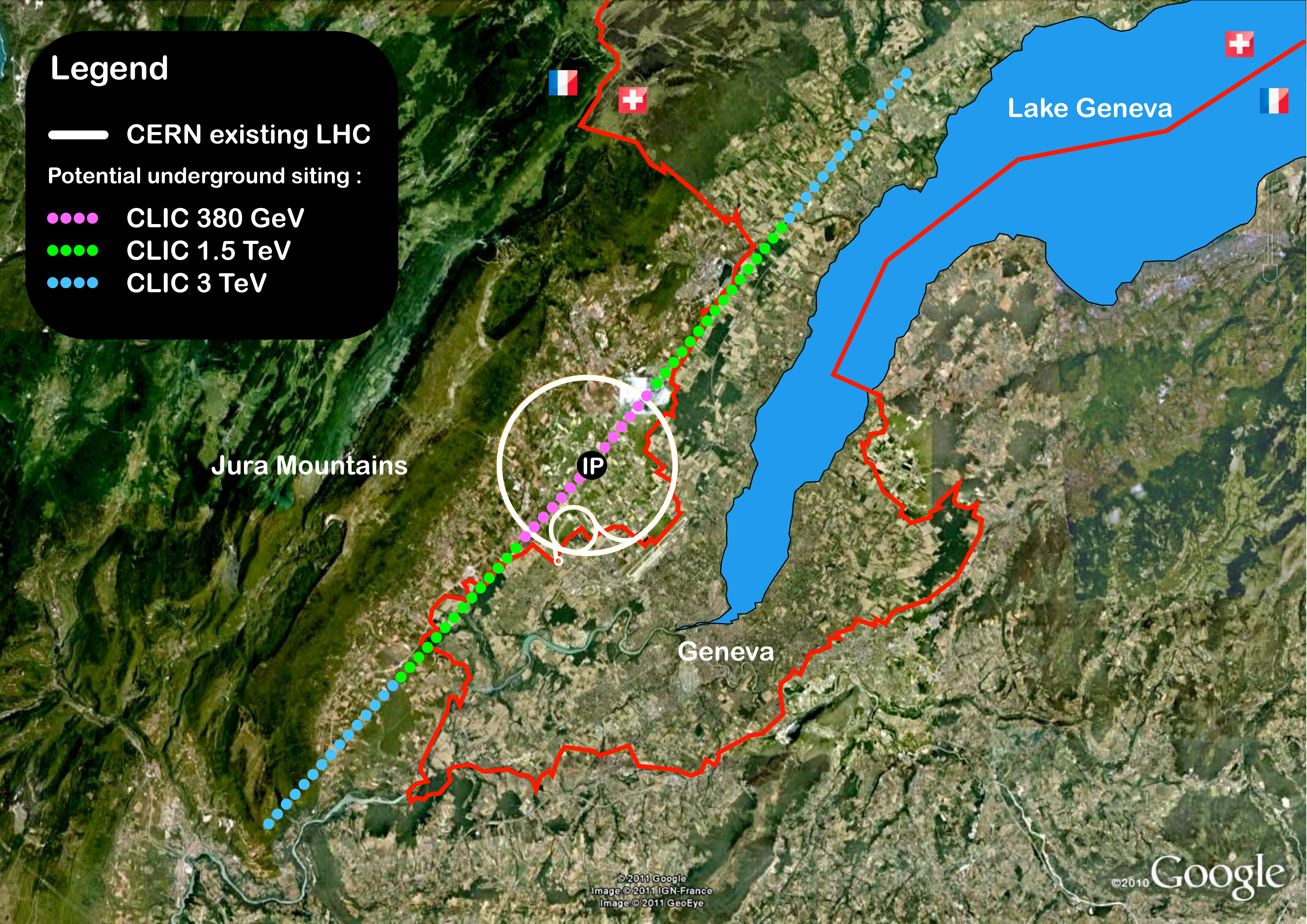}
  \caption{CLIC footprints near CERN, showing the three implementation stages.}
  \label{f:clicfootprints}
\end{figure}

The proposed CLIC staging scenario is described in the following sections. The accelerator is foreseen to be built in three stages with centre-of-mass energies of 380\,GeV, 1.5\,TeV and 3\,TeV.
An implementation plan for the three CLIC stages at CERN is shown in~\autoref{f:clicfootprints}.

\subsection{Description and performance parameters}
The CLIC accelerator can be built in energy stages, re-using existing equipment for subsequent stages.  
At each energy stage the centre-of-mass energy can be tuned to lower values, down to a third of the nominal energy, with limited loss of luminosity performance~\cite{CLICCDR_vol3}.
The key parameters of the present scenario, with stages at 380\,GeV, 1.5\,TeV and 3\,TeV are given in \autoref{t:1}.
In this scenario the first and second stages use a single drive-beam generation complex to feed both linacs, while in the third stage two drive beam complexes are needed. 
The schematic layouts of the CLIC accelerator complex at 380\,GeV and 3\,TeV are shown in \autoref{f:clic380gev} and \autoref{f:clic3tev}.

\begin{table}[t]
\caption{Parameters for the CLIC energy stages. The power consumptions for the 1.5 and 3\,TeV stages are from the CDR; depending on the details of the upgrade they can change at the percent level.}
\label{t:1}
\centering
\begin{tabular}{l l l l l l}
\toprule
\textbf{Parameter}                  & \textbf{Symbol}         & \textbf{Unit}& \textbf{Stage 1} & \textbf{Stage 2} & \textbf{Stage 3} \\
\midrule
Centre-of-mass energy               & $\roots$                &GeV                                        & 380 & 1500 & 3000\\
Repetition frequency                & $f_{\text{rep}}$        &Hz                                         & 50 & 50 & 50\\
Number of bunches per train         & $n_{b}$                 &                                           & 352 & 312 & 312\\
Bunch separation                    & $\Delta\,t$             &ns                                         & 0.5 & 0.5 & 0.5\\
Pulse length                        & $\tau_{\text{RF}}$      &ns                                         &244 &244 &244\\
\midrule
Accelerating gradient               & $G$                     &MV/m                                       & 72 & 72/100 & 72/100\\
\midrule
Total luminosity                    & $\mathcal{L}$           &$10^{34}\;\text{cm}^{-2}\text{s}^{-1}$     & 1.5 & 3.7 & 5.9 \\
Luminosity above 99\% of $\roots$   & $\mathcal{L}_{0.01}$    &$10^{34}\;\text{cm}^{-2}\text{s}^{-1}$     & 0.9 & 1.4 & 2\\
\midrule
Main tunnel length                  &                         &km                                         & 11.4 & 29.0 & 50.1\\
Number of particles per bunch                    & $N$                     &$10^9$                                     & 5.2 & 3.7 & 3.7\\
Bunch length                        & $\sigma_z$              &$\micron$                                  & 70 & 44 & 44\\
IP beam size                        & $\sigma_x/\sigma_y$     &nm                                         & 149/2.9 & $\sim$ 60/1.5 & $\sim$ 40/1\\
Normalised emittance (end of linac) & $\epsilon_x/\epsilon_y$ &nm                                         & 920/20 & 660/20 & 660/20\\
Normalised emittance (at IP)        & $\epsilon_x/\epsilon_y$ &nm                                         & 950/30 & ---    &---\\
Estimated power consumption                        & $P_{\text{wall}}$         &MW                                & 252    & 364    & 589\\
\bottomrule
\end{tabular}
\end{table}

\begin{figure}[t]
  \centering
  \includegraphics[scale=0.82]{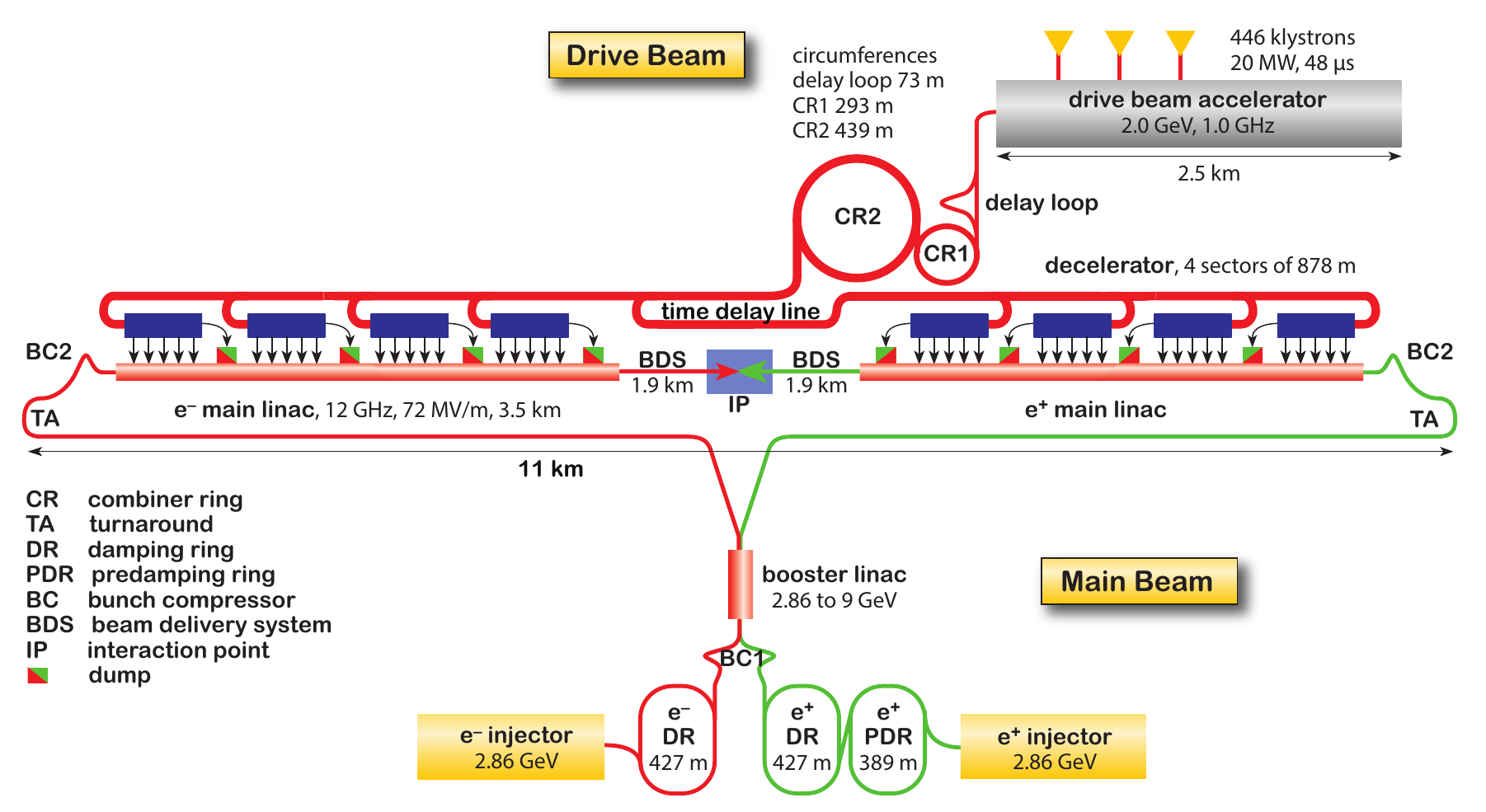}
  \caption{Overview of the CLIC layout at $\sqrt{s}=380$\,GeV. 
  Only one drive beam complex is needed for the first (and second) stage of CLIC.
} 
  \label{f:clic380gev}
\end{figure}

\begin{figure}[t]
  \centering
  \includegraphics[scale=0.82]{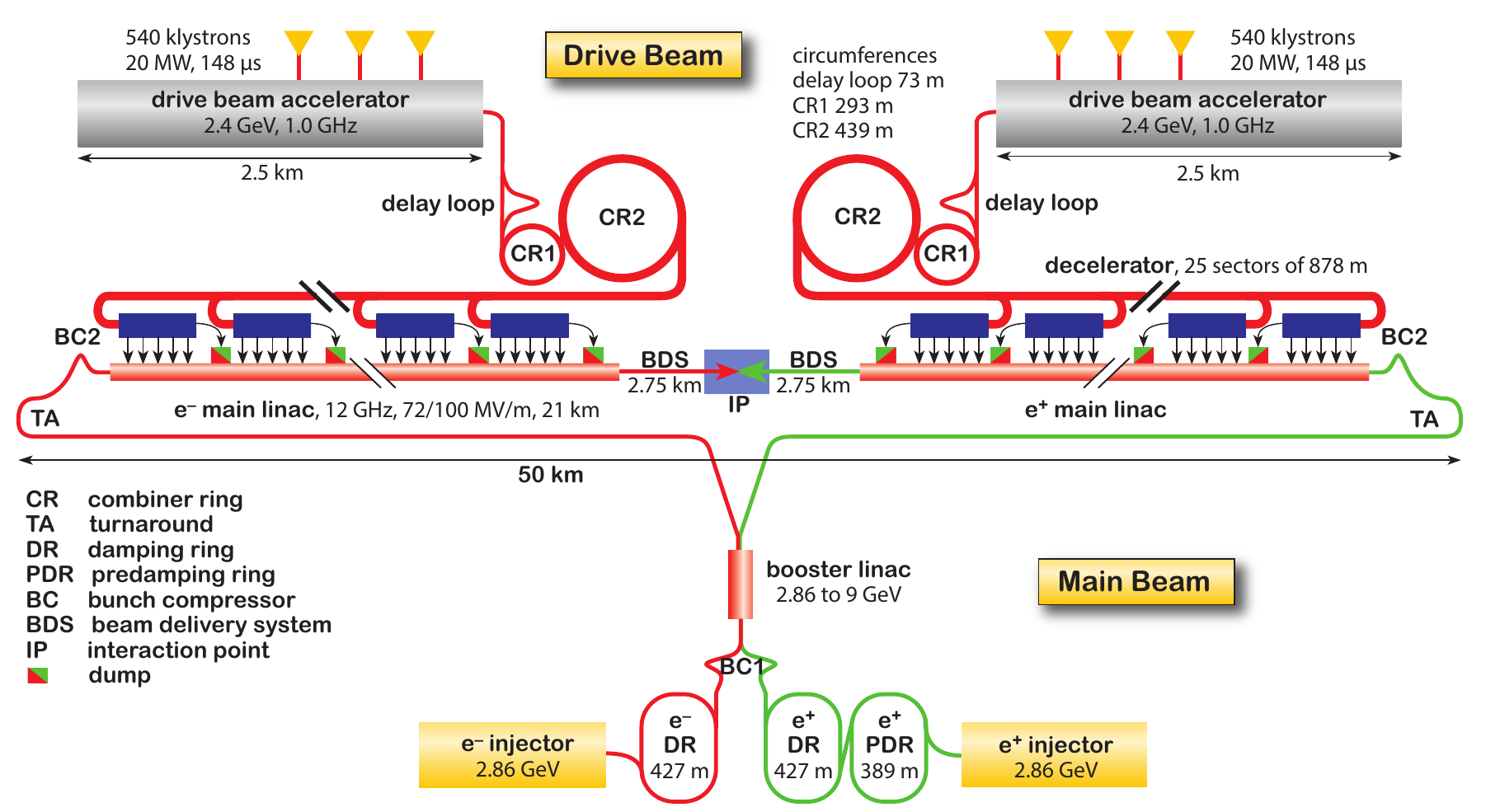}
  \caption{Overview of the CLIC layout at $\sqrt{s}=3$\,TeV.}
  \label{f:clic3tev}
\end{figure}

\subsection{Operating scenario and luminosities}
\label{sec:operationScenario}
The {CLIC} project as outlined is an ambitious long-term
programme, with an initial 7 year construction period~\cite{CLICCDR_vol3} and three energy
stages each lasting 7, 5 and 6 years respectively to achieve the integrated luminosity goals, interrupted by 2-year upgrade periods. The overall duration of the three-stage programme is 22 years from the start of operation.
The operating scenario currently foreseen for the complete CLIC programme is sketched in \autoref{fig:LumiPerYear} and \autoref{fig:integratedLumi} in terms of the luminosity and integrated luminosity as a function of the year of operation.
The duration of each stage is defined by the integrated luminosity targets of 500\,fb$^{-1}$ at 380\,GeV, 1.5\,ab$^{-1}$ at 1.5\,TeV and 3\,ab$^{-1}$ at 3\,TeV collision energy. During the first stage a top threshold scan will be performed near 350\,GeV. For this scan an additional integrated luminosity of 100\,fb$^{-1}$ will be collected during a few months of CLIC operation.

\begin{figure}[ht!]
  \centering
  \includegraphics[width=0.55\textwidth]{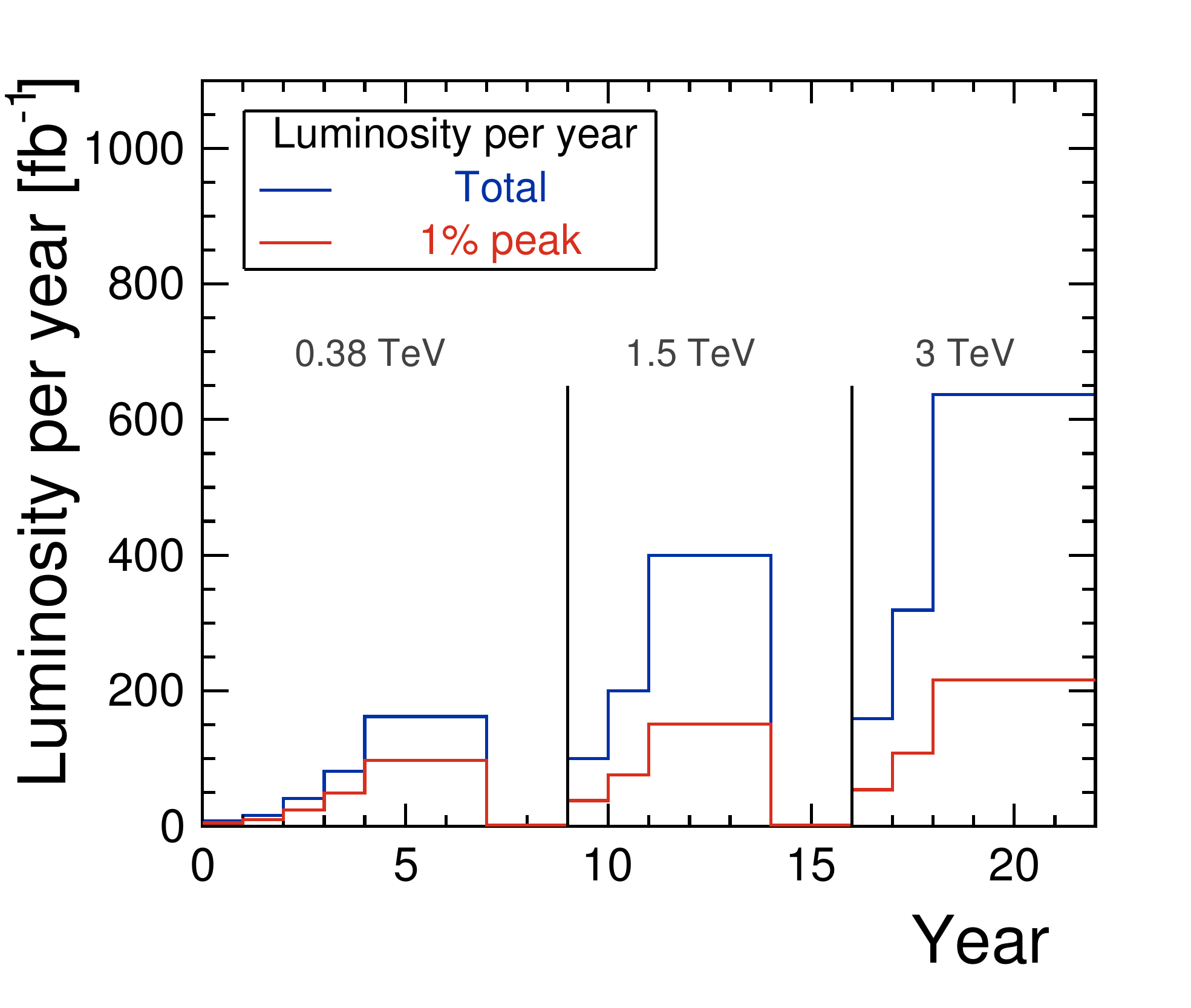}
   \caption{Luminosity per year in the considered staging scenario. 
    Years are counted from the start of beam commissioning.  
    This figure includes luminosity ramp-up of four years (5\%, 10\%, 25\%, 50\%) in the first stage and two years (25\%, 50\%) in subsequent stages.}
  \label{fig:LumiPerYear}
\end{figure}

\begin{figure}[ht!]
  \centering
  \includegraphics[width=0.55\textwidth]{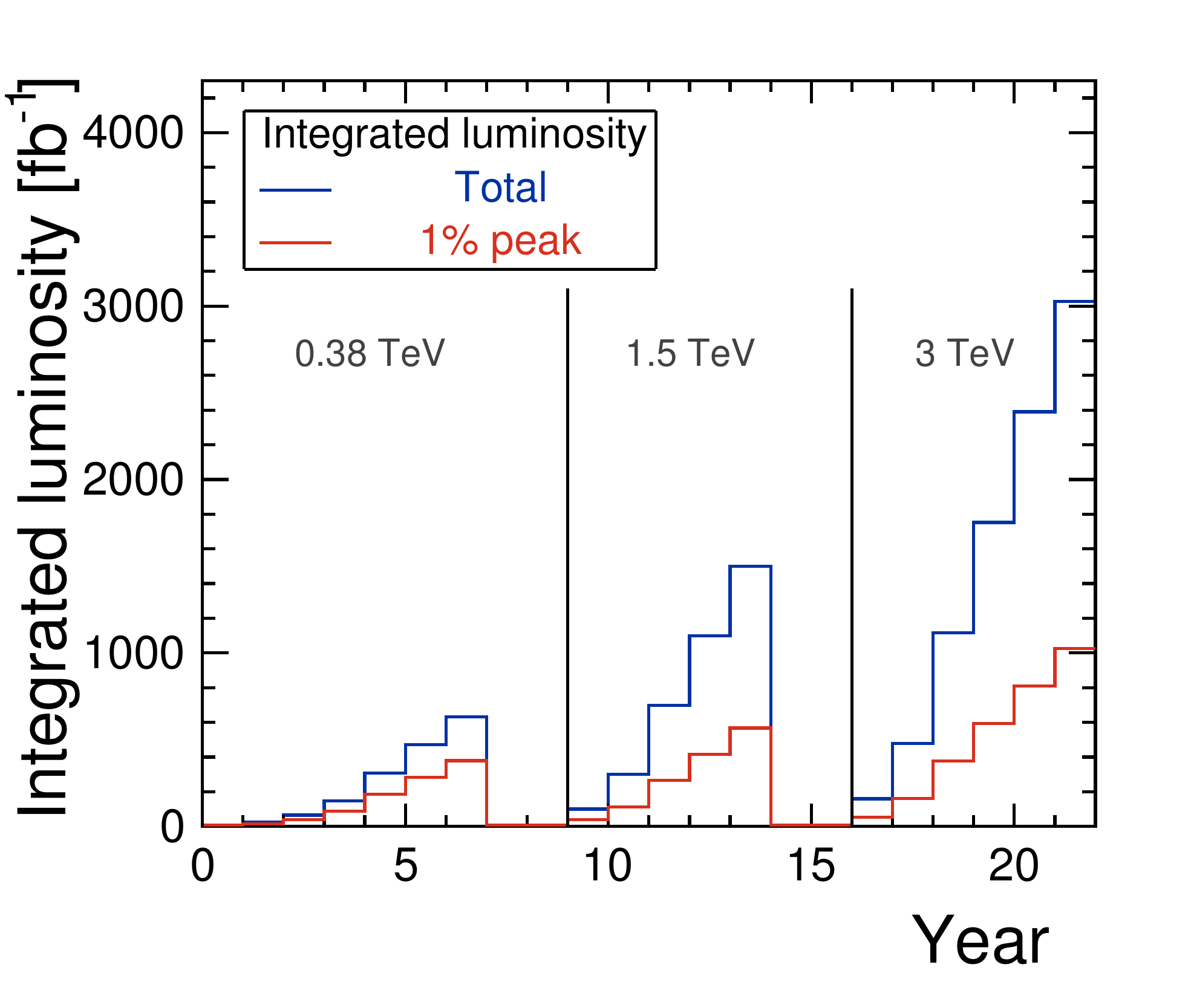}
   \caption{Integrated luminosity in the considered staging scenario. 
    Years are counted from the start of beam commissioning.  
  The luminosity ramp-up corresponds to what is described in \autoref{fig:LumiPerYear}.
}
  \label{fig:integratedLumi}
\end{figure}

\subsection{Power and energy consumption}
\label{sec:PowerEnergy}

The nominal power consumption of CLIC at 380\,GeV centre-of-mass energy has been estimated using the parametric model~\cite{Jeanneret:1599195} derived from the estimates of the CDR~\cite{CLICCDR_vol1}. This yields a total of 252\,MW for all accelerator systems and services, including experimental area and detectors and taking into account network losses for transformation and distribution on site. The breakdowns of this value per domain of the CLIC accelerator complex and per technical system are shown graphically in \autoref{fig:power}. 
Most of the power is used in the drive-beam and main-beam injector complexes, comparatively little in the main linacs. Among the technical systems, the RF represents the major consumer.

\begin{figure}[t!]
\includegraphics[width=0.5\textwidth]{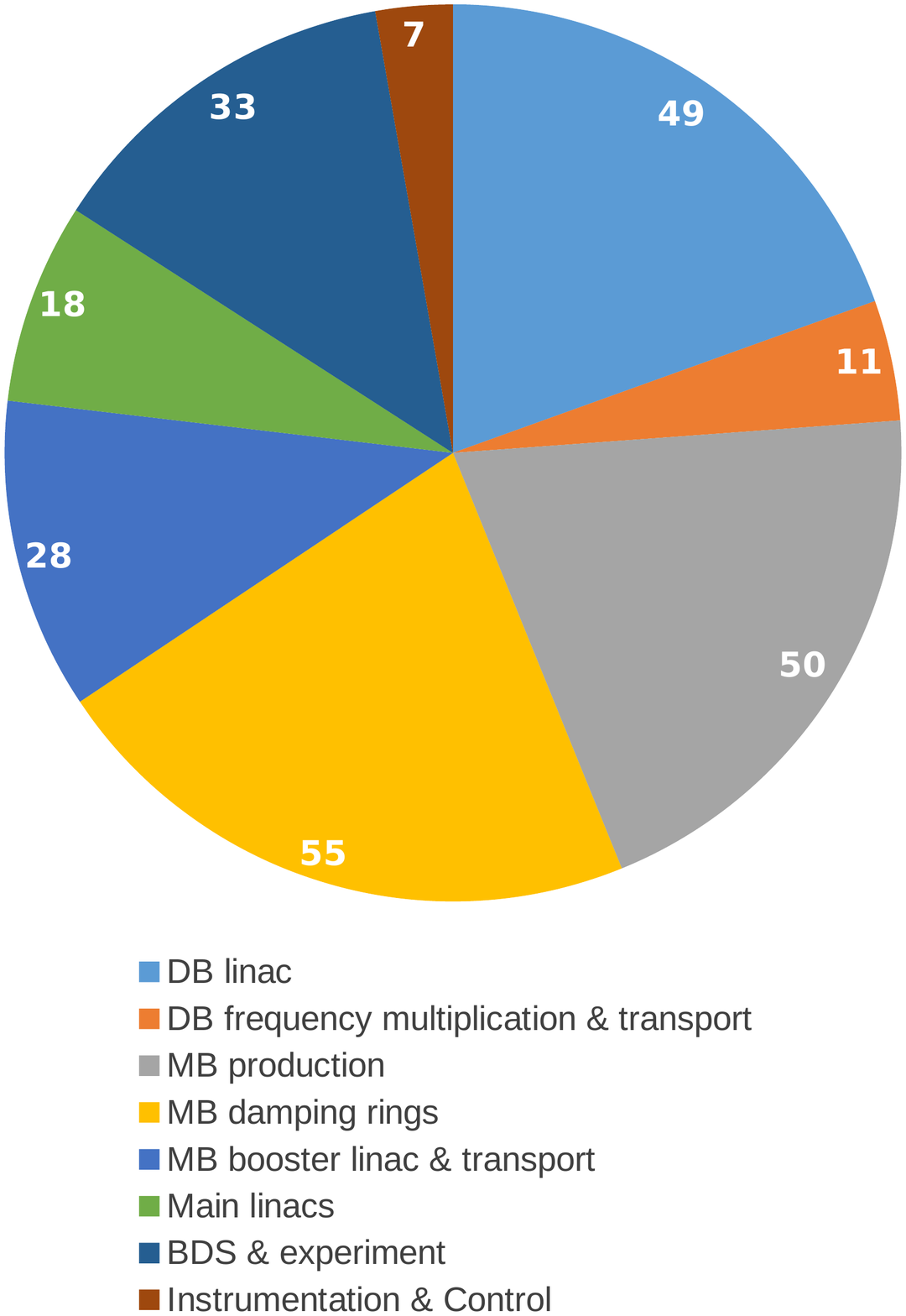}
\includegraphics[width=0.5\textwidth]{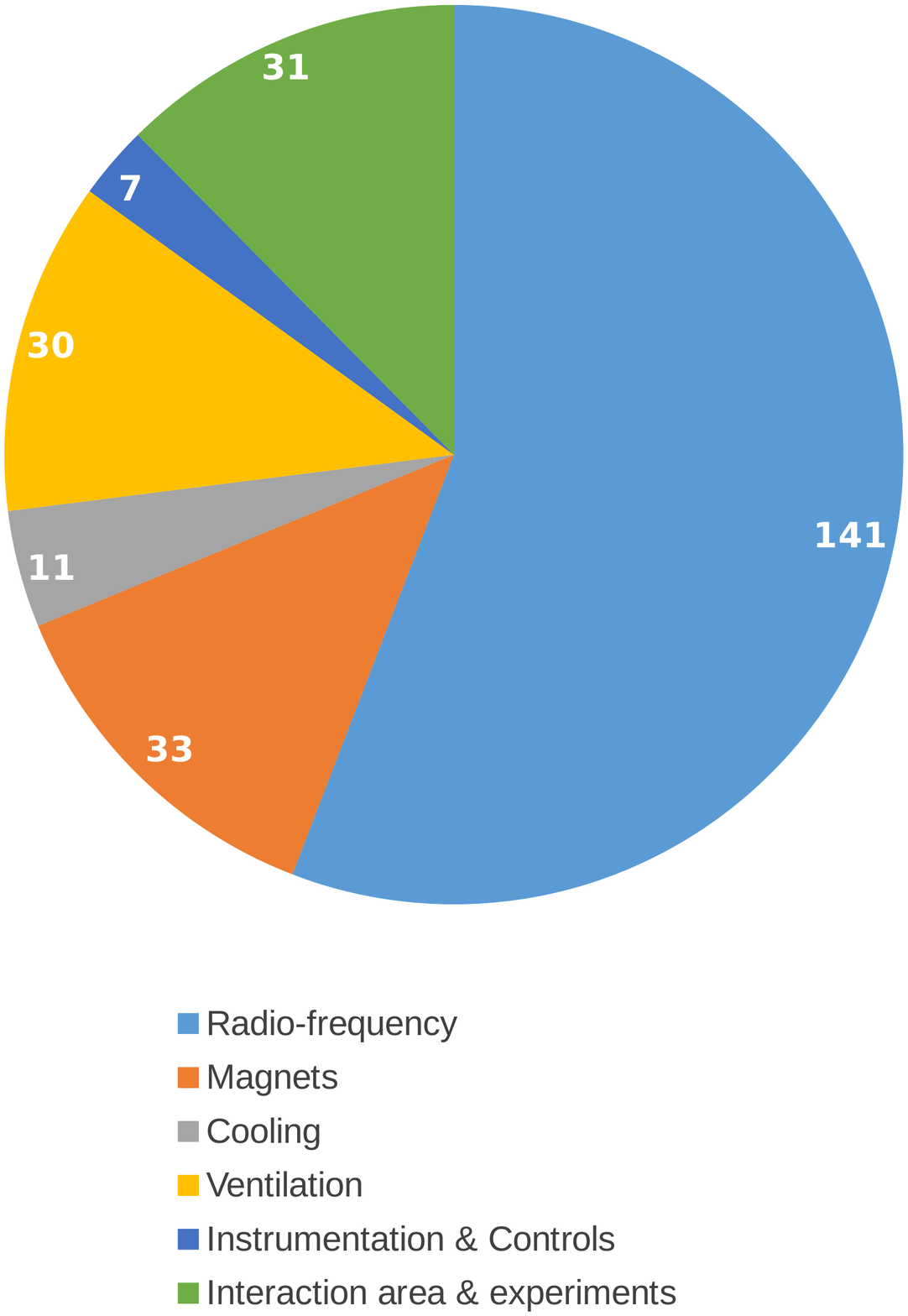}
\caption{Estimated power consumption of CLIC in MW at a centre-of-mass energy of 380\,GeV. 
The contributions add up to a total of 252\,MW. 
Left: breakdown of power consumption between different domains of the accelerator complex. 
Right: breakdown of power consumption between different technical systems.}
\label{fig:power}
\end{figure}

\autoref{tab:power} shows the nominal power consumption of CLIC at the different stages, as well as the residual values for two modes corresponding to short (``waiting for beam'') and long (``stop'') beam interruptions. At any stage the power consumption has a large volatility, allowing CLIC to be operated as a peak-shaving facility for the electrical network, matching not only seasonal, but also daily fluctuations of the demand. This particular feature constitutes a strong asset towards optimal energy management, a necessary approach in view of the large values of power consumption of the CLIC complex during nominal operation.

\begin{table}[t]
  \caption{\label{tab:power}CLIC estimated power consumption for the updated staging scenario.
 Values at the 1.5\,TeV and 3\,TeV centre-of-mass energy stages are taken from the CDR \cite{CLICCDR_vol1}.
}
  \centering
  \begin{tabular}{l l l l }
    \toprule
    $\roots$ [TeV] &
    $P_{\textrm{nominal}}[\textrm{MW}]$ &
    $P_{\textrm{waiting for beam}}[\textrm{MW}]$ &
    $P_{\textrm{stop}}[\textrm{MW}]$ \\
    \midrule
    0.38 & 252 & 168 & 30\\
    1.5  & 364 & 190 & 42\\
    3.0  & 589 & 268 & 58\\
    \bottomrule
  \end{tabular}
\end{table}

Estimating yearly energy consumption from the power numbers requires an annual operational scenario (\autoref{fig:operationSchedule}). In any ``normal'' year, i.e. once CLIC will have been fully commissioned, we consider a 90-day annual shutdown, and an additional 50 days of scheduled maintenance stops (typically 1 day per week and 1 week after every 2 months of running). Out of the remaining 225 days, we assume 80\% availability, i.e. 45 days of fault-induced stops. This leaves 180 days for operation, of which 55 days are allocated to machine development and tuning runs, thus yielding 125 days for physics data taking (``luminosity runs''). This is the assumption used for estimating the build-up of integrated luminosity in~\autoref{fig:integratedLumi}.

\begin{figure}[t!]
\centering
\includegraphics[width=0.7\textwidth]{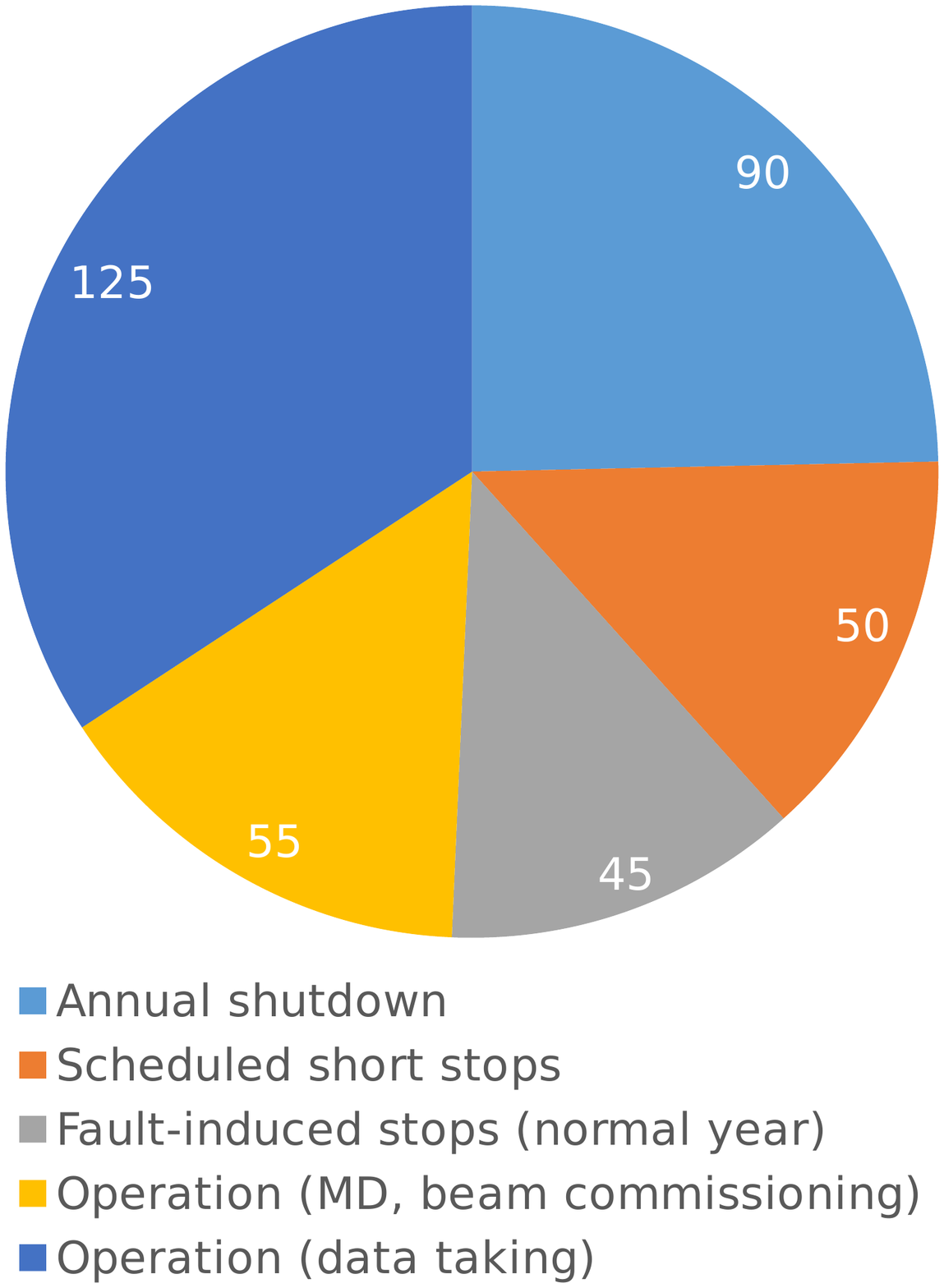}
\caption{Operation schedule in ``normal'' year (days/year).}
\label{fig:operationSchedule}
\end{figure}

For energy consumption, one also has to consider reduced operation in the first years at each energy stage, similar to what was done in the CDR~\cite{CLICCDR_vol1}. For example, at 380\,GeV centre-of-mass energy a single positron target is used for the first three years (10\,MW less with respect to nominal).

During the first year at each centre-of-mass energy, we consider the 180 days of operation to be composed of three periods of 60 days each. In the first period, only one bunch train is produced in order to commission the drive-beam generation complex, and then to commission each decelerator in turn, one at a time. In the second period, the main linacs are commissioned, one at a time. Nominal operation occurs during the third period at nominal power. 
The energy is the power integrated over time, including the period without beam. During operation with beam, an additional down-time of 50\% is taken into account in the first year.

During the second year at each centre-of-mass energy, we consider the 180 days of operation to be composed of two periods of 90 days. In the first period, only one main linac is powered at a time and the additional down-time is taken at 30\%. Nominal conditions are assumed in the second period. The third and following years are nominal in all centre-of-mass energy stages. The evolution of electrical energy consumption over the years is illustrated in \autoref{fig:energyConsuption}. For comparison, CERN's energy consumption in 2015 was approximately 1.3 TWh, of which the accelerator complex used approximately 90\%.

\begin{figure}[t!]
\centering
\includegraphics[width=0.55\textwidth]{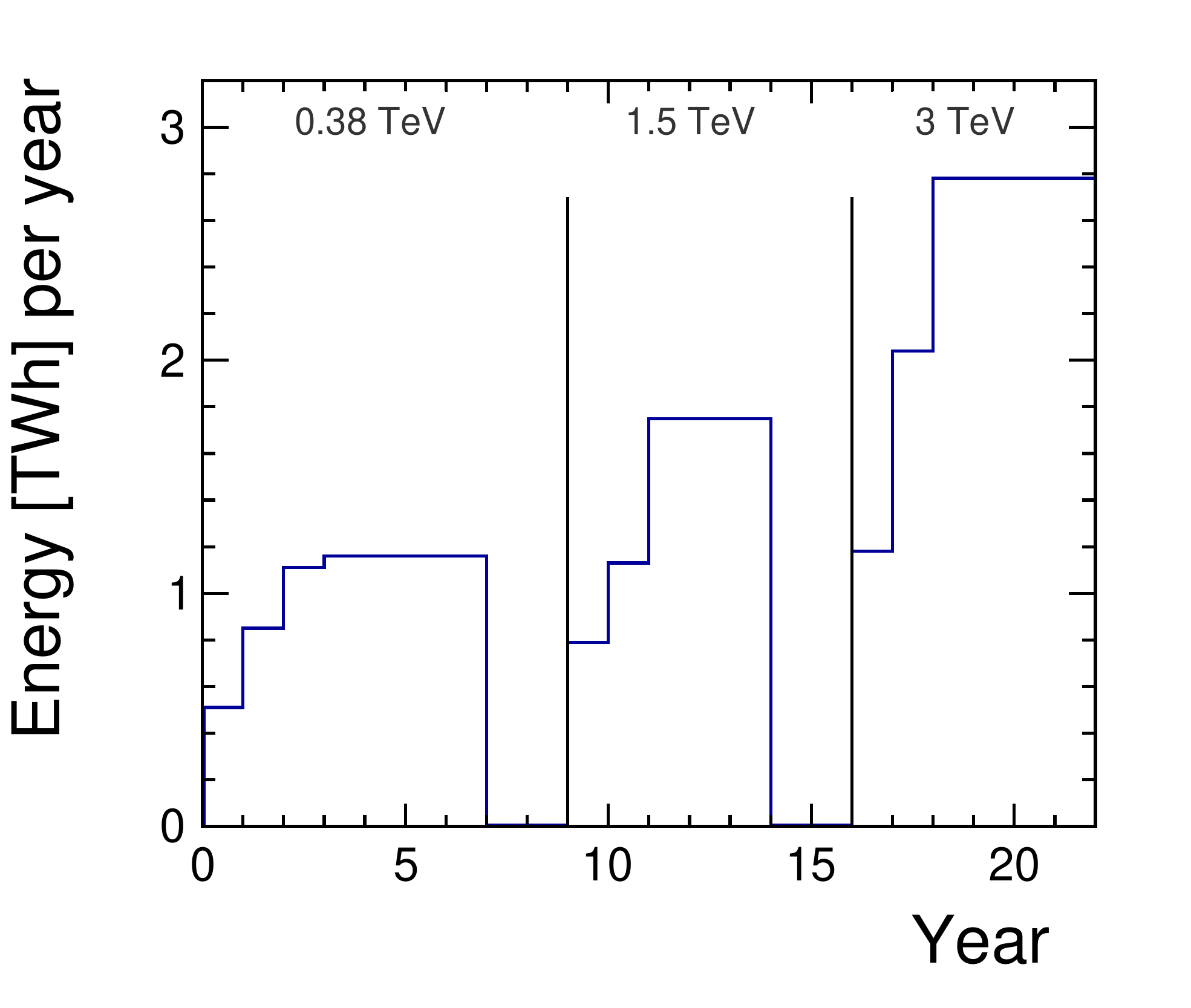}
\caption{Estimated yearly energy consumption of CLIC.}
\label{fig:energyConsuption}
\end{figure}

\subsubsection{Additional energy-saving options}
Several paths of development aiming at saving power and energy have been identified and are under investigation. The first one consists in achieving power sobriety by re-design, possibly trading operation expenditure against investment costs. This is what was done for the re-optimisation of the accelerating structures described under \autoref{sec:optproc}. Other possibilities are lower current density in normal-conducting magnets and cables, use of permanent or ''super-ferric'' magnets~\cite{TunablePermMagnets} and reduction of heat loads to the ventilation system through enhanced use of water cooling. A second path aims at improving efficiency in the use of electrical energy, e.g. by improving network-to-RF power conversion profiting from ongoing developments of more efficient modulators and klystrons \cite{klystron}. Overall, at 380\,GeV the total potential for such power savings is estimated to be around 30\,MW, thus lowering the nominal power consumption to approximately 220\,MW (see \autoref{tab:power}). In addition, waste heat recovery may also constitute an interesting option in view of the large power rejected into water, provided that the heat rejection temperature is high enough to be utilised and that associated needs exist in the neighbourhood.

\subsection{Cost estimate}
The CDR \cite{CLICCDR_vol3} presented value estimates of CLIC at 500\,GeV centre-of-mass energy for two staging scenarios, A and B. In scenario A the 500\,\gev stage deploys dedicated RF structures with a gradient of 80\,MV/m, whereas scenario B uses identical RF structures with 100 MV/m gradient at all energy stages. Scenario A has a better luminosity performance at 500\,\gev, while scenario B is more cost effective. 
These value estimates were built bottom-up from unit costs and quantities for components to subdomains to complete domains, following the work breakdown structure of the project. 
``On-the-shelf'' spare components were not included as they would be charged to the operation budget rather than the construction budget. 
However, ``hot spares'', i.e.\ spare components installed and operational to provide on-line redundancy were part of the value estimates. 
Unit costs were obtained from prices paid for other similar supplies or scaled from them after adequate escalation, and from specific industrial studies when the former were not available.
Uncertainties stemming from technical and commercial risks were also quantified, the latter from a statistical analysis of procurement for the LHC \cite{Lebrun:2010zz}. 
The value estimates were expressed in Swiss francs (CHF) of December 2010. 
``Explicit labour'', i.e. personnel costs not included in the value of supplied components (e.g. laboratory personnel for production follow-up, reception, installation and commissioning of equipment), was estimated globally by scaling from the numbers resulting from LHC construction experience.

As a complete description in the form of a work breakdown structure does not yet exist for the updated baseline at 380\,GeV centre-of-mass energy, the same method cannot be applied and a two-pronged approach is used. 
In most cases, value estimates for each subdomain of CLIC at 380\,GeV centre-of-mass energy are interpolated between values for CLIC 500 A and CLIC 500 B using relevant scaling factors. 
Examples of such cases include scaling main beam injector costs with main beam current, and utility infrastructure costs with nominal power consumption. 
In the specific cases where design changes and simplifications could be identified from the re-baselining study, updated costs were obtained from the analytical model described in \autoref{sec:opt:cost}, or else, corresponding cost differentials were applied to the reference value estimate at 500\,GeV centre-of-mass energy. 
This applies to the suppression of the electron pre-damping ring, to the new configuration of RF in the drive-beam linac and to the corresponding downsizing of its klystron gallery. 
For the sake of comparison with the CDR~\cite{CLICCDR_vol3}, and in view of the significant exchange-rate and purchasing-power-parity fluctuations of the Swiss franc with respect to other European currencies in the past years, it was decided to stay with value estimates expressed in Swiss francs of December 2010. 
Escalation may be applied to these numbers using published Swiss indices or the CERN material budget index.

The results of this exercise are given in \autoref{tab:cost}, yielding a total value of 6690\,MCHF for CLIC at 380\,GeV centre-of-mass energy. 
Comparison with CLIC 500 A and CLIC 500 B~\cite{CLICCDR_vol3} is shown graphically in \autoref{fig:cost}. 

\begin{table}[t!]
 \caption{Value estimate of CLIC at 380\,GeV centre-of-mass energy.}
 \label{tab:cost}
 \centering
 \begin{tabular}{lr}
\toprule
                                               & Value [MCHF of December 2010]\\
\midrule
Main beam production                           & 1245 \\
Drive beam production                          &  974 \\
Two-beam accelerators                          & 2038 \\
Interaction region                             &  132 \\
Civil engineering \& services                  & 2112 \\
Accelerator control \& operational infrastructure  &  216 \\
\midrule
Total                                          & 6690\\
\bottomrule
\end{tabular}
\end{table}

\begin{figure}[t!]
 \includegraphics[width=\textwidth]{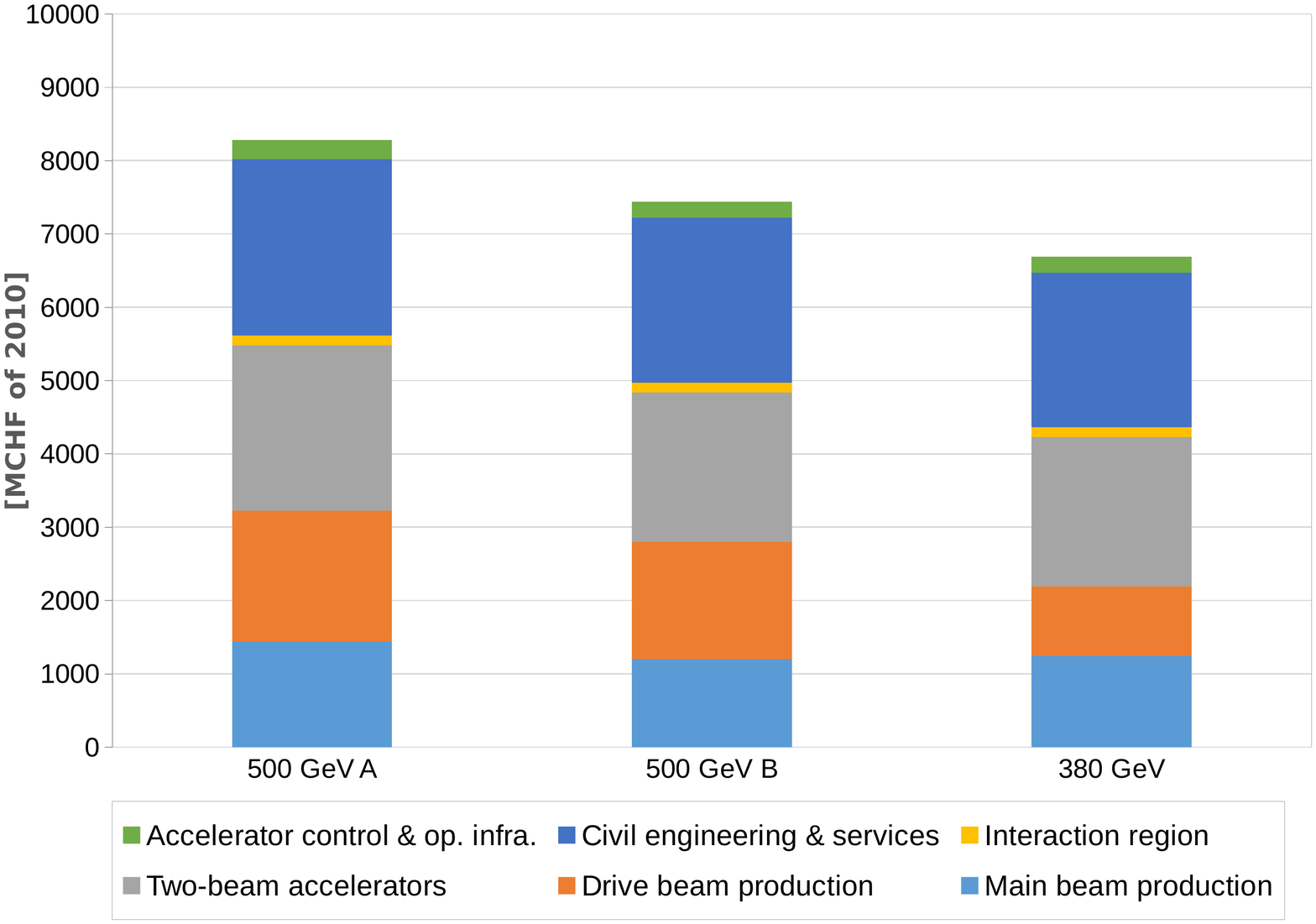}
 \caption{Value estimates of CLIC 500\,GeV A, 500\,GeV B and 380\,GeV, in MCHF of December 2010. The 500\,GeV numbers are taken from the CDR~\cite{CLICCDR_vol3}.}
 \label{fig:cost}
\end{figure}

A complete CLIC cost update is foreseen for 2018--2019, following the establishment of a detailed machine description and work breakdown structure. 
Further design improvements, technical developments and updated industrial quotes will be integrated into the process. 
\section{Alternative klystron-based scenario for the first CLIC stage}
\label{sec:alternative}

\begin{figure}
\centering
\includegraphics[width=0.6\textwidth]{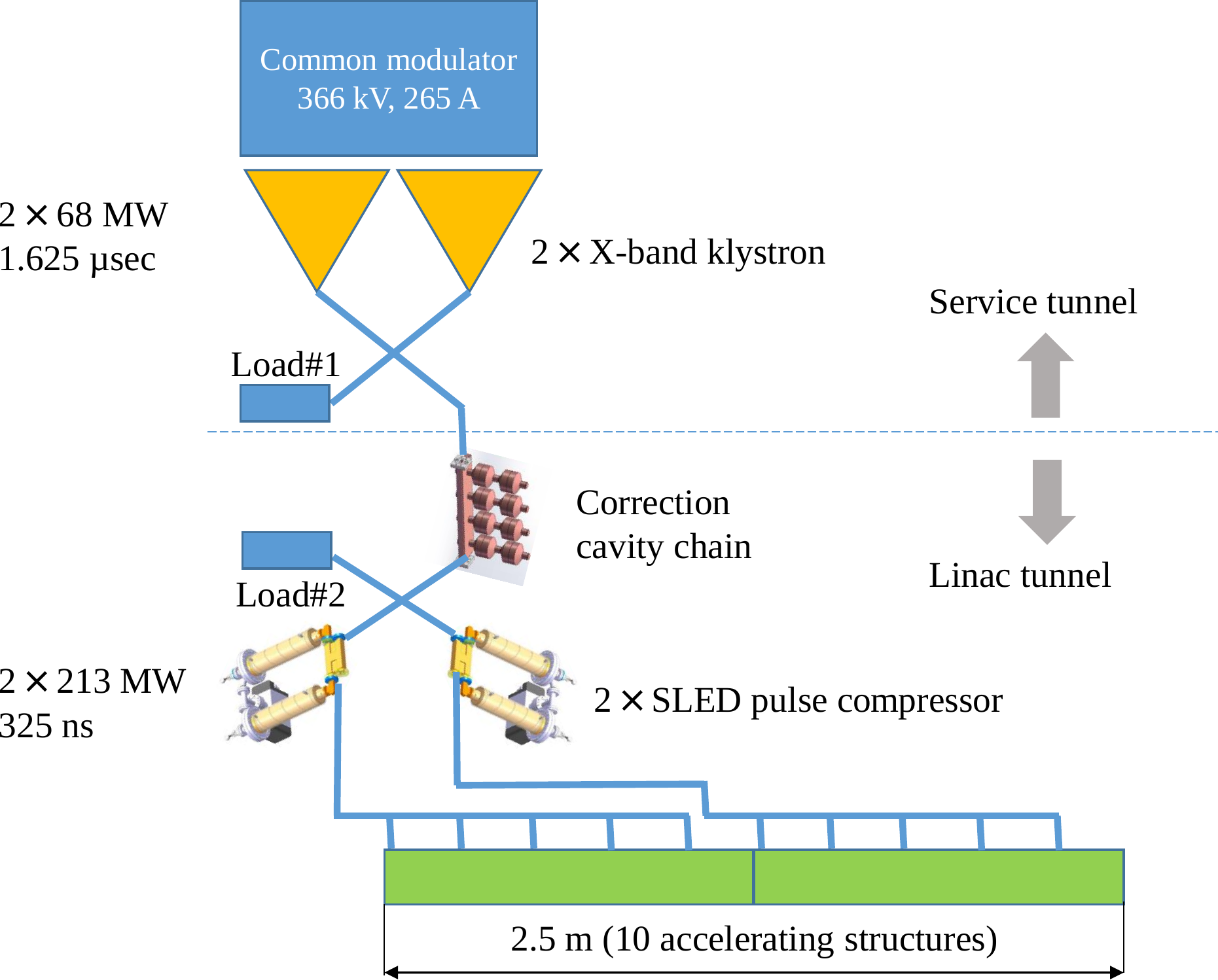}
\caption{
Conceptual Design of an RF unit for a klystron-based CLIC main linac. 
Two klystrons produce RF pulses, which are combined into a single double-power pulse.
The pulse passes the correction cavities chain, which modifies the pulse shape, and is then split into two pulses of half the power in order to feed two SLED pulse compressors.
Each SLED shortens the pulse by increasing the power; two compressors are used to limit the final power in each of them. 
Finally, the pulses are split and distributed into five accelerating structures.
}
\label{f:rfunit}
\end{figure}

An alternative to the CLIC drive-beam scheme is to produce the RF power for the main linac using X-band klystrons.
At the 1.5\,TeV and  3\,TeV stages, the drive-beam scheme will be significantly more efficient and cost effective. At lower
energies the difference is however less pronounced or might even be reversed. 
Therefore the option to power the first energy stage with klystrons has been investigated.
Such an option would offer advantages. The modules with the accelerating structures could be very easily tested at full
specifications in the very same configuration as in the main linac.
In contrast, for the drive-beam option a large facility is required to provide the 100\,A beam that is mandatory to power the main-linac module
prototypes. An important advantage of the drive-beam scheme is that the investment to upgrade the drive-beam complex for the second energy
stage would be relatively small. In contrast, with a klystron-based first stage one would have to invest much more into a new drive-beam complex instead of an upgrade of an operation facility.
Also the klystron scheme requires a second tunnel to house the klystrons, modulators and pulse compressors.

The accelerating structures demand very short pulses with high power. This is not easy to achieve directly with klystrons. Therefore it is foreseen to make use of longer RF pulses from the klystron and to compress them in time before feeding them into the accelerating structures.
The compression increases the power, which allows one to limit the total peak klystron power. However, this leads to some losses of power in the pulse compressors.
Based on this consideration a concept for the basic RF unit of a CLIC energy stage with klystrons has been developed.
It consists of a pair of X-band klystrons, a correction cavities chain, a SLED (SLAC Energy Doubler) pulse compression system~\cite{Farkas:1975az}, and an RF distribution system feeding a number of accelerating structures, see \autoref{f:rfunit}. 

Compared with the current state of the art, the klystrons can profit from recent ideas to increase the efficiency substantially~\cite{c:igor1}. The addition of the correction cavity chain cures power variations along the pulse length. The pulse compression system is based on a novel design, allowing one to reduce the cost with respect to previous solutions~\cite{c:igor2}.

The optimisation procedure for the accelerator parameters has been applied in the same fashion as for the drive-beam
based design, see~\autoref{sec:AcceleratorOptimisation}. Instead of the drive-beam parameters, the number of required RF units has been estimated. The pulse-compressor parameters and
the number of RF units required to feed the main linac are determined using the RF input power and the pulse length in each accelerating structure.
A simplified model is then used to estimate the cost of the accelerator complex. It is based on a specific cost per RF unit and per length of main linac.

Four different parameter sets have been determined and are shown in table~\autoref{t:klystron}. 
First, the cheapest drive beam and klystron-based designs.
\begin{itemize}
\item DB: The cheapest design based on the drive beam.
\item K: The cheapest design based on klystrons.
\end{itemize}
Second, only the structures were considered that can be used with a drive beam in the first stage and that respect the
conditions for an energy upgrade using the drive beam, as discussed above.
In particular the RF pulse length has been fixed to $244\,\text{ns}$.
These structures have the advantage that they could be used either with a drive beam or with klystrons in the first stage. 
So only a single design would need to be developed.
\begin{itemize}
\item DB244: The cheapest design if powered by a drive beam with an RF pulse length fixed to $244\,\text{ns}$.
This is the new parameter set for $380\,\text{GeV}$.
\item K244: The cheapest design if powered by klystrons with an RF pulse length fixed to $244\,\text{ns}$.
\end{itemize}
Each of the four structure designs can be powered either with a drive beam or with klystrons.
For these two options the costs are considered. The difference of these costs compared to the new design (``DB244'', using a drive beam) is given in the table.

The following conclusions can be drawn:
\begin{itemize}
\item The structure design ``DB'' is not consistent with an upgrade using a drive beam. It is only given for comparision.
\item The new structure design and parameter set DB244 is the cheapest drive beam solution that is consistent with the energy upgrade.
It can also be used in a klystron-based design. This would lead to a cost increase of $330\,\text{MCHF}$.
\item The structure ``K244'' can be used with klystrons or drive beam and be upgraded to higher energies similar to ``DB244''.
The cost of the klystron-based option is slightly smaller than in ``DB244'' but the cost for the drive beam-based option increases.
This seems to offer little advantage over option ``DB244''.
\item The structure design ``K'' is the cheapest klystron-based options. In principle,
it could be used with a drive beam in the first energy stage.
However this solution is not consistent with an energy upgrade due to the wrong pulse length. If this structure is powered by klystrons
it is consistent with the energy upgrade since it can accelerate enough bunches with a high enough charge.
This requires to continue to use klystrons for these structures. Using this structure for a klystron-based design reduces the
cost by  about $280\,\text{MCHF}$ compared to the use of the structure ``DB244''.
\end{itemize}
Hence the structure and parameter set ``DB244'' remains the preferred choice for the drive beam-based design. For the klystron-based design
one can consider the same structure at somewhat higher cost. Or one can use a different structure design (``K'') to achieve a lower cost
for the klystron-based design. However, this requires to follow two structure designs for the first stage.

\begin{table}[t!]
\caption{\label{t:klystron}The parameters for the structure designs that are detailed in the text.
}
\centering
\begin{tabular}{l r r r r r r}
\toprule
\textbf{Parameter}                  &\textbf{Symbol}        &\textbf{Unit} &\textbf{DB} &\textbf{K} &\textbf{DB244} &\textbf{K244}\\
\midrule
Frequency                           & $f$                   & GHz          & 12         & 12        & 12            & 12    \\
Acceleration gradient               & $G$                   & MV/m         & 72.5       & 75        & 72            & 79\\
\midrule
RF phase advance per cell           & $\Delta \phi$         & $^{\circ}$   & 120        & 120       & 120           & 120     \\
Number of cells                     & $N_{\text{c}}$        &              & 36         & 28        & 33            & 26\\
First iris radius / RF wavelength   & $a_1/\lambda$         &              & 0.1525     & 0.145     & 0.1625        & 0.15\\
Last iris radius / RF wavelength    & $a_2/\lambda$         &              & 0.0875     & 0.09      & 0.104         & 0.1044\\
First iris thickness / cell length  & $d_1/L_{\text{c}}$    &              & 0.297      & 0.25      & 0.303         & 0.28\\
Last iris thickness / cell length   & $d_2/L_{\text{c}}$    &              & 0.11       & 0.134     & 0.172         & 0.17\\
\midrule
Number of particles per bunch       & $N$                   & $10^9$       & 3.98       & 3.87      & 5.2           & 4.88\\
Number of bunches per train         & $n_{\text{b}}$        &              & 454        & 485       & 352           & 366\\
Pulse length                        & $\tau_{\text{RF}}$    & ns           & 321        & 325       & 244           & 244\\
Peak input power into the structure & $P_{\text{in}}$       & MW           & 50.9       & 42.5      & 59.5          & 54.3\\
\midrule
Cost difference (w.\ drive beam)    & $\Delta C_{\text{w.\ DB}}$ & MCHF   & -50       & (20)    & 0          & (20)\\
Cost difference (w.\ klystrons)     & $\Delta C_{\text{w.\ K}}$  & MCHF   & (120)     & 50      & (330)        & 240\\
\bottomrule
\end{tabular}
\end{table}

The parameters of the RF unit for case K are shown in~\autoref{f:rfunit}. Each of the klystrons produces a 68\,MW RF pulse of 1.625\,$\upmu$s duration,
which is then compressed to a length of 325\,ns to match the required RF pulse length.
This allows the klystrons to operate at longer pulse lengths, which yields a better efficiency.
The compression increases the peak RF power from 68\,MW to 213\,MW per klystron, which reduces the number of klystrons required.
However, approximately 40\% of the power delivered by the klystrons is lost in the compressors, thereby limiting the efficiency.

In conclusion, at 380\,GeV the cost of the klystron-based design is expected to be comparable to the cost of the drive-beam based solution. It is
compatible with an energy upgrade to 3\,TeV based on a drive-beam scheme for the additional stages.
One might therefore consider it as an alternative for the first energy stage. Hence,
a more detailed conceptual design of the klystron-based option will be developed.
This will allow to derive a more robust cost model for this design, leading to a more accurate comparison with the drive-beam based design.

\section{Summary and outlook}
\label{sec:summary}

In this document an updated staging baseline for the CLIC multi-TeV linear $\Pep\Pem$ collider has been described. To achieve optimal luminosity and physics performance, CLIC is foreseen to be built and operated in three centre-of-mass energy stages. Based on the current physics landscape, an optimal choice for these energy stages is defined. Following a comprehensive set of physics benchmark studies, addressing in particular Higgs and top-quark physics, a centre-of-mass energy of 380\,GeV is chosen for the first energy stage with a luminosity goal of $1.5\times10^{34}\,\text{cm}^{-2}\,\text{s}^{-1}$. At this stage, some $15\%$ of the running time will be devoted to a threshold scan of $\PQt\PAQt$ production near 350\,GeV. The second and third stages are proposed at 1.5\,TeV and 3\,TeV, where the second stage corresponds to the maximum centre-of-mass energy reach with a single drive-beam complex. The envisaged integrated luminosities at the three stages are (500+100)\,\fbinv, 1.5\,\abinv and 3\,\abinv, respectively.

In the CLIC Conceptual Design Report, published in 2012, the feasibility of the accelerator technology was demonstrated, based on design studies, component development, extensive hardware tests and full system tests. These included successful two-beam accelerator tests at the CLIC test facility at CERN~(CTF3), reaching acceleration gradients well above 100\,MV/m. The studies towards the CDR focused on a fully optimised 3\,TeV collider, while the lower energy stages were not yet optimised at the same level. The CDR studies have led to a detailed understanding of the CLIC accelerator. As a result simulation tools have been established, allowing to derive performance parameters such as luminosity, cost and power as a function of a comprehensive set of component parameters.

The document describes how these simulation tools have been used to optimise CLIC for the three proposed energy stages. Special attention is given to the optimisation of the 380\,GeV stage in a way that prepares for a cost-effective upgrade path to the higher-energy stages and maximises the re-use of equipment. In this scenario the RF pulse length is the same (244\,ns) at all stages. The optimisation studies show that an RF structure design which is very similar to the one used in recent high-gradient test campaigns can be maintained for the upgrades to the higher energy stages. These structures will operate at 100\,MV/m, while the structure proposed for the 380\,GeV stage (and for the beginning of the linac at the higher energy stages) will operate at 72\,MV/m.

The simulation tools provide estimates of cost and power levels for those elements of the complex that are directly addressed in the optimisation. These include major elements in the drive beam generation complex, the drive beam frequency multiplication, the drive beam distribution system, the main beam injector complex and the main linac. Together with non-variable cost and power components already presented in the CDR (e.g. for the beam delivery system and the experimental area), estimates of the cost, power and energy consumption for the 380\,GeV stage are derived. The estimated power, determined in this way, amounts to 252\,MW for operation at 380\,GeV centre-of-mass energy, while the scaled cost of the first stage is estimated at 6.7\,BCHF.

A klystron-based option is presented for the first stage at 380\,GeV. Profiting from ongoing developments towards more efficient modulators and klystrons, a klystron-based design is expected to become competitive in cost for the first stage, as compared to the drive-beam option. 
It will nevertheless be compatible with a drive-beam scheme for the higher energy CLIC stages. As a potentially cheaper option, the klystron scheme will be subject for further studies in the coming years.

Physics performance studies have shown that the proposed staged CLIC facility offers a large scope of precision physics opportunities: 
\begin{itemize}
\item	The first stage at $\sqrt{s}=380$\,GeV provides accurate measurements of the properties of the Higgs boson, in particular the $g_{\PH\PZ\PZ}$ coupling (to 0.8\%) and couplings to other major decay channels, the Higgs mass (at the $\sim$100\,MeV level) and furthermore the top-quark mass (at the $\sim$50\,MeV level) and kinematic properties of $\PQt\PAQt$ production. 
\item	Subsequent high-energy running at $\sqrt{s}=1.5$\,TeV and $\sqrt{s}=3$\,TeV allows to accumulate large samples of Higgs boson decays, providing a range of Higgs boson couplings at the ${\cal{O}}(1\%)$ level.
\item	High-energy CLIC operation allows for the measurement of the top Yukawa coupling at the level of 4\% and the Higgs trilinear self-coupling parameter at the 10\% level. 
\item	CLIC will enable the discovery of new particles through direct detection, typically up to the 1.5\,TeV kinematic limit in the case of pair production; it will allow to measure their masses with an accuracy of ${\cal{O}}(1\%)$ and their production and decay modes, as well as their quantum numbers, to high accuracy. 
\item	CLIC offers indirect sensitivity to BSM physics through precision measurements, where, for example, $\PZpr$ and Higgs compositeness models can be probed up to scales of approximately 40\,TeV and 70\,TeV, respectively.
\end{itemize}
The full physics programme covering the three energy stages would span a period of 22 years, with 5 to 7 years of running at each of the stages, including luminosity ramp-up in the earlier years and 2-year upgrade periods in between the stages.\\
 
In preparation for the next update of the European strategy for particle physics in 2019--2020, the CLIC accelerator project aims at presenting a Project Implementation Plan. 
This document will include a final description of the accelerator and detector parameters, cost and power estimates, site studies, the staging scenario and the construction schedule. 

The activities foreseen between now and 2019 focus on R\&D areas that are most critical for such a project plan:
\begin{itemize}
\item	X-band structure development, optimisation and testing;
\item	Integrated design for the initial configuration and the subsequent stages;
\item	Implementation planning including cost and power optimisation studies, for example of accelerating structures, magnets and RF systems;
\item	System tests at CERN and other facilities, for example at light-sources, ATF2 and FACET; 
\item	Technical developments addressing as a priority those parts that have high relevance for the cost, power or system performances. 
\end{itemize}

The studies described above are carried out in a collaborative effort involving the CLIC/CTF3 institutes, in line with the available expertise and resources within the collaboration.

The associated detector and physics studies are pursued by the CLICdp collaboration. The CLICdp studies for the next European Strategy Update focus principally on:
\begin{itemize}
\item	Studies of the CLIC physics potential, closely following the evolution of LHC results and with particular emphasis on Higgs physics, top-quark physics and BSM phenomena;
\item Detector optimisation studies driven by the physics aims;
\item Technology demonstrators for the most challenging detector elements (the vertex detector, the silicon tracker and the fine-grained calorimetry).
\end{itemize}

\begin{figure}[t]
  \centering
\includegraphics[width=0.9\columnwidth]{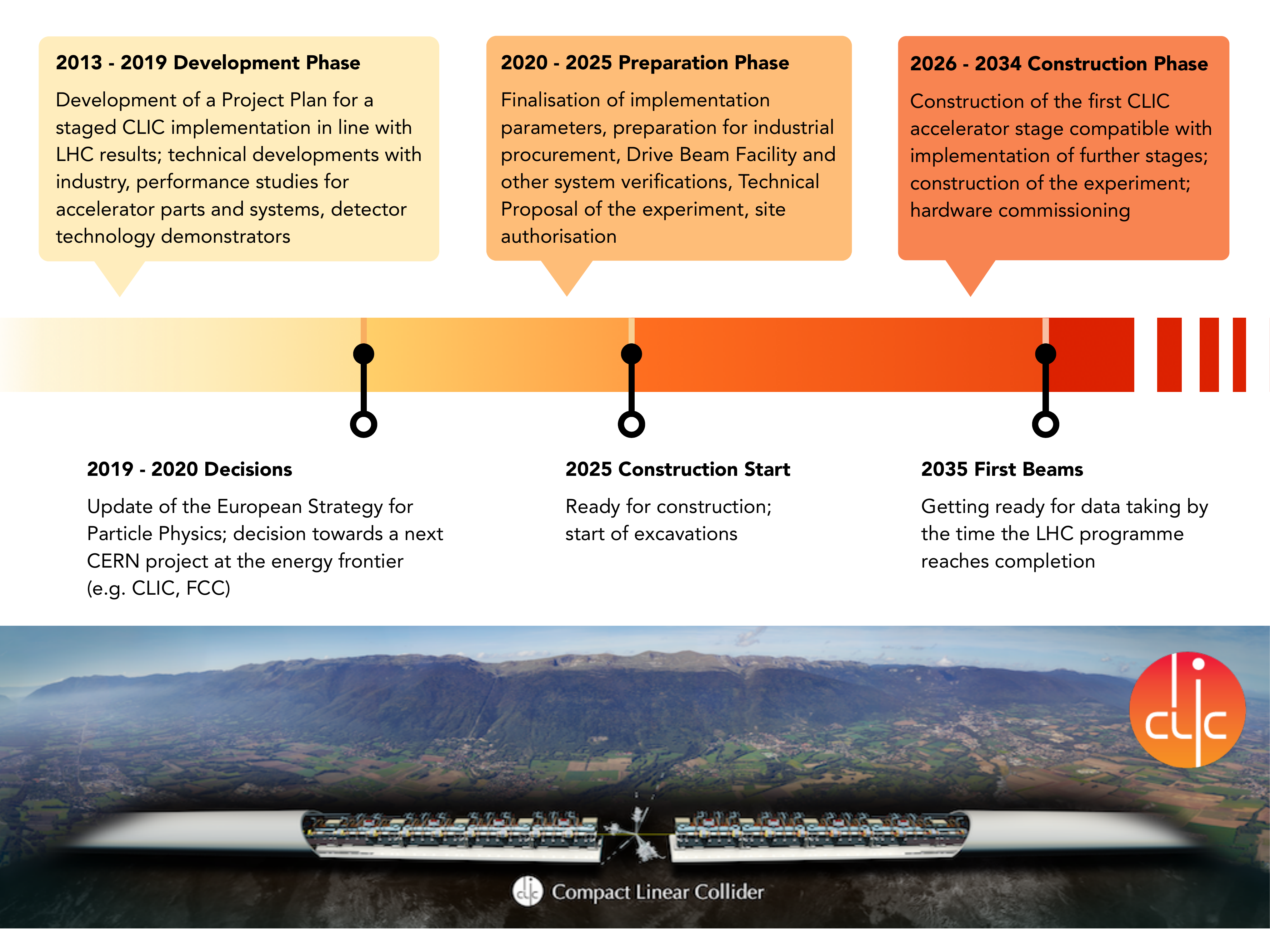}
 \caption{An outline of the CLIC project timeline from the current development phase up to future first beams at the 380\,\gev CLIC energy stage.
\label{fig:CLIC:strategy}}
\end{figure}

By the next European Strategy Update, the LHC physics results and the technical developments for future accelerators at the energy frontier, principally electron-positron and proton-proton options, are expected to have reached sufficient maturity to allow selection of the most appropriate future collider option with physics capabilities complementing the LHC.

If CLIC is chosen the project implementation will require an initial Preparation Phase of $\sim$5 years prior to construction start by 2025. This phase will focus on industrial build-up, larger-scale system verifications, risk and cost reduction, as well as developments towards the Technical Proposal for the detector. The governance structure and the international collaboration agreements for the construction and operation will be set up during this time. Site authorisations will also be established during this period. Preliminary site studies show that CLIC can be implemented underground near CERN, with the central main beam and drive beam injector complexes on the CERN campus. 

The construction of the first CLIC energy stage could start around 2025. As illustrated in \autoref{fig:CLIC:strategy}, CLIC would be ready to produce first beams by 2035, by the time LHC operation approaches completion. Construction of further stages could be launched following 2-3 years of operation at the first stage, drawing on all experiences acquired at that time.

\clearpage
\section*{Acknowledgements}
This work was supported by 
Comisi\'{o}n Nacional de Investigaci\'{o}n Cient\'{\i}fica y Tecnol\'{o}gica (CONICYT), Chile;
the Ministry of Education, Youth and Sports, Czech Republic, under Grant INGO II-LG 14033;
the ``Research internationalization'' program of the European Regional Development Fund and Estonian Research Council Grant PUT 57;
the DFG cluster of excellence ``Origin and Structure of the Universe'' of Germany;
the German-Israel Foundation (GIF), the Israel Science Foundation (ISF), the I-CORE programme of VATAT and ISF and the European Union's Horizon 2020 Research and Innovation programme under Grant Agreement no.654168;
the Research Council of Norway;
the Polish Ministry of Science and Higher Education under contract nr 3501/H2020/2016/2;
the National Science  Centre, Poland, HARMONIA project, under contract UMO-2015/18/M/ST2/00518;
the EC HIGGSTOOLS project, under contract PITN-GA-2012-316704;
the Romanian agencies UEFISCDI and ROSA;
the Ministry of Education, Science, and Technological Development of the Republic of Serbia, through the national project OI171012;
the Secretar\'{\i}a de Estado de Investigaci\'{o}n, Desarrollo e Innovaci\'{o}n and Programa Consolider-Ingenio 2010, Spain;
the Swedish Research Council under contract 2014-6360;
the UK Science and Technology Facilities Council (STFC);
the U.S.\ Department of Energy contract DE-AC02-06CH11357.
This work benefited from services provided by the ILC Virtual Organisation, supported by the national resource providers of the EGI Federation.

% references
\clearpage
\printbibliography[title=References]

\end{document}